\documentclass[onecolumn,showpacs,preprintnumbers]{revtex4}
\usepackage{mathrsfs}
\usepackage{graphicx}
\usepackage{dcolumn}
\usepackage{bm}
\usepackage{epsfig}
\usepackage{amsmath,amssymb}
\usepackage{multirow}
\usepackage{placeins}
\usepackage{caption}
\usepackage{array}
\usepackage{float}
\usepackage{subfigure}

\newcommand{\PreserveBackslash}[1]{\let\temp=\\#1\let\\=\temp}

\newcolumntype{C}[1]{>{\PreserveBackslash\centering}p{#1}}

\newcolumntype{R}[1]{>{\PreserveBackslash\raggedleft}p{#1}}

\newcolumntype{L}[1]{>{\PreserveBackslash\raggedright}p{#1}}

\begin{document}
\title{Systematic analysis of single heavy baryons $\Lambda_{Q}$, $\Sigma_{Q}$ and $\Omega_{Q}$}
\author{Guo-Liang Yu$^{1}$}
\email{yuguoliang2011@163.com}
\author{Zhen-Yu Li$^{2}$}
\author{Zhi-Gang Wang$^{1}$}
\email{zgwang@aliyun.com}
\author{Lu Jie$^{1}$}
\author{Yan Meng$^{1}$}

\affiliation{$^1$ Department of Mathematics and Physics, North China
Electric Power University, Baoding 071003, People's Republic of
China\\$^2$ School of Physics and Electronic Science, Guizhou Education University, Guiyang 550018, People's Republic of
China}
\date{\today }

\begin{abstract}
Motivated by great progresses in experiments in searching for the heavy baryons, we systematically analyze the mass spectra and root mean square radius of single heavy baryons $\Lambda_{Q}$, $\Sigma_{Q}$ and $\Omega_{Q}$. The calculations of the mass spectra are carried out in the frame work of relativized quark model, where the baryon is regarded as a three-body system of quarks. Our results show that the mass of single heavy baryon with $\lambda$-mode is lower than those of the $\rho$-mode and $\lambda$-$\rho$ mixing mode, which indicates that the lowest state is dominated by the $\lambda$-mode. Basing on this research, we systematically calculate the mass spectra and the root mean square radius of the baryons with $\lambda$ excited mode. With these predicated mass spectra, the Regge trajectories in the ($J$,$M^{2}$) plane are constructed, and the slopes, intercepts of the Regge trajectories are obtained by linear fitting. It is found that all available experimental data are well reproduced by model predictions and fit nicely to the constructed Regge trajectories.
\end{abstract}


\maketitle

\begin{Large}
\textbf{1 Introduction}
\end{Large}

In the field of heavy baryon physics, scientists have made great progresses in experiments as well as in theories, which makes the mass spectra of heavy baryon families become more and more abundant. In the past few decades, almost all of the 1$S$
single heavy baryons have been well established and some of the 1$P$ states $\Lambda_{c}(2595)$\cite{2595}, $\Lambda_{c}(2625)$\cite{2625}, $\Lambda_{b}(5912)$ and $\Lambda_{b}(5920)$\cite{59121,59122} have also been well observed in experiments and been confirmed in theory\cite{article2A,article2B}.
Besides these states, many other single heavy baryons have also been discovered by Belle, BABAR, CLEO and LHCb, such as $\Lambda_{c}(2765)$\cite{LambdaC2765}, $\Lambda_{c}(2940)$\cite{LambdaC29401,LambdaC29402,LambdaC29403}, $\Lambda_{b}(6072)$\cite{LambdaB60721}, $\Lambda_{b}(6146)$\cite{LambdaB6146}, $\Lambda_{b}(6152)$\cite{LambdaB6146}, $\Sigma_{c}(2800)$\cite{2800}, $\Sigma_{b}(6097)$\cite{6097}, $\Omega_{c}(3000)$, $\Omega_{c}(3050)$, $\Omega_{c}(3065)$, $\Omega_{c}(3090)$, $\Omega_{c}(3119)$\cite{OmegaC3000}, $\Omega_{b}(6330)$, $\Omega_{b}(6316)$, $\Omega_{b}(6350)$ and $\Omega_{b}(6340)$\cite{OmegaB6316}. Some of these baryons may belong to the low-lying $S$- or $P$-wave states, while some of them can be assigned to $D$-wave baryons and all of these assignments needs further confirmation in more ways. In order to identify their quantum numbers and assign each baryon a suitable position in the mass spectra, it is necessary to systematically investigate single heavy baryon spectroscopy.

In the past decades, the single heavy baryons have been extensively investigated by many theoretical methods/models, including various quark model\cite{GI,quam1,quam2,quam3,quam4,quam5,quam6,quam7,quam8,quam9,quam10,quam11,quam12,quam13,quam14,quam15,quam16,quam17,quam18,quam19,quam20,quam21,quam22,quam23,quam24,quam25,quam26,quam27}, the heavy hadron chiral perturbation theory\cite{chiral1,chiral2,chiral3,chiral4,chiral5,chiral6}, lattice QCD\cite{Lattice1,Lattice2,Lattice3,Lattice4}, light cone QCD sum rules\cite{LCsum1,LCsum2,LCsum3,LCsum4,LCsum5,LCsum6,LCsum7,LCsum8}, QCD sum rules\cite{Sum1,Sum2,Sum3,Sum4,Sum5,Sum6,Sum7,Sum8,Sum9,Sum10,Sum11,Sum12,Sum13,WZG2,WZG3,WZG5,WZG1,WZG4,WZG8,GLY1} and relativistic flux tube model\cite{Fluxtube}. To our knowledge, only Ref.\cite{quam2} focused on the mass spectra of the heavy baryons from ground states to the high excited states systematically in the quark-diquark picture. In this literature, heavy baryons are considered in the heavy-quark-light-diquark approximation, which can simplify the complicated relativistic three-body problem. Under this quark-diquark picture, the initial three-body problem is reduced to two-step two-body calculations. Another important work was carried out by Roberts and Pervin\cite{quam3}. With the non-relativized quark model, they studied various excited states of single heavy baryons by solving three quark systems explicitly. On the other hand, a Gaussian expansion method(GEM) was introduced in the non-relativized quark model to solve the three-quark system in Ref.\cite{quam4}. In the calculations of the matrix elements of the Hamiltonian of three-body system, particularly when complicated interactions are employed, integrations over all of the radial and angular coordinates become laborious even with GEM. In Ref.\cite{quam4}, they developed an infinitesimally-shifted Gaussian(ISG) basis function to simplify this process.

In this work, we extend the method of ISG basis function to the relativized quark model to study the mass spectra of the single heavy baryons. The relativized quark model was first developed by Godfrey and Isgur to study the
mass spectra of mesons\cite{GI}, and later was extended by Capstick and Isgur to investigate baryon spectra\cite{quam1}. Recently, this quark model has also been used to investigate the mass of tetraquark states\cite{LV1,LV2,LV3}. It was found that GEM is one of the best method for three and four body bound states\cite{quam4}, which can help us to obtain precise energy eigenvalues of excited states. Especially, taking ISG function as the basis function has the advantage of simplifying the calculation of matrix elements\cite{quam4}. The first goal of our present work is to calculate the mass spectra and root mean square radius of the excited heavy baryons up to rather high orbital and radial excitations. With these results, we will construct the heavy baryon Regge trajectories in the ($J$,$M^{2}$) planes and determine their Regge slopes and intercepts.

The paper is organized as follows. In Section II, we present the the relativized quark model of heavy baryons based on the method of ISG basis functions. With this method, we systematically investigate the mass spectra and root mean square radius of single heavy baryons $\Lambda_{Q}$, $\Sigma_{Q}$ and $\Omega_{Q}$. In Sec III the heavy baryon Regge trajectories in the ($J$, $M^{2}$) plane are constructed with the predicted mass spectra, and slopes, intercepts of parent and daughter trajectories are also obtained by linear fitting. And Sec IV is reserved for our conclusions.

\begin{Large}
\textbf{2 Phenomenological method adopted in this work}
\end{Large}

\begin{large}
\textbf{2.1 The Jacobi coordinate and relativized quark model}
\end{large}

In this work, the single heavy baryon is regarded as a three-body system which has two light quarks($u$, $d$ or $s$ quarks) and one heavy quark($c$ or $b$ quark) inside. In order to express internal motions of the quarks in this three-body system, we commonly introduce three sets of Jacobi coordinates as in Fig.1. Each set of Jacobi coordinate is called a channel($c$) which is defined as,
\begin{figure}[h]
\centering
\includegraphics[height=4.5cm,width=16cm]{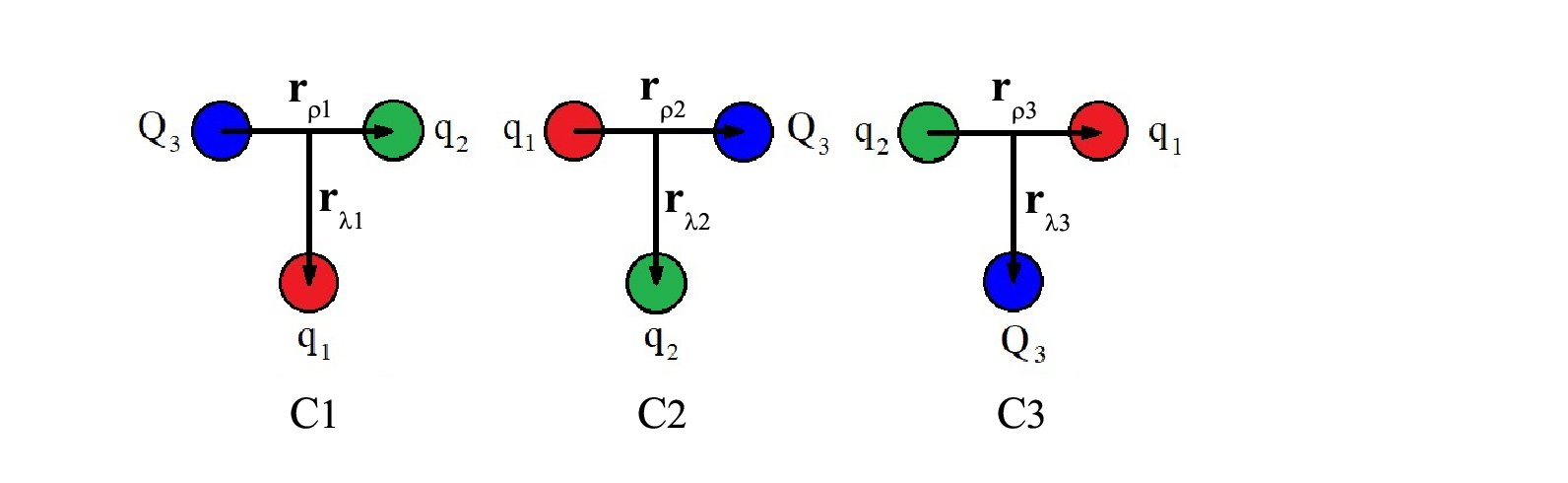}
\caption{ Jacobi coordinates for the three body system.}
\end{figure}
\begin{eqnarray}
& \boldsymbol{r}_{\lambda}=\boldsymbol{r}_{k}-\frac{m_{i}\boldsymbol{r}_{i}+m_{j}\boldsymbol{r}_{j}}{m_{i}+m_{j}} & \\
& \boldsymbol{r}_{\rho}=\boldsymbol{r}_{i}-\boldsymbol{r}_{j}&
\end{eqnarray}
\begin{table*}[h]
\begin{tabular}{m{1.0cm}<{\centering}m{1.0cm}<{\centering}m{1.0cm}<{\centering}m{1.0cm}<{\centering}}
\hline
Channel&i&j&k\\
\hline\hline
$c=1$&2&3&1\\
\hline
$c=2$&3&1&2\\
\hline
$c=3$&1&2&3\\
\hline
\end{tabular}\caption{The quark assignments ($i$,$j$,$k$) for the Jacobi coordinates}
\end{table*}
where assignments of ($i$,$j$,$k$) are given in Table I. The Jacobi coordinate of $c=3$ can be expressed in terms of the coordinate $c=1$ or $c=2$,
\begin{eqnarray}
&\boldsymbol{r}_{\rho_{3}}=\alpha_{31(32)}^{r}\boldsymbol{r}_{\rho_{1(2)}}+\beta_{31(32)}^{r}\boldsymbol{r}_{\lambda_{1(2)}} \\
&\boldsymbol{r}_{\lambda_{3}}=\gamma_{31(32)}^{r}\boldsymbol{r}_{\rho_{1(2)}}+\delta_{31(32)}^{r}\boldsymbol{r}_{\lambda1(2)}
\end{eqnarray}
where the transforming coefficients for $c:3\rightarrow 1$ and $3\rightarrow 2$ are,

$\alpha_{31}^{r}=-\frac{m_{3}}{m_{2}+m_{3}}$,$\beta_{31}^{r}=1$,$\gamma_{31}^{r}=-\frac{m_{2}(m_{1}+m_{2}+m_{3})}{(m_{1}+m_{2})(m_{2}+m_{3})}$,$\delta_{31}^{r}=-\frac{m_{1}}{m_{1}+m_{2}}$,

$\alpha_{32}^{r}=-\frac{m_{3}}{m_{1}+m_{3}}$,$\beta_{32}^{r}=-1$,$\gamma_{32}^{r}=\frac{m_{1}(m_{1}+m_{2}+m_{3})}{(m_{1}+m_{2})(m_{1}+m_{3})}$,$\delta_{32}^{r}=-\frac{m_{2}}{m_{1}+m_{2}}$.

In the following, we give a brief introduction to the Hamiltonian of relativized quark model. The relativistic Hamiltonian for a three-body system can be written as\cite{GI,quam1},
\begin{eqnarray}
\widehat{H}=\sum_{i=1}^{3}(p_{i}^{2}+m_{i}^{2})^{1/2}+\sum_{i<j}\widetilde{H}_{ij}^{\mathrm{conf}}+\sum_{i<j}\widetilde{H}_{ij}^{\mathrm{hyp}}+\sum_{i<j}\widetilde{H}_{ij}^{\mathrm{so}}
\end{eqnarray}
where the first term is the relativistic kinetic energy term, $\widetilde{H}^{\mathrm{conf}}$ is the spin-independent potential which contains a linear confining potential $\widetilde{S}(r_{ij})$ and the one-gluon
exchange potential $G^{\prime}(r_{ij})$,
\begin{eqnarray}
\widetilde{H}^{\mathrm{conf}}_{ij}=\widetilde{S}(r_{ij})+G^{\prime}(r_{ij})
\end{eqnarray}
with
\begin{eqnarray}
\widetilde{S}(r_{ij})=-\frac{3}{4}\textbf{\emph{F}}_{i}\cdot\textbf{\emph{F}}_{j}\Big[b r_{ij}\big[\frac{e^{-\sigma_{ij}^{2}r_{ij}^{2}}}{\sqrt{\pi}\sigma_{ij} r_{ij}}+\big(1+\frac{1}{2\sigma_{ij}^{2}r_{ij}^{2}}\big)\frac{2}{\sqrt{\pi}} \int^{\sigma_{ij} r_{ij}}_{0}e^{-x^{2}}dx\big]+c\Big]
\end{eqnarray}
\begin{eqnarray}
\sigma_{ij}=\sqrt{s^{2}\Big[\frac{2m_{i}m_{j}}{m_{i}+m_{j}}\Big]^{2}+\sigma_{0}^{2}\Big[\frac{1}{2}\big(\frac{4m_{i}m_{j}}{(m_{i}+m_{j})^{2}}\big)^{4}+\frac{1}{2}\Big]}
\end{eqnarray}
The $\textbf{\emph{F}}_{i}\cdot\textbf{\emph{F}}_{j}$ stands for the color matrix and $F_{n}$ reads
\begin{equation}
F_{n}=\left\{
      \begin{array}{l}
       \frac{\lambda_{n}}{2} \quad \mathrm{for} \, \mathrm{quarks}, \\
        -\frac{\lambda_{n}^{*}}{2} \quad    \mathrm{for} \, \mathrm{antiquarks} \\
      \end{array}
      \right.
\end{equation}
with $n=1,2\cdots8$. It should be noticed that the one-gluon exchange potential $G^{\prime}(r_{ij})$ is achieved by introducing momentum-dependent factors,
\begin{eqnarray}
G^{\prime}(r_{ij})=\Big(1+\frac{p^{2}_{ij}}{E_{i}E_{j}}\Big)^{\frac{1}{2}}\widetilde{G}(r_{ij})\Big(1+\frac{p^{2}_{ij}}{E_{i}E_{j}}\Big)^{\frac{1}{2}}
\end{eqnarray}
with
\begin{eqnarray}
\widetilde{G}(r_{ij})=\textbf{\emph{F}}_{i}\cdot\textbf{\emph{F}}_{j}\mathop{\sum}\limits_{k=1}^{3}\frac{2\alpha_{k}}{3\sqrt{\pi}r_{ij}}\int^{\tau_{k}r_{ij}}_{0}e^{-x^{2}}dx
\end{eqnarray}

and $\tau_{k}=\frac{1}{\sqrt{\frac{1}{\sigma_{ij}^{2}}+\frac{1}{\gamma_{k}^{2}}}}$.

In, Eq.(5), $\widetilde{H}^{\mathrm{hyp}}$ is the color-hyperfine interaction which includes  tensor interaction and the contact interaction,
\begin{eqnarray}
\widetilde{H}^{\mathrm{hyp}}_{ij}=\widetilde{H}^{\mathrm{tensor}}_{ij}+\widetilde{H}_{ij}^{\mathrm{c}}
\end{eqnarray}
with
\begin{eqnarray}
\widetilde{H}^{\mathrm{tensor}}_{ij}=-\Big(\frac{\textbf{S}_{i}\cdot \textbf{r}_{ij}\textbf{S}_{j}\cdot \textbf{r}_{ij}/r_{ij}^{2}-\frac{1}{3}\textbf{S}_{i}\cdot\textbf{S}_{j}}{m_{i}m_{j}}\Big)\times\Big(\frac{\partial^{2}}{\partial r_{ij}^{2}}-\frac{1}{r_{ij}}\frac{\partial}{\partial r_{ij}}\Big)\widetilde{G}_{ij}^{\mathrm{t}},
\end{eqnarray}
\begin{eqnarray}
\widetilde{H}^{\mathrm{c}}_{ij}=\frac{2\textbf{S}_{i}\cdot\textbf{S}_{j}}{3m_{i}m_{j}}\bigtriangledown^{2}\widetilde{G}_{ij}^{\mathrm{c}}
\end{eqnarray}
For the spin-orbit interaction, it can be divided into two parts which can be written as,
\begin{eqnarray}
\widetilde{H}^{\mathrm{so}}_{ij}=\widetilde{H}^{\mathrm{so(v)}}_{ij}+\widetilde{H}_{ij}^{\mathrm{so(s)}},
\end{eqnarray}
with
\begin{eqnarray}
\widetilde{H}^{\mathrm{so(v)}}_{ij}=\frac{\textbf{S}_{i}\cdot \textbf{L}_{ij}}{2m_{i}^{2}r_{ij}}\frac{\partial \widetilde{G}^{\mathrm{so(v)}}_{ii}}{\partial r_{ij}}+\frac{\textbf{S}_{j}\cdot \textbf{L}_{ij}}{2m_{j}^{2}r_{ij}}\frac{\partial \widetilde{G}^{\mathrm{so(v)}}_{jj}}{\partial r_{ij}}+\frac{(\textbf{S}_{i}+\textbf{S}_{j})\cdot \textbf{L}_{ij}}{m_{i}m_{j}r_{ij}}\frac{1}{r_{ij}}\frac{\partial \widetilde{G}^{\mathrm{so(v)}}_{ij}}{\partial r_{ij}}
\end{eqnarray}
and
\begin{eqnarray}
\widetilde{H}^{\mathrm{so(s)}}_{ij}=-\frac{\textbf{S}_{i}\cdot \textbf{L}_{ij}}{2m_{i}^{2}r_{ij}}\frac{\partial \widetilde{S}^{\mathrm{so(s)}}_{ii}}{\partial r_{ij}}-\frac{\textbf{S}_{j}\cdot \textbf{L}_{ij}}{2m_{j}^{2}r_{ij}}\frac{\partial \widetilde{S}^{\mathrm{so(s)}}_{jj}}{\partial r_{ij}}
\end{eqnarray}
In Eqs.(13),(14),(16) and (17), $\widetilde{G}^{\mathrm{t}}_{ij}$, $\widetilde{G}^{\mathrm{c}}_{ij}$, $\widetilde{G}^{\mathrm{so(v)}}_{ij}$ and $\widetilde{S}^{\mathrm{so(s)}}_{ii}$ are achieved from $\widetilde{G}(r_{ij})$ and $\widetilde{S}(r_{ij})$ by introducing momentum-dependent factors,
\begin{eqnarray}
G^{\mathrm{t}}_{ij}=\Big(\frac{m_{i}m_{j}}{E_{i}E_{j}}\Big)^{\frac{1}{2}+\epsilon_{\mathrm{t}}}\widetilde{G}(r_{ij})\Big(\frac{m_{i}m_{j}}{E_{i}E_{j}}\Big)^{\frac{1}{2}+\epsilon_{\mathrm{t}}}
\end{eqnarray}
\begin{eqnarray}
G^{\mathrm{c}}_{ij}=\Big(\frac{m_{i}m_{j}}{E_{i}E_{j}}\Big)^{\frac{1}{2}+\epsilon_{\mathrm{c}}}\widetilde{G}(r_{ij})\Big(\frac{m_{i}m_{j}}{E_{i}E_{j}}\Big)^{\frac{1}{2}+\epsilon_{\mathrm{c}}}
\end{eqnarray}
\begin{eqnarray}
G^{\mathrm{so(v)}}_{ij}=\Big(\frac{m_{i}m_{j}}{E_{i}E_{j}}\Big)^{\frac{1}{2}+\epsilon_{\mathrm{so(v)}}}\widetilde{G}(r_{ij})\Big(\frac{m_{i}m_{j}}{E_{i}E_{j}}\Big)^{\frac{1}{2}+\epsilon_{\mathrm{so(v)}}}
\end{eqnarray}
\begin{eqnarray}
\widetilde{S}^{\mathrm{so(s)}}_{ii}=\Big(\frac{m_{i}^{2}}{E_{i}^{2}}\Big)^{\frac{1}{2}+\epsilon_{\mathrm{so(s)}}}\widetilde{S}(r_{ij})\Big(\frac{m_{i}^{2}}{E_{i}^{2}}\Big)^{\frac{1}{2}+\epsilon_{\mathrm{so(s)}}}
\end{eqnarray}
with $E_{i}=\sqrt{m_{i}^{2}+p_{ij}^{2}}$, and $\epsilon_{\mathrm{t}}$, $\epsilon_{\mathrm{c}}$, $\epsilon_{\mathrm{so(v)}}$ and $\epsilon_{\mathrm{so(s)}}$ are free parameters that are discussed in Table II. The $p_{ij}$ is the magnitude of the momentum of either of the quarks in the $ij$ center-of-mass frame and these relative momentums $p_{23}$, $p_{13}$ and $p_{12}$ correspond to $p_{\rho_{1}}$, $p_{\rho_{2}}$ and
$p_{\rho_{3}}$ in three channels.

\begin{large}
\textbf{2.2 Wave function of single heavy baryon}
\end{large}

In quark model scheme, the full wave function for a definite baryon is built from a linear superposition of the channel wave function, which can be written as,
\begin{eqnarray}
\Psi_{full}^{JM}=\sum_{c\alpha}C_{c,\alpha}\Psi_{JM}^{c}(\boldsymbol{r}_{\rho_{c}},\boldsymbol{r}_{\lambda_{c}})
\end{eqnarray}
where the index $\alpha$ denotes $n_{\rho}$,$n_{\lambda}$,$l_{\rho}$,$l_{\lambda}$,$L$,$s$,$j$, and $c$ denotes the Jacobi coordinate channel with $c=1,2,3$ as illustrated in Fig.1.
$\Psi_{JM}^{c}(\boldsymbol{r}_{\rho_{c}},\boldsymbol{r}_{\lambda_{c}})$ is the wave function for channel $c$ and it is described in terms of antisymmetric color-singlet wave function $\phi_{\mathrm{color}}$, multiplying flavor $\phi_{\mathrm{flavor}}$, spin and space wave function $\psi_{JM}^{c}$
\begin{eqnarray}
\Psi_{JM}^{c}(\boldsymbol{r}_{\rho_{c}},\boldsymbol{r}_{\lambda_{c}})=\phi_{\mathrm{color}}\otimes \phi_{\mathrm{flavor}}\otimes\psi_{JM}^{c}
\end{eqnarray}
with
\begin{eqnarray}
\phi_{\mathrm{color}}=\frac{1}{\sqrt{6}}\big(rgb-rbg+gbr-grb+brg-bgr\big),
\end{eqnarray}
\begin{eqnarray}
\psi_{JM}^{c}=\Big[\big[[\chi_{1/2}(q)\chi_{1/2}(q)]_{s,m_{s}}\Phi^{c}_{l_{\rho},l_{\lambda},L}\big]_{j,m_{j}}\chi_{1/2}(Q)\Big]_{JM},
\end{eqnarray}
where $r$, $g$, $b$ are the color of the quark and $\chi_{1/2}$ is the spin wave function.
Within the flavor SU(3) subgroups, the single heavy baryons belong either to a sextet($6_{F}$) of flavor symmetric states, or an
antitriplet($\overline{3}_{F}$) of flavor antisymmetric states. For the $\Sigma_{Q}$ and $\Omega_{Q}$ baryons, they belong to flavor-symmetric sextet $\phi_{6_{F}}$, while $\Lambda_{Q}$ baryon is flavor-antisymmetric antitriplet $\phi_{\overline{3}_{F}}$. In Eq.(25), $\Phi^{c}_{l_{\rho},l_{\lambda},L}$ is the spatial wave function which is constructed from the wave functions of the two Jacobi coordinates $\rho$ and $\lambda$, and takes the form
\begin{eqnarray}
\Phi^{c}_{l_{\rho},l_{\lambda},L}=\Big[\phi_{l_{\rho}m_{l_{\rho}}}(\boldsymbol{r}_{\rho_{c}})\phi_{l_{\lambda}m_{l_{\lambda}}}(\boldsymbol{r}_{\lambda_{c}})\Big]_{L} \quad c=1,2,3
\end{eqnarray}
In this work, the spatial wave function of a three-body system is expanded in terms of a set of Gaussian basis functions. Thus, $\phi_{l_{\rho}m_{l_{\rho}}}(\boldsymbol{r}_{\rho_{c}})$ and $\phi_{l_{\lambda}m_{l_{\lambda}}}(\boldsymbol{r}_{\lambda_{c}})$ in Eq.(26) are written as,
\begin{eqnarray}
\phi_{n_{\rho}l_{\rho}m_{l_{\rho}}}(\boldsymbol{r}_{\rho_{c}})=N_{n_{\rho}l_{\rho}}r_{\rho_{c}}^{l_{\rho}}e^{-\nu_{n_{\rho}}r_{\rho_{c}}^{2}}Y_{l_{\rho}m_{l_{\rho}}}(\hat{\boldsymbol{r}}_{\rho_{c}})
\end{eqnarray}
\begin{eqnarray}
\phi_{n_{\lambda}l_{\lambda}m_{l_{\lambda}}}(\boldsymbol{r}_{\lambda_{c}})=N_{n_{\lambda}l_{\lambda}}r_{\lambda_{c}}^{l_{\lambda}}e^{-\nu_{n_{\lambda}}r_{\lambda_{c}}^{2}}Y_{l_{\lambda}m_{l_{\lambda}}}(\hat{\boldsymbol{r}}_{\lambda_{c}})
\end{eqnarray}
with
\begin{eqnarray}
N_{n_{\rho}l_{\rho}}=\sqrt{\frac{2^{l_{\rho}+2}(2\nu_{n_{\rho}})^{l_{\rho}+3/2}}{\sqrt{\pi}(2l_{\rho}+1)!!}}
\end{eqnarray}
\begin{eqnarray}
N_{n_{\lambda}l_{\lambda}}=\sqrt{\frac{2^{l_{\lambda}+2}(2\nu_{n_{\lambda}})^{l_{\lambda}+3/2}}{\sqrt{\pi}(2l_{\lambda}+1)!!}}
\end{eqnarray}
\begin{eqnarray}
\nu_{n_{\rho}}=\frac{1}{r_{n_{\rho}}^{2}}, \quad r_{n_{\rho}}=r_{a}\Big[\frac{r_{amax}}{r_{a }}\Big]^{\frac{n-1}{n_{max}-1}} \quad(n=1,\cdots,n_{max})
\end{eqnarray}
\begin{eqnarray}
\nu_{n_{\lambda}}=\frac{1}{r_{n_{\lambda}}^{2}}, \quad r_{n_{\lambda}}=r_{b}\Big[\frac{r_{bmax}}{r_{b }}\Big]^{\frac{n-1}{n_{max}-1}} \quad(n=1,\cdots,n_{max})
\end{eqnarray}
where $r_{a}$, $r_{b}$, $r_{amax}$ and $r_{bmax}$ are the Gaussian range parameters, which take the values as $r_{a}=r_{b}$=0.18, $r_{amax}=r_{bmax}$=15. The $n_{max}$ is the maximum number of the Gaussian basis functions which is determined as 10 in Section 3.1.
In momentum space, the Gaussian basis functions can be written as,
\begin{eqnarray}
\phi_{n_{\rho}l_{\rho}m_{l_{\rho}}}(\boldsymbol{p}_{\rho_{c}})=N^{\prime}_{n_{\rho}l_{\rho}}p_{\rho_{c}}^{l_{\rho}}e^{-\frac{p_{\rho_{c}}^{2}}{4\nu_{n_{\rho}}}}Y_{l_{\rho}m_{l_{\rho}}}(\hat{\boldsymbol{p}}_{\rho_{c}})
\end{eqnarray}
\begin{eqnarray}
\phi_{n_{\lambda}l_{\lambda}m_{l_{\lambda}}}(\boldsymbol{p}_{\lambda_{c}})=N^{\prime}_{n_{\lambda}l_{\lambda}}p_{\lambda_{c}}^{l_{\lambda}}e^{-\frac{p_{\lambda_{c}}^{2}}{4\nu_{n_{\lambda}}}}Y_{l_{\lambda}m_{l_{\lambda}}}(\hat{\boldsymbol{p}}_{\lambda_{c}})
\end{eqnarray}
with
\begin{eqnarray}
N^{\prime}_{n_{\rho}l_{\rho}}=(-i)^{l_{\rho}}\sqrt{\frac{2^{l_{\rho}+2}}{\sqrt{\pi}(2\nu_{n_{\rho}})^{l_{\rho}+3/2}(2l_{\rho}+1)!!}}
\end{eqnarray}
\begin{eqnarray}
N^{\prime}_{n_{\lambda}l_{\lambda}}=(-i)^{l_{\lambda}}\sqrt{\frac{2^{l_{\lambda}+2}}{\sqrt{\pi}(2\nu_{n_{\lambda}})^{l_{\lambda}+3/2}(2l_{\lambda}+1)!!}}
\end{eqnarray}

In the heavy quark limit, one heavy quark within the heavy baryon system is decoupled from two light quarks. Under this picture, the interactions
which depend on the spin of the heavy quark disappear. Thus, two states whose quantum number are $J=j+1/2$ and $J=j-1/2$ will be degenerate, where $j$ denotes the light-spin-component. It can be seen from Fig.1 that the picture of Jacobi coordinate channel $3$ properly reflects the characteristic of the heavy quark symmetry.
As it is shown in Fig.1, the degree of freedom between two light quarks is commonly called the $\rho$-mode, while the degree between the center of mass of two light quarks and the heavy quark is called the $\lambda$-mode.
In the SU(3) limit in the light quark sector, the $\lambda$- and $\rho$-modes are mixed. While in the heavy quark limit, it was indicated by Ref.\cite{quam4} that the excitation energy of the $\lambda$-mode is lower than that of the $\rho$-mode for the $P$-wave baryon. They drew the conclusion that the lowest state of a $P$-wave single heavy baryon is dominated by the $\lambda$-mode.
The first motivation of our present work is to further study the mass spectra of different excitation modes, $\lambda$-mode, $\rho$-mode and $\lambda$-$\rho$ mixing mode, with higher orbital excitation. Basing on these above considerations, we perform our calculations in the channel $3$ and rewrite the full wave function of a single heavy baryon as,
\begin{eqnarray}
\Psi_{full}^{JM}=\sum_{n_{_{\rho}},n_{_{\lambda}}}C_{n_{_{\rho}},n_{_{\lambda}}}\Psi_{JM}^{3}(\boldsymbol{r}_{\rho_{3}},\boldsymbol{r}_{\lambda_{3}})
\end{eqnarray}

\begin{large}
\textbf{2.3 Evaluations of the matrix elements}
\end{large}

In the calculations of the Hamiltonian matrix elements of three-body system, particularly when
complicated interactions are employed, integrations over all of the radial and angular coordinates become
laborious even with Gaussian basis functions. This process can be simplified by introducing the ISG basis functions. For Jacobi coordinates $\rho$ and $\lambda$ in channel $3$, these new sets of basis functions can be written as\cite{IGEM},
\begin{eqnarray}
\phi_{n_{\rho}l_{\rho}m_{l_{\rho}}}(\boldsymbol{r}_{\rho})=N_{n_{\rho}l_{\rho}}\lim_{\varepsilon\rightarrow 0}\frac{1}{(\nu_{n_{\rho}}\varepsilon)^{l_{\rho}}}\sum_{k=1}^{k_{max}}C_{l_{\rho}m_{l_{\rho}},k}e^{-\nu_{n_{\rho}}(\boldsymbol{r}_{\rho}-\varepsilon \textbf{D}_{l_{\rho}m_{l_{\rho}},k})^{2}}
\end{eqnarray}
\begin{eqnarray}
\phi_{n_{\lambda}l_{\lambda}m_{l_{\lambda}}}(\boldsymbol{r}_{\lambda})=N_{n_{\lambda}l_{\lambda}}\lim_{\varepsilon\rightarrow 0}\frac{1}{(\nu_{n_{\lambda}}\varepsilon)^{l_{\lambda}}}\sum_{K=1}^{K_{max}}C_{l_{\lambda}m_{l_{\lambda}},K}e^{-\nu_{n_{\lambda}}(\boldsymbol{r}_{\lambda}-\varepsilon \textbf{D}_{l_{\lambda}m_{l_{\lambda}},K})^{2}}
\end{eqnarray}
where $\varepsilon$ is the shifted distance of the Gaussian basis. Taking the limit $\varepsilon\rightarrow 0$ is to be carried out after the matrix elements have been calculated analytically.

All of the evaluated matrix elements are presented in \textbf{Appendixes} \textbf{B.1}-\textbf{B.5}. Finally, the mass spectra can be obtained by solving the generalized eigenvalue problem,
\begin{flalign}
\sum_{j=1}^{n_{max}^{2}}\Big(H_{ij}-EN_{ij}\Big)C_{j}=0, \quad (i=1-n_{max}^{2})
\end{flalign}
Where $H_{ij}$ denotes the matrix element in the total color-flavor-spin-spatial base, $E$ is the eigenvalue, $C_{j}$ stands for the corresponding eigenvector, and $N_{ij}$ is the overlap matrix elements of the Gaussian functions, which arises from the nonorthogonality of the bases and can be expressed as,
\begin{flalign}
\notag
N_{ij}\equiv \langle\phi_{n_{\rho_{a}}l_{\rho_{a}}m_{l_{\rho_{a}}}}|
\phi_{n_{\rho_{b}}l_{\rho_{b}}m_{l_{\rho_{b}}}}\rangle\times
\langle\phi_{n_{\lambda_{a}}l_{\lambda_{a}}m_{l_{\lambda_{a}}}}
|\phi_{n_{\lambda_{b}}l_{\lambda_{b}}m_{l_{\lambda_{b}}}}\rangle & \\
=\Big(\frac{2\sqrt{\nu_{n_{\rho_{a}}}\nu_{n_{\rho_{b}}}}}{\nu_{n_{\rho_{a}}}+\nu_{n_{\rho_{b}}}}\Big)^{l_{\rho_{a}}+3/2}\times
\Big(\frac{2\sqrt{\nu_{n_{\lambda_{a}}}\nu_{n_{\lambda_{b}}}}}{\nu_{n_{\lambda_{a}}}+\nu_{n_{\lambda_{b}}}}\Big)^{l_{\lambda_{a}}+3/2} &
\end{flalign}

\begin{Large}
\textbf{3 Mass spectra and root mean square radius}
\end{Large}

\begin{large}
\textbf{3.1 Numerical stabilities and $\lambda$-modes}
\end{large}
\begin{table*}[h]
\begin{ruledtabular}\caption{Relevant parameters of the relativized quark model}
\begin{tabular}{c c c c c c}
$m_{u}/m_{d}$(MeV)& $m_{s}$(MeV) &$m_{c}$(MeV)&$m_{b}$(MeV)&$\alpha_{1}$ & $\alpha_{2}$\\
$220$ &$419$ &$1628$& $4997$ & $0.25$ & $0.15$ \\ \hline
 $\alpha_{3}$ &$\gamma_{1}$(GeV)&$\gamma_{2}$(GeV)&$\gamma_{3}$(GeV) &$b$(GeV$^{2}$)& $c$(MeV) \\
$0.20$ &$\frac{1}{2}$& $\sqrt{10}/2$ & $\sqrt{1000}/2$ & $0.14$ &$-198$ \\ \hline
$\sigma_{0}$(GeV)&$s$&$\epsilon_{\mathrm{c}}$ & $\epsilon_{\mathrm{(so)v}}$& $\epsilon_{\mathrm{t}}$ &$\epsilon_{\mathrm{(so)s}}$ \\
$1.8$& $1.55$ & $-0.168$ & $-0.035$ &$0.025$ &$0.055$ \\ \hline
\end{tabular}
\end{ruledtabular}
\end{table*}
\begin{figure}[h]
\begin{minipage}[h]{0.45\linewidth}
\centering
\includegraphics[height=5cm,width=7cm]{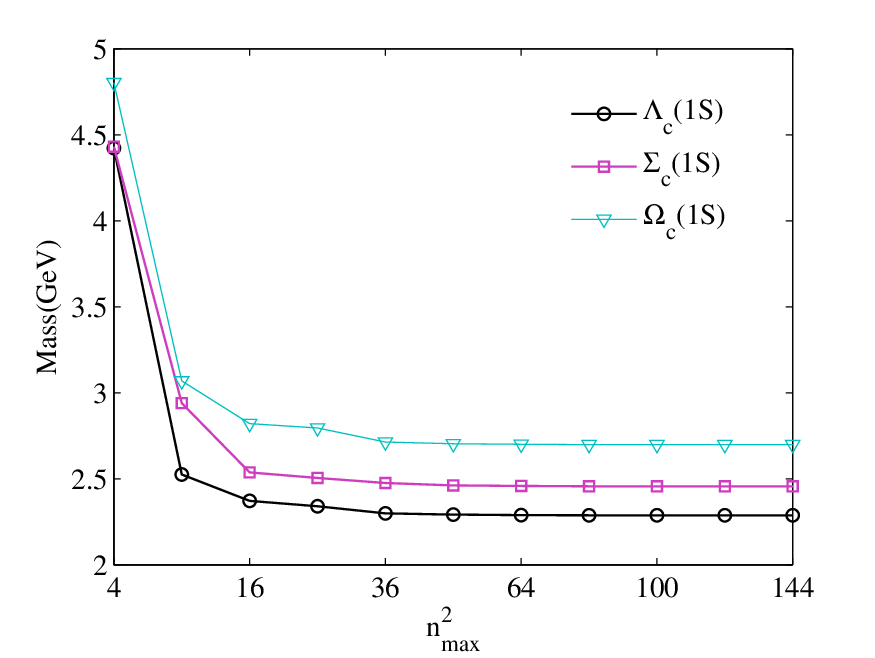}
\caption{Convergence of the energy of the lowest $\Lambda_{c}$, $\Sigma_{c}$ and $\Omega_{c}$ for
increasing the number of bases functions.}
\end{minipage}
\hfill
\begin{minipage}[h]{0.45\linewidth}
\centering
\includegraphics[height=5cm,width=7cm]{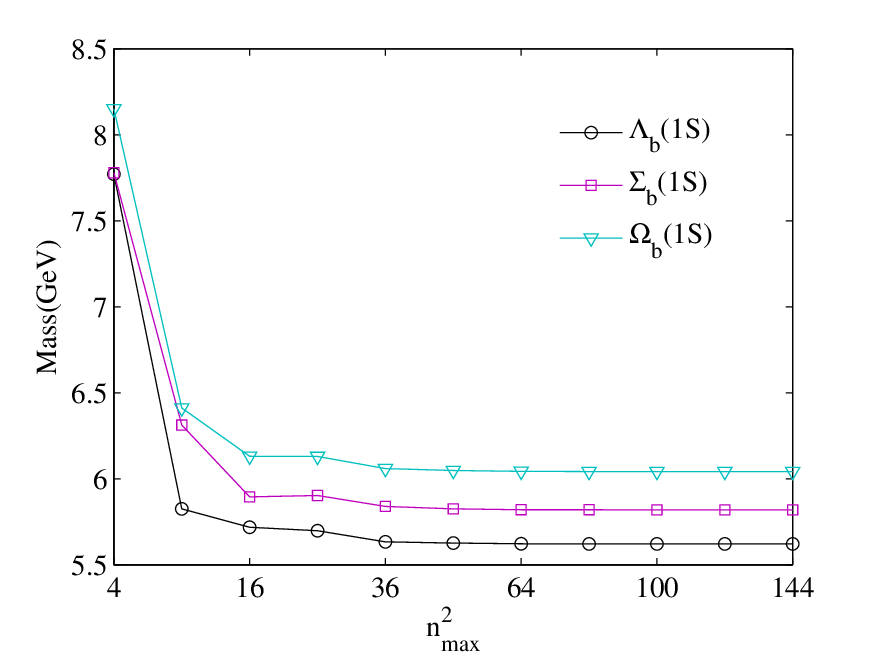}
\caption{Convergence of the energy of the lowest $\Lambda_{b}$, $\Sigma_{b}$ and $\Omega_{b}$ for
increasing the number of bases functions.}
\end{minipage}
\end{figure}

Most of the parameters used in this work are taken from the original reference\cite{quam1} and they are all collected in Table II for convenience. To reproduce the lowest-lying baryon mass spectra, we repair the values of parameters $b$ and $c$ in the line confining potential from $0.18$ and $-253$ to $0.14$ and -$198$. In the relativized quark model, the stability of the numerical results depends strongly on the number of bases. Theoretically, the number of bases should be large enough to guarantee approximate completeness, otherwise the numerical results are not reliable. In order to investigate the convergence and stability of the numerical results, we plot the eigen-energy of the lowest lying $\Lambda_{Q}(\frac{1}{2}^{+})$, $\Sigma_{Q}(\frac{1}{2}^{+})$ and $\Omega_{Q}(\frac{1}{2}^{+})$ baryons in Figs. 2-3. We can see that the eigenvalues decrease with the basis number and converge to a stable value when the $n_{max}^{2}=10^2$. Thus, it is enough for us to obtain the mass spectra with $10^2$ Gaussian bases in present work.
\begin{figure}[h]
\begin{minipage}[h]{0.45\linewidth}
\centering
\includegraphics[height=5cm,width=7cm]{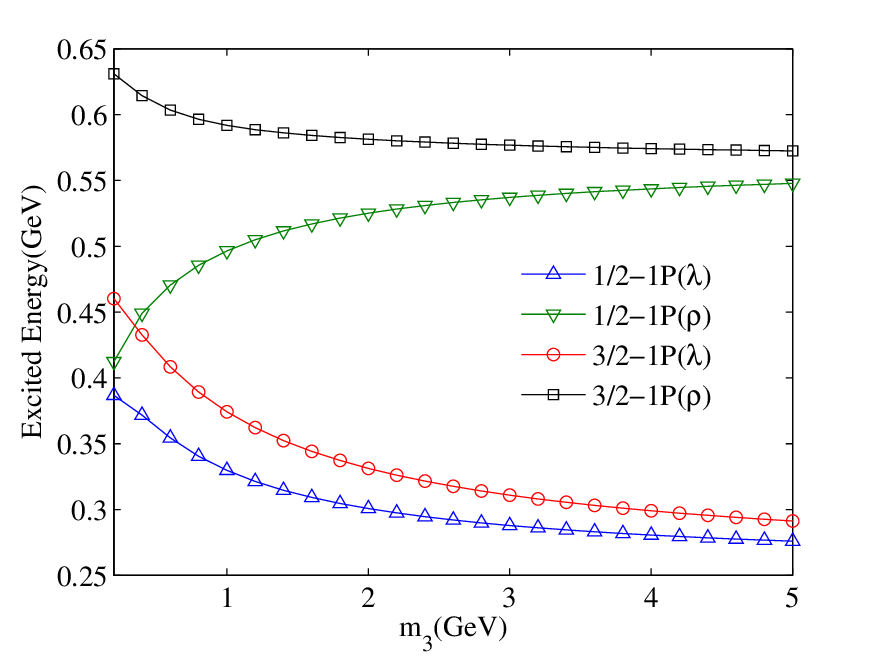}
\caption{Heavy quark mass dependence of excited energy of $1P$ $\Lambda_{Q}$($\frac{1}{2}^{-}$,$\frac{3}{2}^{-}$) with $\lambda$ mode and $\rho$ mode}
\end{minipage}
\hfill
\begin{minipage}[h]{0.45\linewidth}
\centering
\includegraphics[height=5cm,width=7cm]{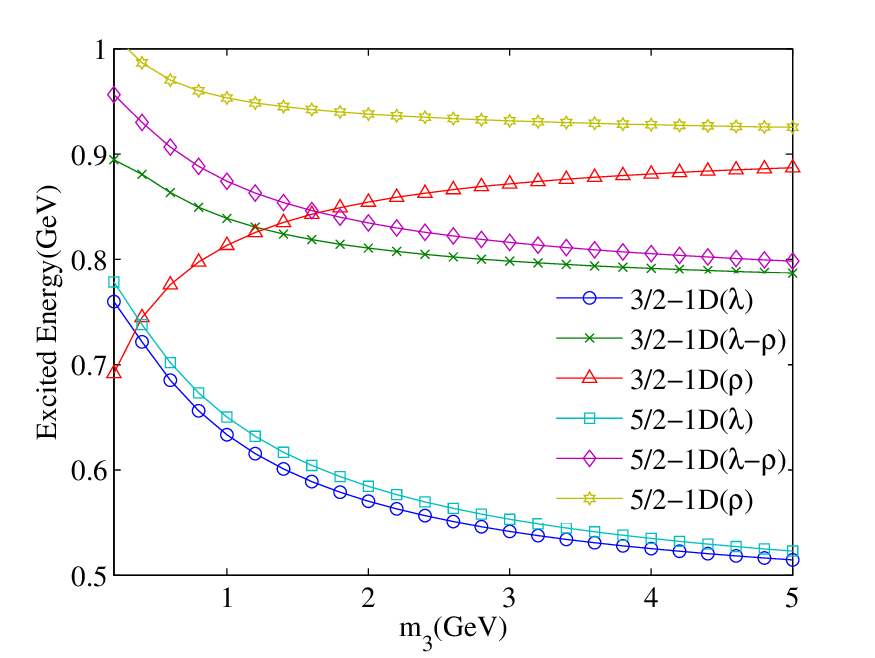}
\caption{Heavy quark mass dependence of excited energy of $1D$ $\Lambda_{Q}$($\frac{3}{2}^{+}$,$\frac{5}{2}^{+}$) with $\lambda$ mode ,$\rho$ mode and $\lambda$-$\rho$ mixing mode}
\end{minipage}
\end{figure}
\begin{figure}[h]
\begin{minipage}[h]{0.45\linewidth}
\centering
\includegraphics[height=5cm,width=7cm]{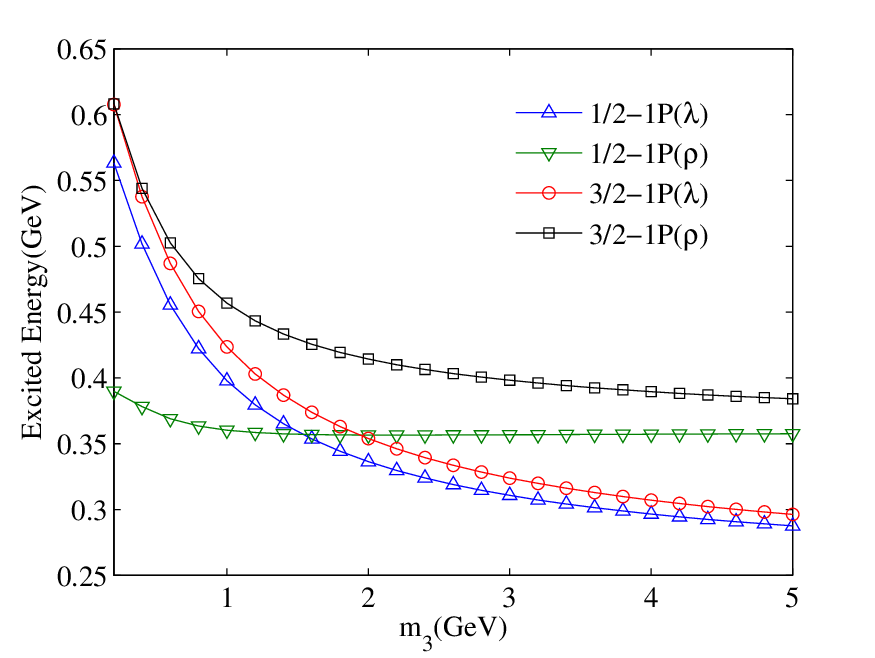}
\caption{Heavy quark mass dependence of excited energy of $1P$ $\Sigma_{Q}$($\frac{1}{2}^{-}$,$\frac{3}{2}^{-}$) with $\lambda$ mode and $\rho$ mode}
\end{minipage}
\hfill
\begin{minipage}[h]{0.45\linewidth}
\centering
\includegraphics[height=5cm,width=7cm]{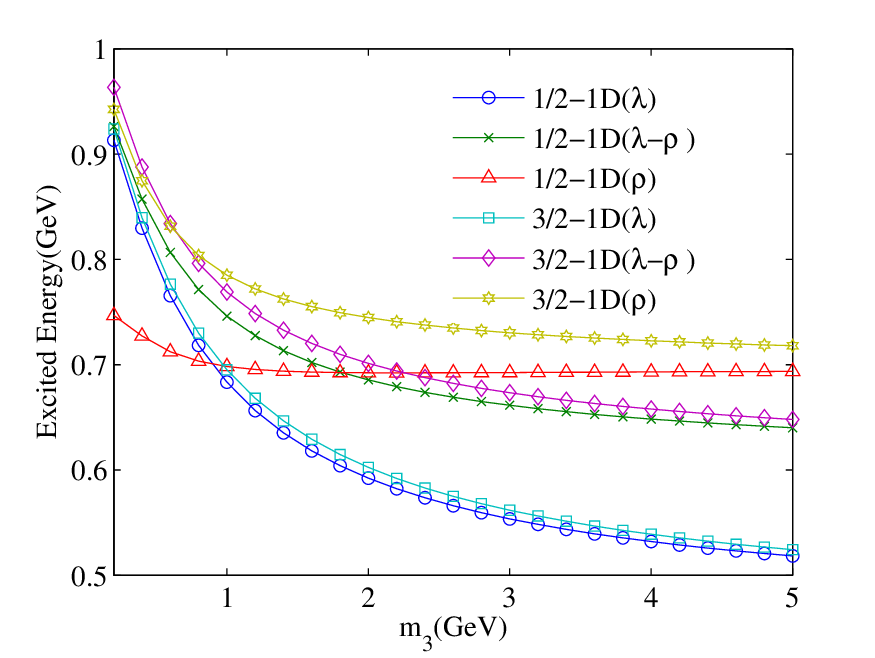}
\caption{Heavy quark mass dependence of excited energy of $1D$ $\Sigma_{Q}$($\frac{1}{2}^{+}$,$\frac{3}{2}^{+}$) with $\lambda$ mode ,$\rho$ mode and $\lambda$-$\rho$ mixing mode}
\end{minipage}
\end{figure}
\begin{figure}[h]
\begin{minipage}[h]{0.45\linewidth}
\centering
\includegraphics[height=5cm,width=7cm]{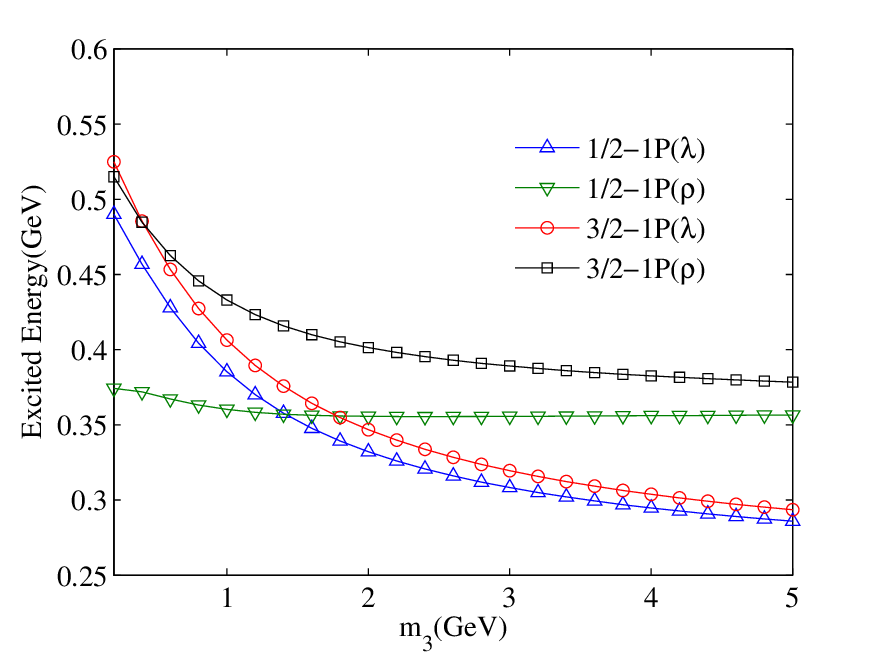}
\caption{Heavy quark mass dependence of excited energy of $1P$ $\Omega_{Q}$($\frac{1}{2}^{-}$,$\frac{3}{2}^{-}$) with $\lambda$ mode and $\rho$ mode.}
\end{minipage}
\hfill
\begin{minipage}[h]{0.45\linewidth}
\centering
\includegraphics[height=5cm,width=7cm]{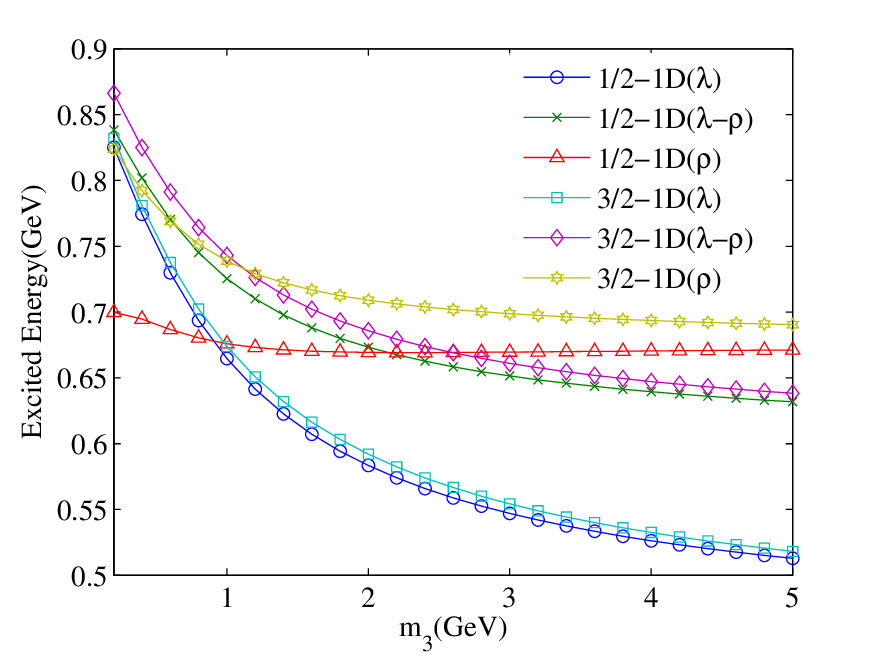}
\caption{Heavy quark mass dependence of excited energy of $1D$ $\Omega_{Q}$($\frac{1}{2}^{+}$,$\frac{3}{2}^{+}$) with $\lambda$ mode, $\rho$ mode and $\lambda$-$\rho$ mixing mode}
\end{minipage}
\end{figure}

In the scheme of channel $3$, the orbital angular momentum between the two light quarks is denoted by $l_{\rho}$,
while the angular momentum between the light diquarks and the heavy quark is denoted by $l_{\lambda}$. For $P$-wave baryons, there are two orbital excitation modes $\lambda$- and $\rho$-mode with ($l_{\rho}$,$l_{\lambda}$)=($0$,$1$) and ($1$,$0$) respectively. While there are three excitation modes for $D$-wave baryon with ($l_{\rho}$,$l_{\lambda}$)=($0$,$2$), ($2$,$0$) and ($1$,$1$), which are called the $\lambda$-mode, $\rho$-mode and $\lambda$-$\rho$ mixing mode, respectively. For higher angular excited states, they are similar with the $D$-wave baryon, which also have three excitation modes. To investigate the mass spectra of these different excitation modes, we change the heavy quark mass $m_{Q}$, from $0.2$ GeV to $5.0$ GeV and analyze their excitation energies. Figs. 4-9 show the results of $\Lambda_{Q}$, $\Sigma_{Q}$ and $\Omega_{Q}$ systems. We can see that as $m_{Q}$ becomes large, the $\lambda$-mode appears lower in excited energy than both the $\rho$-mode and $\lambda$-$\rho$ mixing mode for these baryons. This indicates the lowest state becomes dominated by the $\lambda$-mode as $m_{Q}$ becomes large whether for $P$- or $D$-wave baryon. In Figs. 4-5, the $\lambda$ dominance is seen at about $m_{Q}=0.4$ GeV for $\Lambda_{Q}$ baryon and this feature appears at $m_{Q}\geq1.0$ GeV for $D$-wave $\Sigma_{Q}$ and $\Omega_{Q}$(see Figs. 7 and 9). For $P$-wave $\Sigma_{Q}$ and $\Omega_{Q}$ systems(see Figs. 6 and 8), although $\lambda$ and $\rho$ excited modes are separated more slowly, the $\lambda$ dominance is also seen at about $m_{Q}\geq1.6$ GeV.

From these figures, the heavy quark spin symmetry(HQS) is also displayed. For $P$-wave $\Lambda_{Q}$ as an example(see Fig. 4), when $m_{Q}$ increases from $0.2$ GeV to $5.0$ GeV, the splitting of the spin-doublet ($\frac{1}{2}^{-}$,$\frac{3}{2}^{-}$) of $\lambda$ excitation mode decreases from $80$ MeV to $10$ MeV. And this behavior is more obvious for $P$-wave $\Lambda_{Q}$ baryon with $\rho$-mode(see Fig. 4).

\begin{large}
\textbf{3.2 Mass spectra of $\Lambda_{Q}$, $\Sigma_{Q}$ and $\Omega_{Q}$}
\end{large}

From these above discussions, we know that each state of the single heavy baryons is dominated and characterized by the $\lambda$-mode. Thus, we obtain the mass spectra of the $\Lambda_{Q}$, $\Sigma_{Q}$, and $\Omega_{Q}$ baryons with $\lambda$ excitation mode.
Our predictions for mass spectra and root mean square radius  are shown in \textbf{Appendix A.1}(Tables III-VIII). In the first two columns of these tables we give the baryon quantum numbers ($l_{\rho}$  $l_{\lambda}$ $L$ $s$ $j$) and $nL$($J^{P}$), while in the remaining columns our results, experimental data and the results from other models are shown.

\begin{center}
\begin{large}
\textbf{A. $\Lambda_{Q}$ states}
\end{large}
\end{center}

It can be seen from Tables III-IV, the available experimental data for $\Lambda_{Q}$ states are well reproduced by the quark model in present work. There are a few observed states, $\Lambda_{c}(2765)$, $\Lambda_{c}(2940)$ and $\Lambda_{b}(6070)$, whose spin and parity have not been confirmed in the latest PDG\cite{article2A,article2B}. Our prediction for the mass of 2$S$($\frac{1}{2}^{+}$) state is $2.764$ GeV, which is very close to the experimental data for $\Lambda_{c}(2765)$. Thus, it is reasonable to describe $\Lambda_{c}(2765)$ as a 2$S$($\frac{1}{2}^{+}$) state. In the latest PDG, the spin parity of $\Lambda_{c}(2940)$ was suggested to be $J^{P}=\frac{3}{2}^{-}$, but was not confirmed. Model predictions for 2$P$($\frac{1}{2}^{-}$) and 2$P$($\frac{3}{2}^{-}$) states are $2.988$ GeV and $3.013$ GeV, respectively. By comparing with the experimental data, the 2$P$($\frac{1}{2}^{-}$) state seems to be a better candidate for $\Lambda_{c}(2940)$ than 2$P$($\frac{3}{2}^{-}$). The predicted mass for 2$P$($\frac{1}{2}^{-}$) baryon is about $50$ MeV higher than the experimental data. This notable exception to this state is also seen in the predictions in Ref.\cite{quam2}. For $\Lambda_{b}(6070)$, its quantum numbers were suggested to be 2$S$($\frac{1}{2}^{+}$) in Refs.\cite{60701,WZG2}. Our prediction for this state is $6.041$ GeV, which is $29$ MeV lighter than the experimental data. Under the uncertainties of the model, this result is still within the realm of validity for model like these.

Besides of these above states, some other low-lying $\Lambda_{Q}$ states are also predicted in present work, which have not been observed in experiments. First, it is the $2P$-wave $\Lambda_{c}$($\frac{3}{2}^{-}$) state which belongs to a spin-doublet ($\frac{1}{2}^{-}$,$\frac{3}{2}^{-}$) and its model predicted mass is $3.013$ GeV. In addition, there are two $\Lambda_{b}$ model states that still remain to be found in experiments. They are the partners of the $2P$-wave doublet ($\frac{1}{2}^{-}$,$\frac{3}{2}^{-}$) of $\Lambda_{c}$, and their masses are predicted to be $6.238$ GeV and $6.249$ GeV, respectively.

\begin{center}
\begin{large}
\textbf{B. $\Sigma_{Q}$ states}
\end{large}
\end{center}

The lowest states of $S$-wave $\Sigma_{c}$ and $\Sigma_{b}$ baryons with the quantum numbers $J^{P}=\frac{1}{2}^{+}$ and $J^{P}=\frac{3}{2}^{+}$ have been observed and confirmed in experiments\cite{article2A}.
In Tables V-VI, it can be seen that model predictions for these $S$-wave $\Sigma_{Q}$ baryons also agree well with the experimental data. In addition, quark model predicts two spin doublets ($\frac{1}{2}^{-}$,$\frac{3}{2}^{-}$), ($\frac{3}{2}^{-}$,$\frac{5}{2}^{-}$) and a singlet $\frac{1}{2}^{-}$ for $1P$-wave $\Sigma_{Q}$ baryons. The previously observed $\Sigma_{c}(2800)$ and $\Sigma_{b}(6097)$ states are good candidates for the $P$-wave
states. We can see from Tables V-VI that predicted masses of these lowest $P$-wave states are very close to each other and they are all comparable with the experimental data of $\Sigma_{c}(2800)$ or $\Sigma_{b}(6097)$. It is shown in Table V that there are two states whose masses are closer to the experimental result of $\Sigma_{c}(2800)$, namely, the states $1P$($\frac{1}{2}^{-}$)$_{j=1}$ with a mass $2.809$ GeV and $1P$($\frac{3}{2}^{-}$)$_{j=2}$ with $2.802$ GeV. In Table VI, two $\Sigma_{b}$ states $1P$($\frac{1}{2}^{-}$)$_{j=1}$ and $1P$($\frac{3}{2}^{-}$)$_{j=2}$ have masses $6.107$ GeV and $6.104$ GeV respectively, both excellently match to the experimental state $\Sigma_{b}(6097)$. Now, it is difficult for us to confirm the quantum numbers of $\Sigma_{c}(2800)$ and $\Sigma_{b}(6097)$ only through the mass spectra. If the strong decay behaviors of these $P$-wave $\Sigma_{Q}$ systems were analyzed, better assignments will be suggested.

It can be seen from Tables V-VI, more efforts are needed in searching for $\Sigma_{Q}$ baryons in experiments because many model-predicted states are still missing. Besides of $\Sigma_{c}(2800)$ and $\Sigma_{b}(6097)$, the other $1P$ states have good potentials to be observed within the mass spectra predicted in present work. On the other hand, it is shown in Tables V-VI that the splitting between the $1P$ and $2S$ state is about $100$ MeV, while the splitting between the five $1P$ states ranges from several to $30$ MeV. We hope these information will be helpful in searching for these $2S$ and $1P$ $\Sigma_{Q}$ states in future experiments. Finally, the $1D$-wave $\Sigma_{Q}$ states are most likely to be observed in forthcoming experiments as well. The predicted masses range from $3.072$ GeV to $3.084$ GeV for $1D$-wave $\Sigma_{c}$ states and from $6.338$ GeV to $6.346$ GeV for $\Sigma_{b}$ ones.

\begin{center}
\begin{large}
\textbf{C. $\Omega_{Q}$ states}
\end{large}
\end{center}

For $\Omega_{Q}$ systems, there are only three states, the lowest $S$-wave $\Omega_{c}$ doublet ($\frac{1}{2}^{+}$,$\frac{3}{2}^{+}$) and $1S$-wave $\Omega_{b}$($\frac{1}{2}^{+}$), that have been discovered and confirmed\cite{article2A}. In Tables VII-VIII, it can be seen that the known $\Omega_{Q}$ states are well reproduced by the quark model. However, the predicted $1S$($\frac{3}{2}^{+}$) $\Omega_{b}$ state with a mass of $6.069$ GeV still remains to be found. In the last few years, several new $\Omega_{c}$ and $\Omega_{b}$ states were observed and they are $\Omega_{c}(3000)$, $\Omega_{c}(3050)$, $\Omega_{c}(3066)$, $\Omega_{c}(3090)$, $\Omega_{c}(3119)$\cite{OmegaC3000}, $\Omega_{b}(6316)$, $\Omega_{b}(6330)$, $\Omega_{b}(6340)$ and $\Omega_{b}(6350)$\cite{OmegaB6316}. Model prediction for the $\Omega_{c}$ state of $2S$($\frac{1}{2}^{+}$) is $3.150$ GeV, which is close to $\Omega_{c}(3119)$. If model uncertainties are considered, $\Omega_{c}(3119)$ can be interpreted as a $2S$($\frac{1}{2}^{+}$) state. The other four $\Omega_{c}$ baryons can be assigned as the lowest $P$-wave states because their measured masses are compatible with model predictions in Table VII. For the $1P$-wave $\Omega_{c}$ doublets ($\frac{1}{2}^{-}$, $\frac{3}{2}^{-}$) and ($\frac{3}{2}^{-}$, $\frac{5}{2}^{-}$), it is showed in Table VII that the splitting of these doublets are $17$ MeV and $28$ MeV, respectively. In comparison with the experimental data, the possible assignments for these four experimental $\Omega_{c}$ states are ($\Omega_{c}(3000)$, $\Omega_{c}(3050)$)$=$($\frac{1}{2}^{-}$, $\frac{3}{2}^{-}$)$_{j=1}$ and ($\Omega_{c}(3066)$, $\Omega_{c}(3090)$)=($\frac{3}{2}^{-}$, $\frac{5}{2}^{-}$)$_{j=2}$. For $\Omega_{b}$ baryons, quark model predicts four $1P$-wave states and their situation is similar with the $\Omega_{c}$ baryons. The splitting of spin-doublets ($\frac{1}{2}^{-}$, $\frac{3}{2}^{-}$)$_{j=1}$ and ($\frac{3}{2}^{-}$, $\frac{5}{2}^{-}$)$_{j=2}$ are $7$ MeV and $13$ MeV(see Table VIII). These four experimental $\Omega_{b}$ states are good candidates of these $P$-wave doublets, which can be assigned as ($\Omega_{b}(6315)$, $\Omega_{b}(6330)$)$=$($\frac{1}{2}^{-}$, $\frac{3}{2}^{-}$)$_{j=1}$ and ($\Omega_{b}(6340)$, $\Omega_{b}(6350)$)=($\frac{3}{2}^{-}$, $\frac{5}{2}^{-}$)$_{j=2}$. Actually, different collaborations had different conclusions about the quantum numbers of these $\Omega_{c}$ and $\Omega_{b}$ states\cite{OmegaC30001,OmegaC30002,OmegaC30003,OmegaC30004,OmegaC30005,OmegaC30006,OmegaC30007,OmegaB63301,OmegaB63302,OmegaB63303} and more theoretical and experimental efforts are needed to assign a proper place in spectra for these $P$-wave $\Omega_{Q}$ systems.

In quark model scheme, there are two $2S$-wave excitations whose masses are predicted to be $6.446$ GeV and $6.466$ GeV, respectively. The other collaborations also reported the same results as ours\cite{quam2}, which suggests that these $\Omega_{b}$ states have potentials to be observed at this mass spectra in forthcoming experiments. Finally, the $1D$-wave $\Omega_{Q}$ states also have the possibilities to be found in experiments. The masses are predicted to be around $3.304$ GeV $\sim$ $3.315$ GeV for $\Omega_{c}$($1D$) states and $6.555$ GeV $\sim$ $6.562$ GeV for $\Omega_{b}$($1D$) ones.

\begin{large}
\textbf{3.3 Root mean square radius and radial density distributions}
\end{large}

Besides of the root mean square radius, we also calculated the radial density distributions which are defined as,
\begin{eqnarray}
\notag
\omega(r_{\rho})=\int|\Psi(\textbf{r}_{\rho},\textbf{r}_{\lambda})|^{2}d\textbf{r}_{\lambda}d\Omega_{\rho} \\
\omega(r_{\lambda})=\int|\Psi(\textbf{r}_{\rho},\textbf{r}_{\lambda})|^{2}d\textbf{r}_{\rho}d\Omega_{\lambda}
\end{eqnarray}
where $\Omega_{\rho}$ and $\Omega_{\lambda}$ are the solid angles spanned by vectors $\textbf{r}_{\rho}$ and $\textbf{r}_{\lambda}$, respectively. As an example, we have shown the results about  radial density distributions of $\Lambda_{Q}$ baryons in Figs. 10-13.
\begin{figure}[H]
  \centering
   \subfigure[]{
   \begin{minipage}{3.3cm}
   \centering
   \includegraphics[width=4cm]{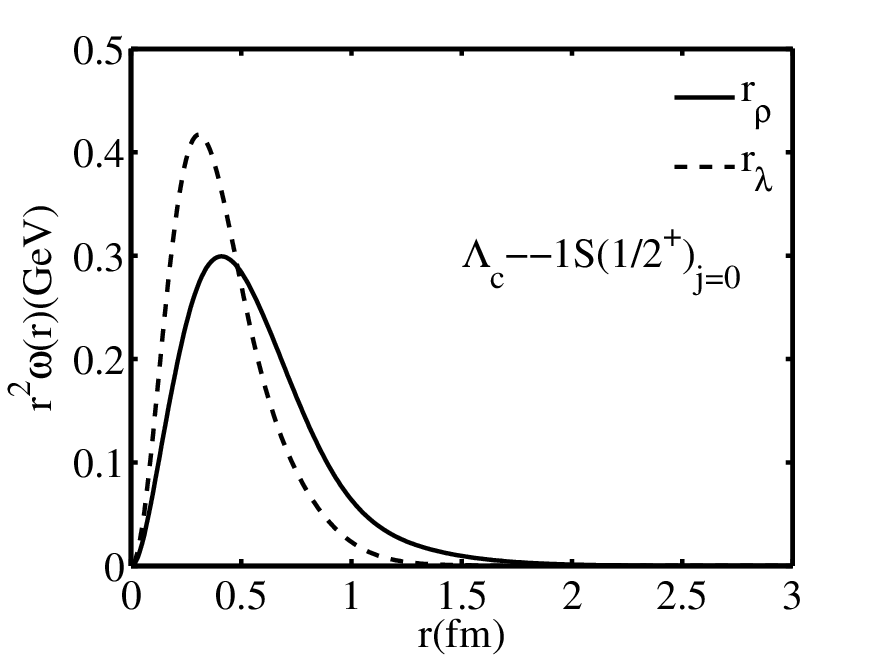}
  \end{minipage}
  }
 \subfigure[]{
   \begin{minipage}{3.3cm}
   \centering
   \includegraphics[width=4cm]{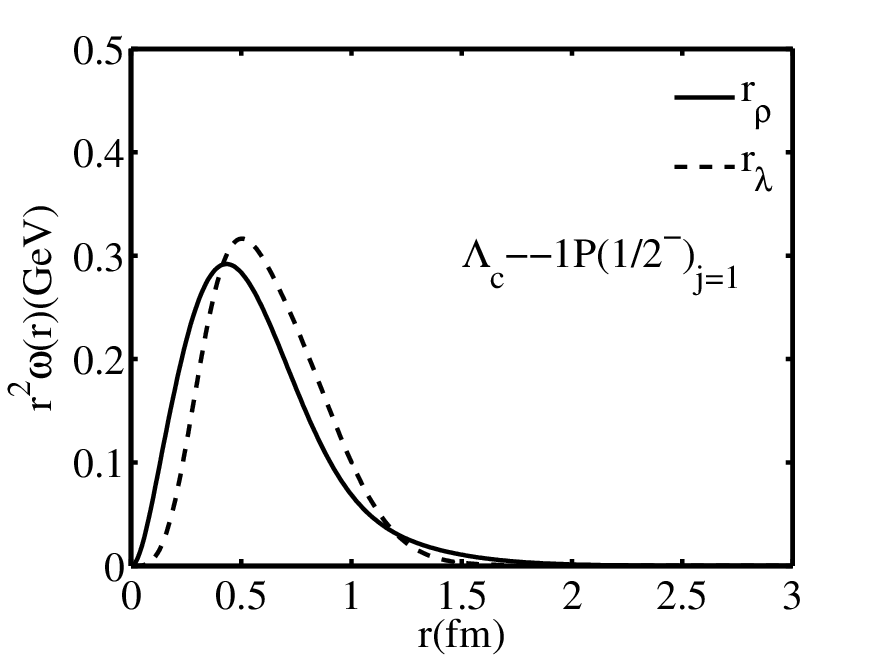}
  \end{minipage}
  }
   \subfigure[]{
   \begin{minipage}{3.3cm}
   \centering
   \includegraphics[width=4cm]{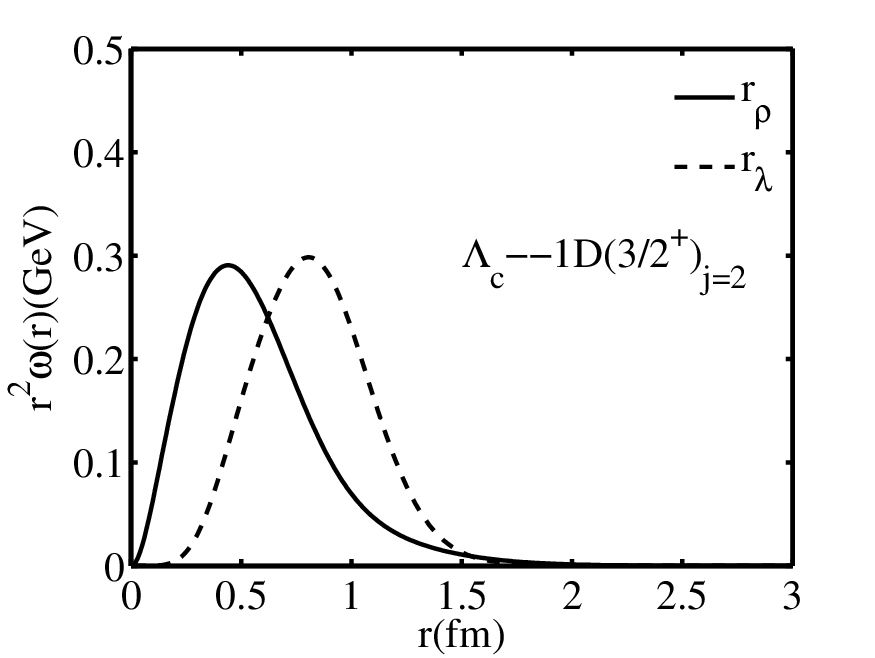}
  \end{minipage}
  }
    \subfigure[]{
   \begin{minipage}{3.3cm}
   \centering
   \includegraphics[width=4cm]{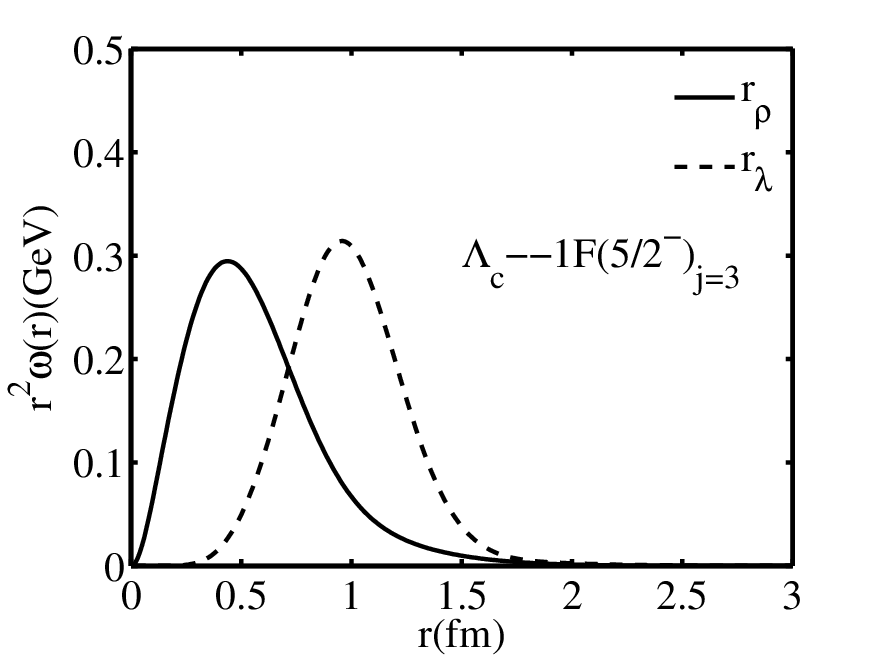}
  \end{minipage}
  }
  \caption{Radial density distributions for some $1S-1F$ states in the $\Lambda_{c}$ families}
\end{figure}
\begin{figure}[ht]
  \centering
   \subfigure[]{
   \begin{minipage}{3.3cm}
   \centering
   \includegraphics[width=4cm]{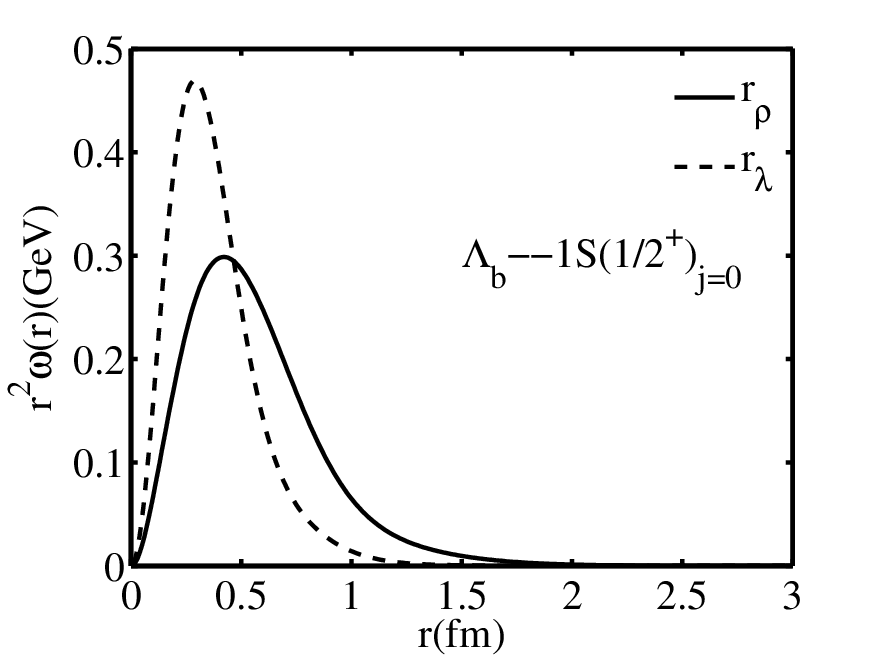}
  \end{minipage}
  }
 \subfigure[]{
   \begin{minipage}{3.3cm}
   \centering
   \includegraphics[width=4cm]{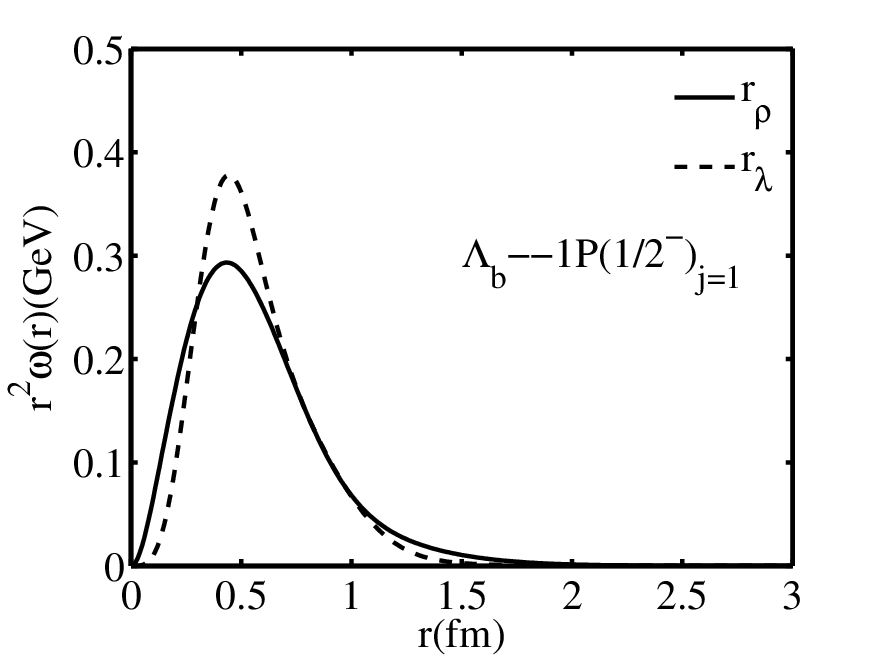}
  \end{minipage}
  }
   \subfigure[]{
   \begin{minipage}{3.3cm}
   \centering
   \includegraphics[width=4cm]{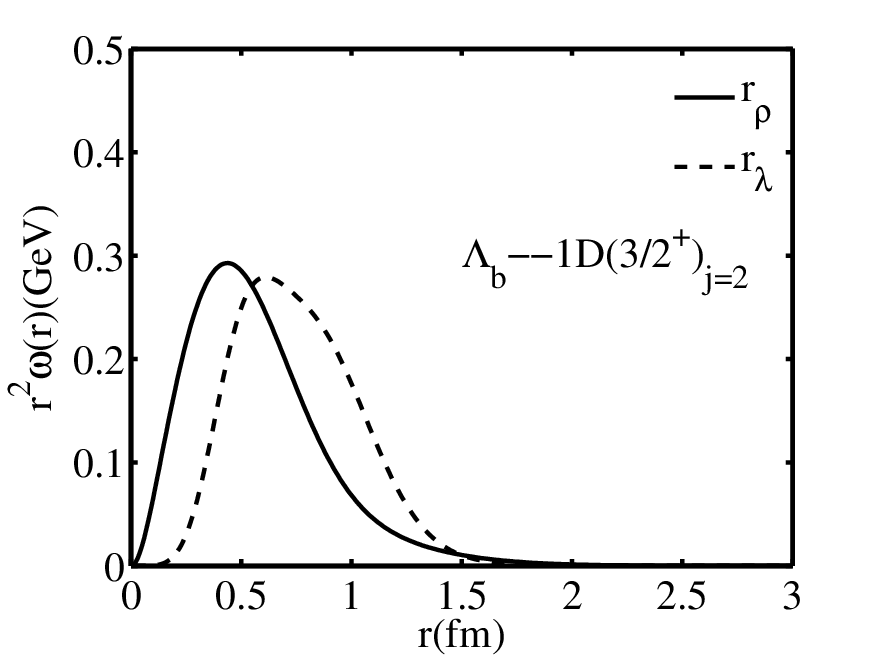}
  \end{minipage}
  }
    \subfigure[]{
   \begin{minipage}{3.3cm}
   \centering
   \includegraphics[width=4cm]{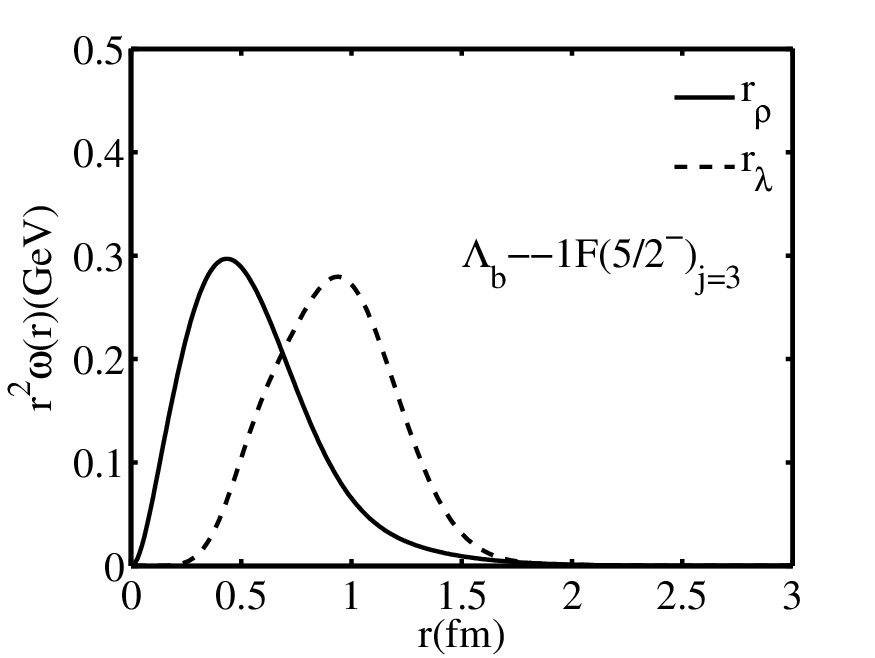}
  \end{minipage}
  }
  \caption{Radial density distributions for some $1S-1F$ states in the $\Lambda_{b}$ families}
\end{figure}
\begin{figure}[H]
  \centering
   \subfigure[]{
   \begin{minipage}{3.9cm}
   \centering
   \includegraphics[width=4.5cm]{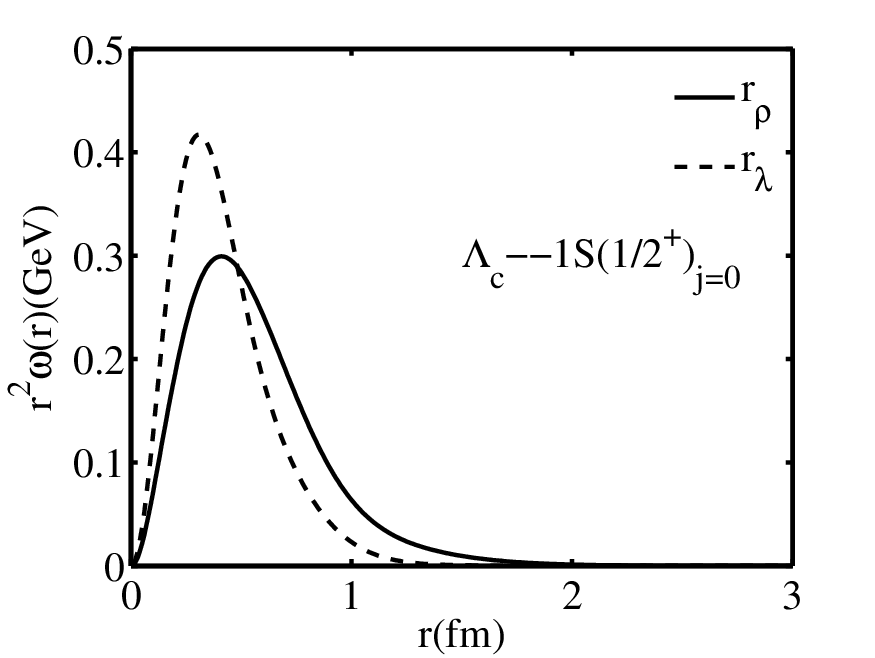}
  \end{minipage}
  }
 \subfigure[]{
   \begin{minipage}{3.9cm}
   \centering
   \includegraphics[width=4.5cm]{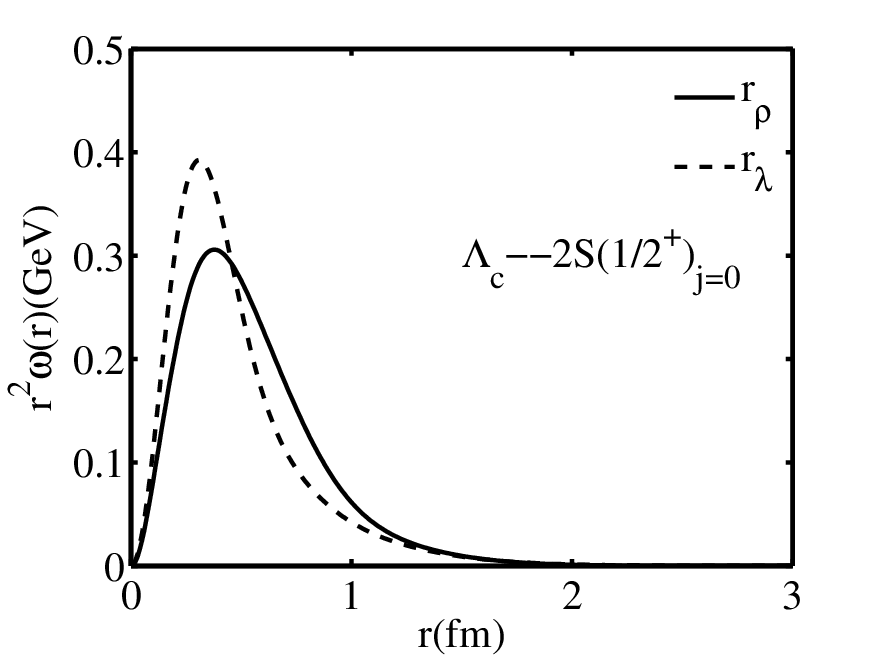}
  \end{minipage}
  }
   \subfigure[]{
   \begin{minipage}{3.9cm}
   \centering
   \includegraphics[width=4.5cm]{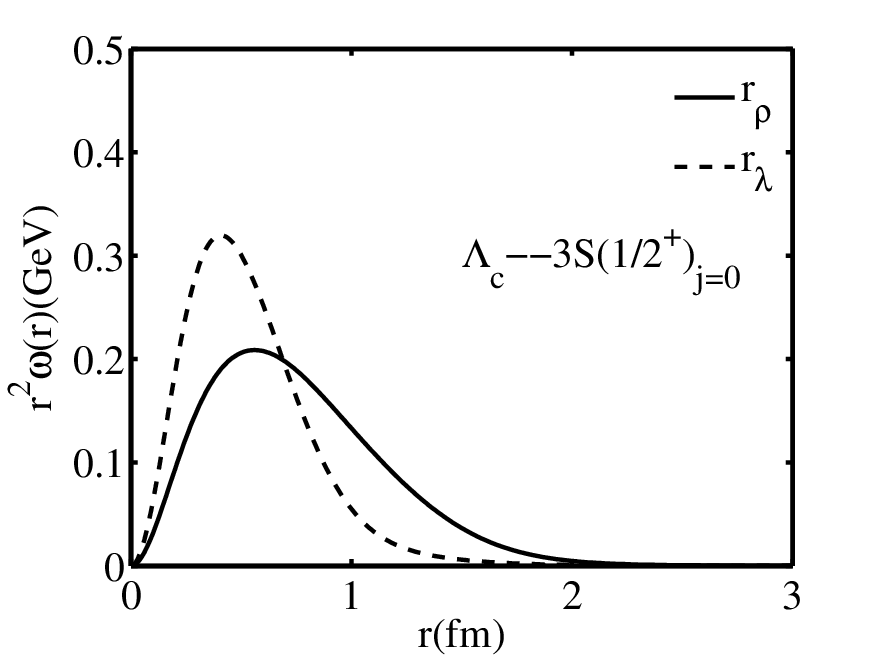}
  \end{minipage}
  }
  \caption{Radial density distributions for $1S\sim3S$ states in the $\Lambda_{c}$ families.}
\end{figure}
\begin{figure}[H]
  \centering
   \subfigure[]{
   \begin{minipage}{3.9cm}
   \centering
   \includegraphics[width=4.5cm]{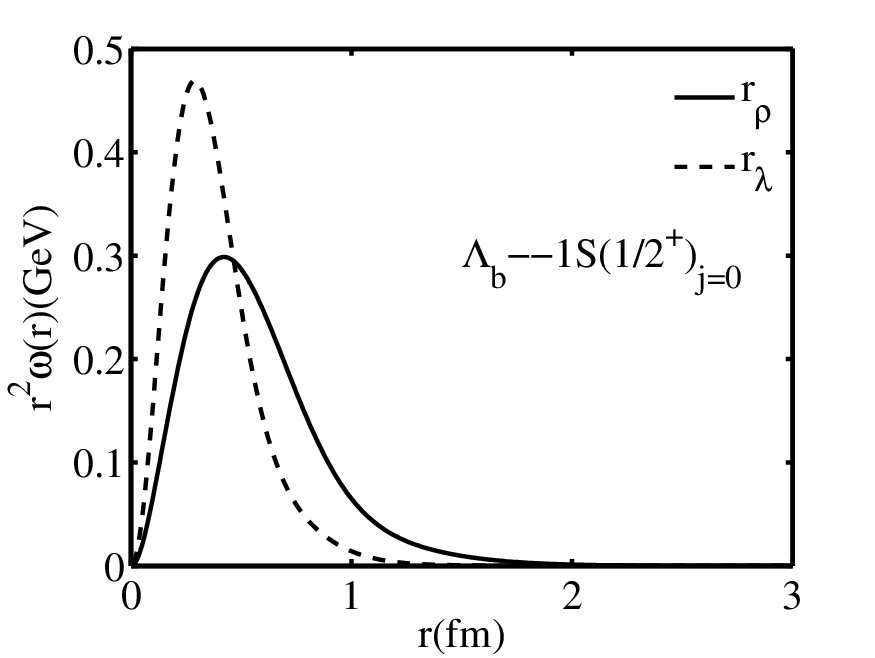}
  \end{minipage}
  }
 \subfigure[]{
   \begin{minipage}{3.9cm}
   \centering
   \includegraphics[width=4.5cm]{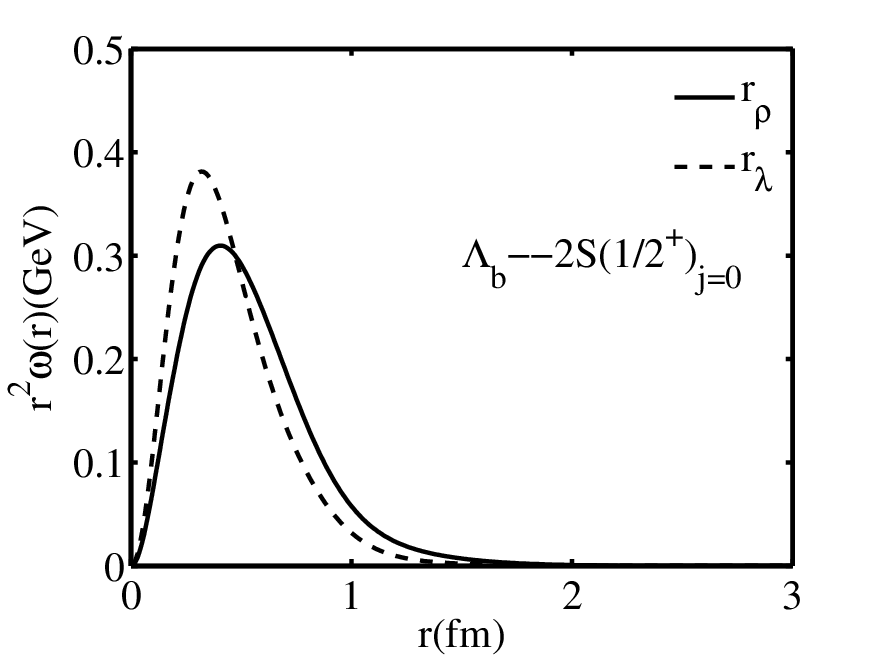}
  \end{minipage}
  }
   \subfigure[]{
   \begin{minipage}{3.9cm}
   \centering
   \includegraphics[width=4.5cm]{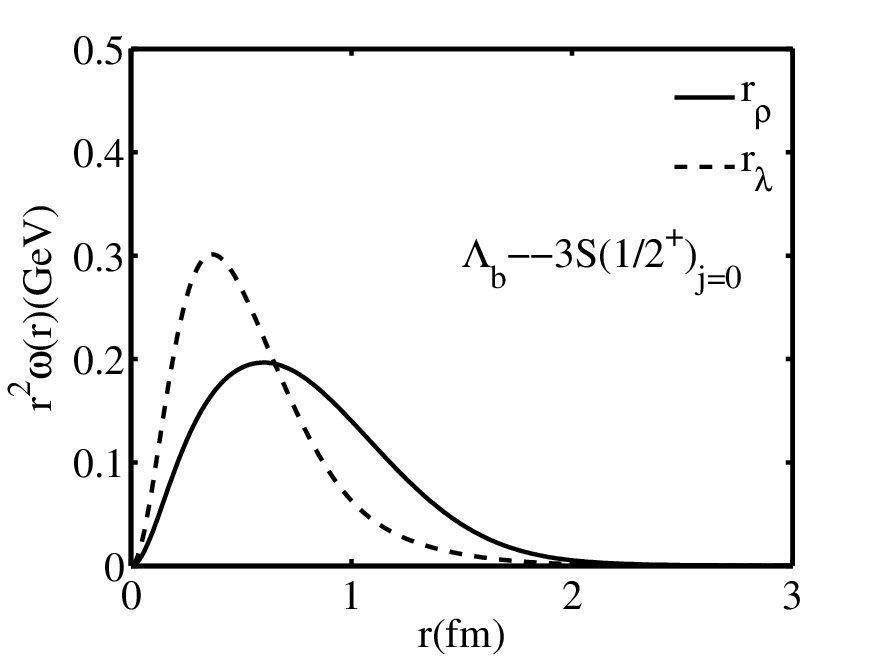}
  \end{minipage}
  }
  \caption{Radial density distributions for $1S\sim3S$ states in the $\Lambda_{b}$ families.}
\end{figure}
From Tables III-VIII, we can see the root mean square radius $\langle r_{\rho}\rangle_{\frac{1}{2}}$ and $\langle r_{\lambda}\rangle_{\frac{1}{2}}$ of the $1S$ states are 0.39$\sim$0.44 fm and 0.51$\sim$0.63 fm, respectively. For the states with same radial quantum number $n$, the $\langle r_{\lambda}\rangle_{\frac{1}{2}}$ becomes larger when the orbital angular momentum $L$ increases. However, $\langle r_{\rho}\rangle_{\frac{1}{2}}$ remains almost unchanged. This feature is consistent with the results shown in Figs. 10-11, where the peak of $\omega(r_{\lambda})$ shifts outward with $L$ increment and $\omega(r_{\rho})$ changes little. For the states with same angular momentum $L$, both $\langle r_{\rho}\rangle_{\frac{1}{2}}$ and $\langle r_{\lambda}\rangle_{\frac{1}{2}}$ increase with radial quantum numbers $n$. And the peaks of their density distributions also shift outward obviously as shown in Figs. 12-13.

It is known that the larger the root mean square radius become, the looser the baryons will be. We can see from Tables III-VIII, the mean square radius of the experimentally established baryons are almost less than 0.8 fm. If we roughly take this value as a criterion, all of the predicted states with radius less than 0.8 fm have potentials to be discovered in forthcoming experiments. 

\begin{Large}
\textbf{4 Regge trajectories of single heavy baryons $\Lambda_{Q}$, $\Sigma_{Q}$ and $\Omega_{Q}$}
\end{Large}

The well-known Regge theory was first developed by T.Regge in 1959\cite{Regge1,Regge2}, which preceded the QCD. In 1961, Chew and Frautschi extended this theory and found mesons and baryons lie on linear trajectories of the ($J$,$M^{2}$) plane\cite{Regge3,Regge4}, where $J$ is the angular momentum and $M$ is the mass of hadron. The Regge theory is very successful in studying the strong interactions at high energies and it is an indispensable tool in phenomenological studies. Now, it has been widely used to study the hadronic spectra and was commonly called Regge trajectory.
Actually, there have been lots of studies based on QCD theory to understand the Regge trajectory\cite{Regge5,Regge6,Regge7,Regge8}. Among them, the most simple and straightforward one was proposed by Nambu in 1978\cite{Regge9,Regge10}, where the relation of $J$ and $M^{2}$ was described as,
\begin{eqnarray}
J=\frac{M^{2}}{2\pi \sigma}+c
\end{eqnarray}
where, $\sigma$ is the string tension and $c$ is a constant.

In the present work, we obtained the masses of both orbitally and radially excited charmed and bottom baryons up to rather high excitation numbers, which makes it easy for us to construct the baryon Regge trajectories in ($J$,$M^{2}$) plane.
First, both charmed and bottom baryons are classified into two groups which have natural and unnatural parities with $P=(-1)^{J-1/2}$ and $P=(-1)^{J+1/2}$\cite{quam2}, respectively. Then, with the predicted mass spectra by quark model, we plot the Regge trajectories in the ($J$,$M^{2}$) plane for $\Lambda_{Q}$, $\Sigma_{Q}$ and $\Omega_{Q}$ systems in Figs. 14-19. Here, we use the following definition about the ($J$,$M^{2}$) Regge trajectories\cite{quam2},
\begin{eqnarray}
J=\alpha M^{2}+\alpha_{0}
\end{eqnarray}
where $\alpha$ and $\alpha_{0}$ are slope and intercept. In these figures, the masses calculated in our model are denoted by diamonds and the experimental data are shown by inverted triangle with particle names. The straight lines are obtained by linear fitting of the predicted values. The fitted slopes and intercepts of the Regge trajectories are listed in \textbf{Appendix A.2}(Table IX).

\begin{figure}[h]
\begin{minipage}[h]{0.45\linewidth}
\centering
\includegraphics[height=5cm,width=7cm]{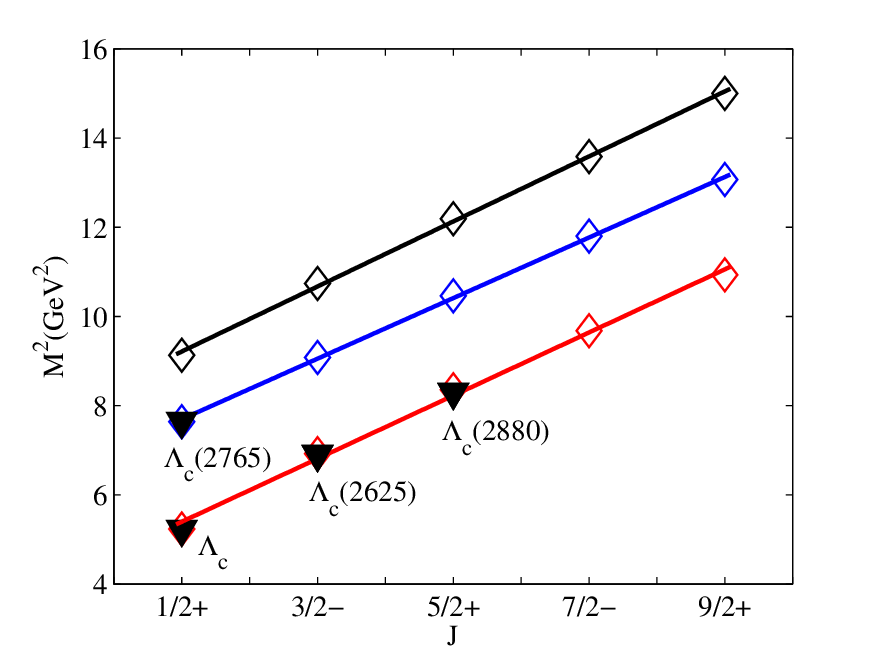}
{(a)}
\end{minipage}
\hfill
\begin{minipage}[h]{0.45\linewidth}
\centering
\includegraphics[height=5cm,width=7cm]{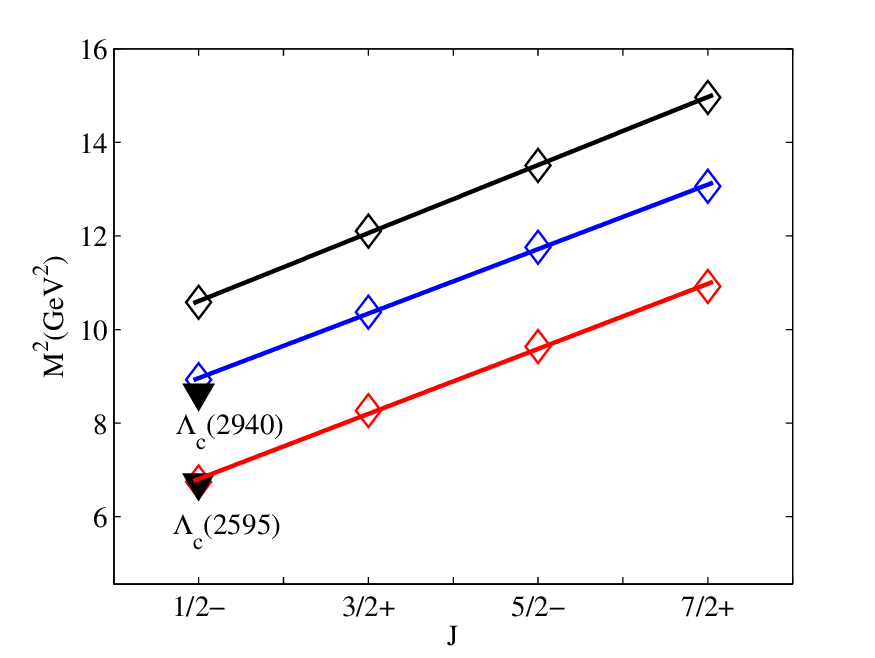}
{(b)}
\end{minipage}
\caption{Parent and daughter ($J$,$M^{2}$) Regge trajectories for the $\Lambda_{c}$ baryons with natural (a) and unnatural (b) parities.
Diamonds are predicted masses. Available experimental data are given by inverted triangle with particle names.}
\end{figure}
\begin{figure}[h]
\begin{minipage}[h]{0.45\linewidth}
\centering
\includegraphics[height=5cm,width=7cm]{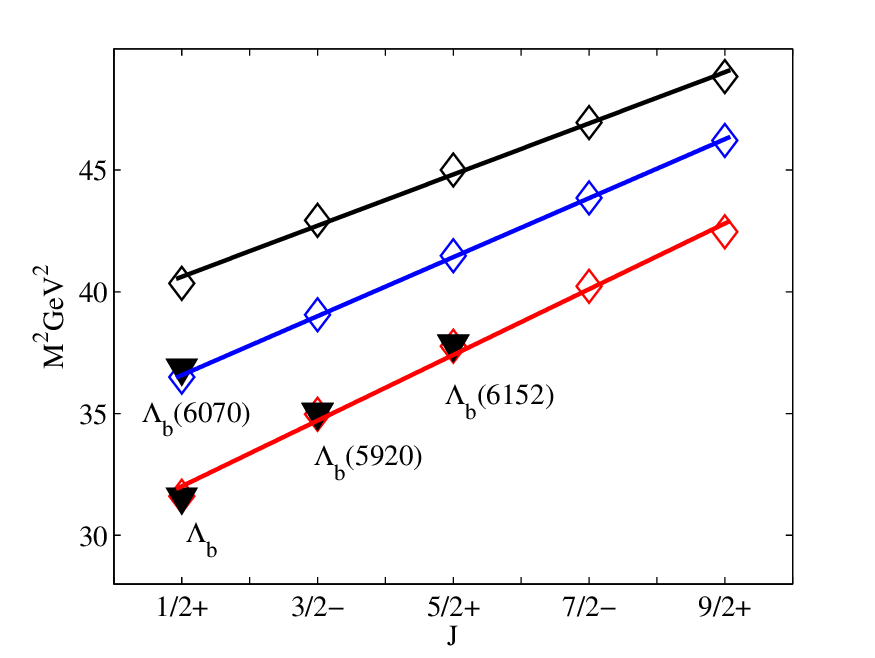}
{(a)}
\end{minipage}
\hfill
\begin{minipage}[h]{0.45\linewidth}
\centering
\includegraphics[height=5cm,width=7cm]{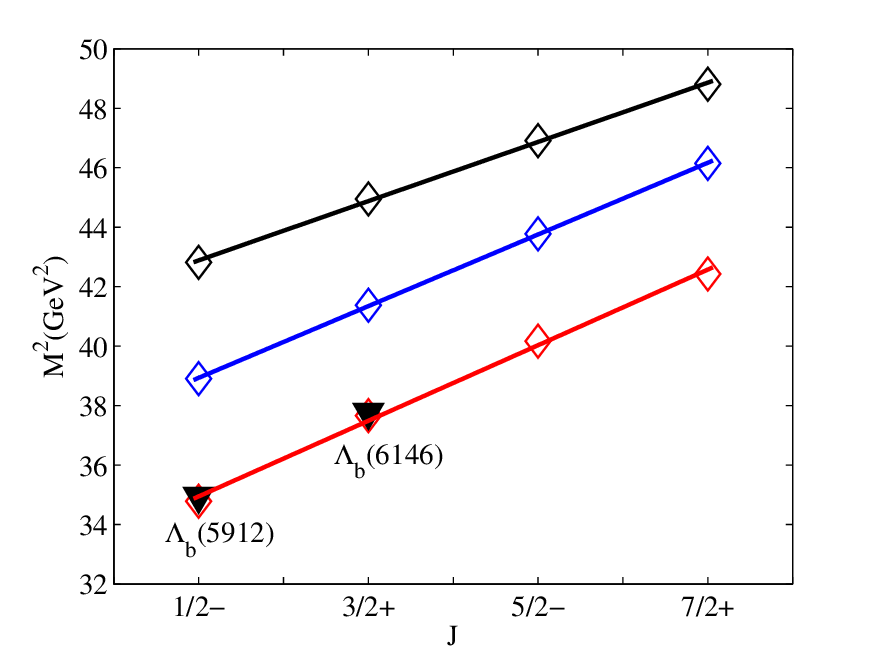}
{(b)}
\end{minipage}
\caption{Same as in FIG.10 for the $\Lambda_{b}$ baryons}
\end{figure}
\begin{figure}[h]
\begin{minipage}[h]{0.45\linewidth}
\centering
\includegraphics[height=5cm,width=7cm]{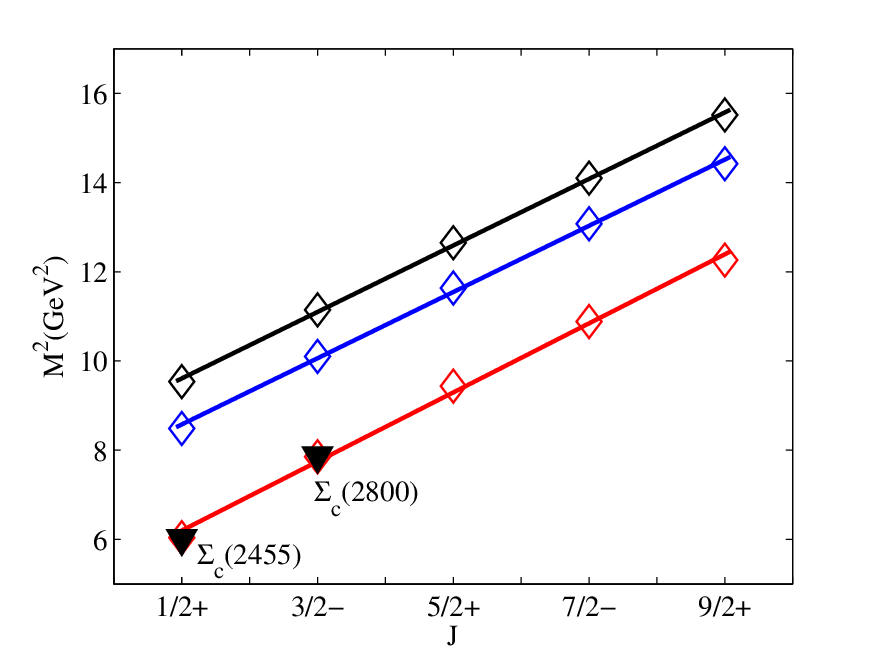}
{(a)}
\end{minipage}
\hfill
\begin{minipage}[h]{0.45\linewidth}
\centering
\includegraphics[height=5cm,width=7cm]{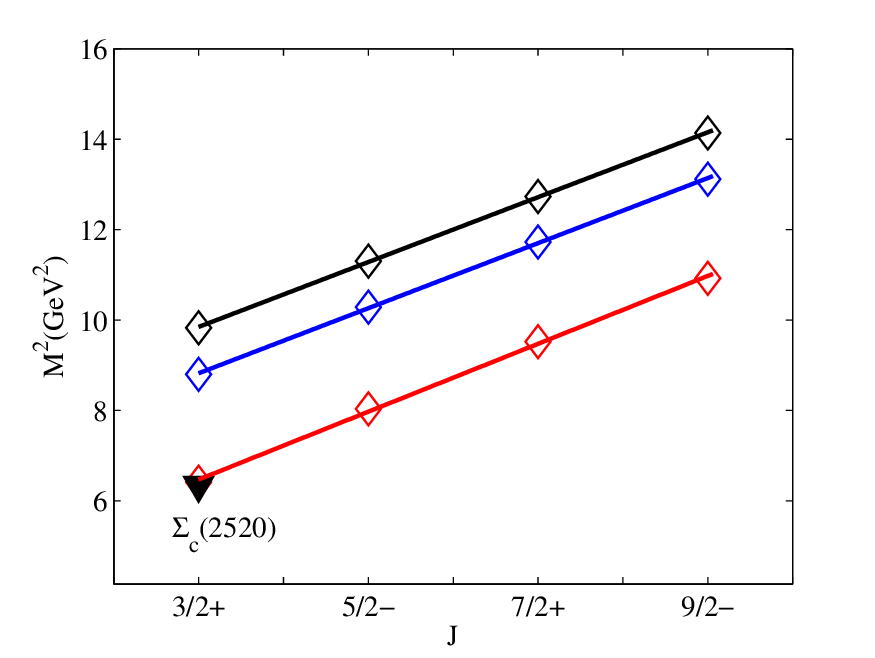}
{(b)}
\end{minipage}
\caption{Same as in FIG.10 for the $\Sigma_{c}$ baryons}
\end{figure}
\begin{figure}[h]
\begin{minipage}[h]{0.45\linewidth}
\centering
\includegraphics[height=5cm,width=7cm]{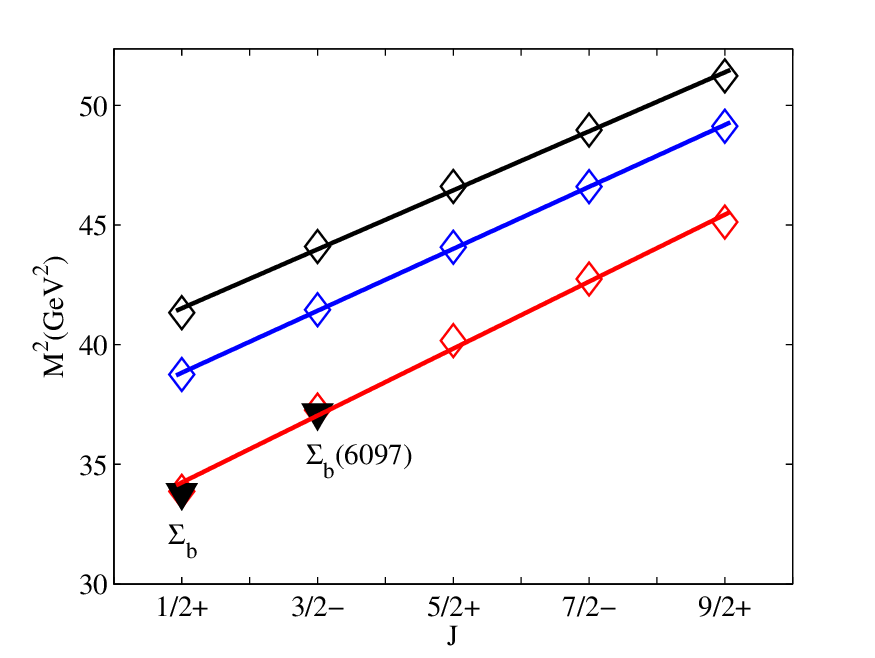}
{(a)}
\end{minipage}
\hfill
\begin{minipage}[h]{0.45\linewidth}
\centering
\includegraphics[height=5cm,width=7cm]{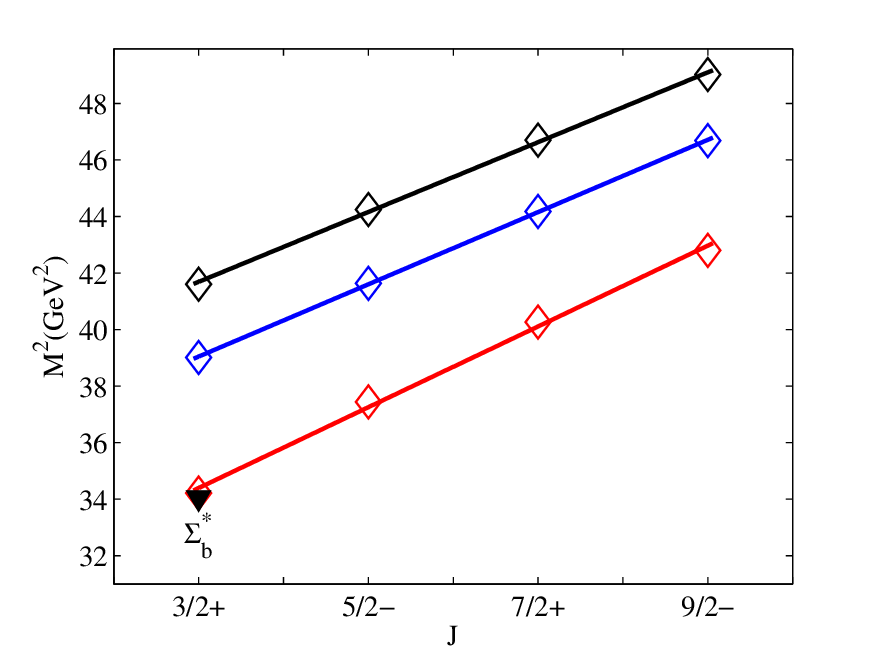}
{(b)}
\end{minipage}
\caption{Same as in FIG.10 for the $\Sigma_{b}$ baryons}
\end{figure}
\begin{figure}[h]
\begin{minipage}[h]{0.45\linewidth}
\centering
\includegraphics[height=5cm,width=7cm]{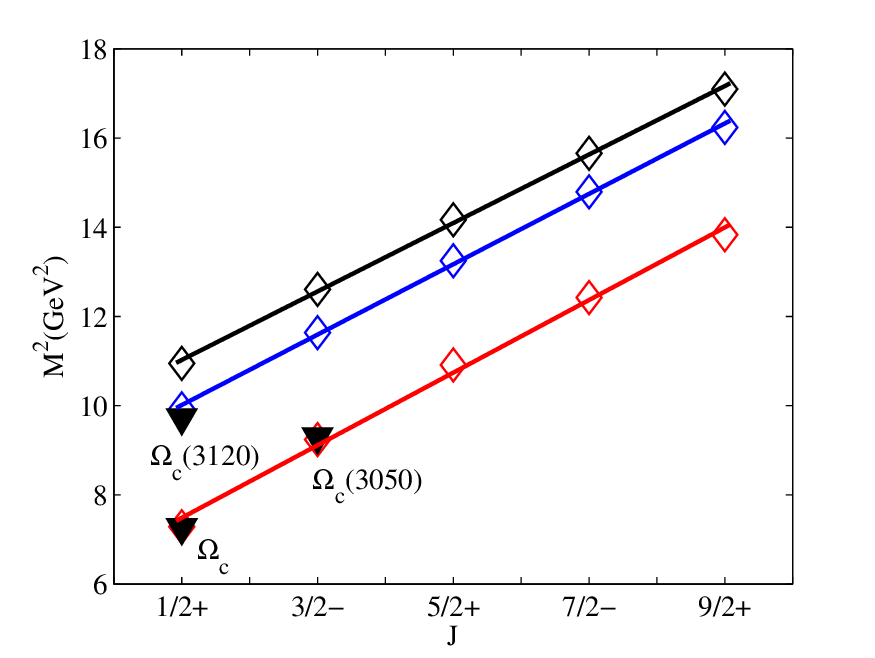}
{(a)}
\end{minipage}
\hfill
\begin{minipage}[h]{0.45\linewidth}
\centering
\includegraphics[height=5cm,width=7cm]{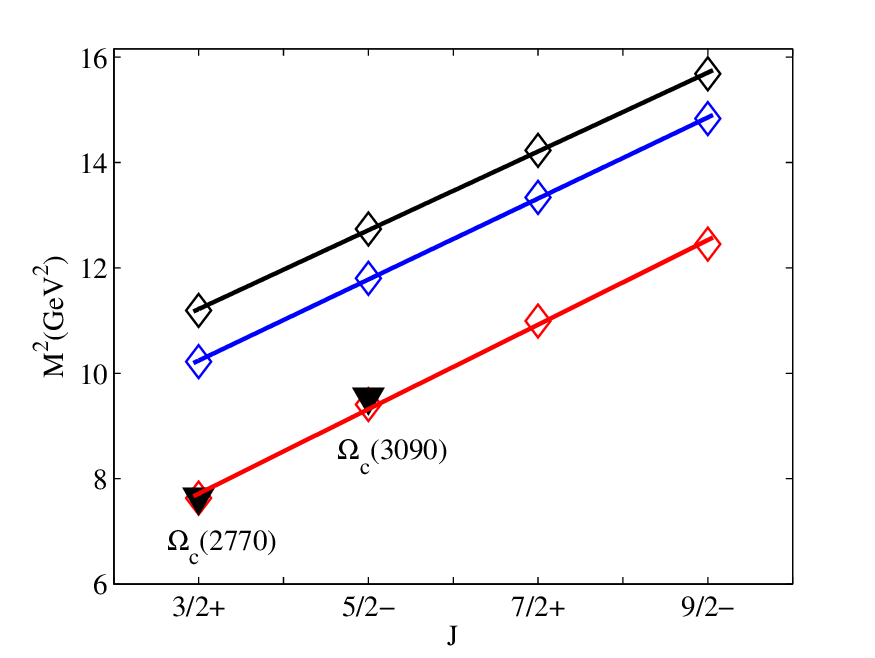}
{(b)}
\end{minipage}
\caption{Same as in FIG.10 for the $\Omega_{c}$ baryons}
\end{figure}
\begin{figure}[h]
\begin{minipage}[h]{0.45\linewidth}
\centering
\includegraphics[height=5cm,width=7cm]{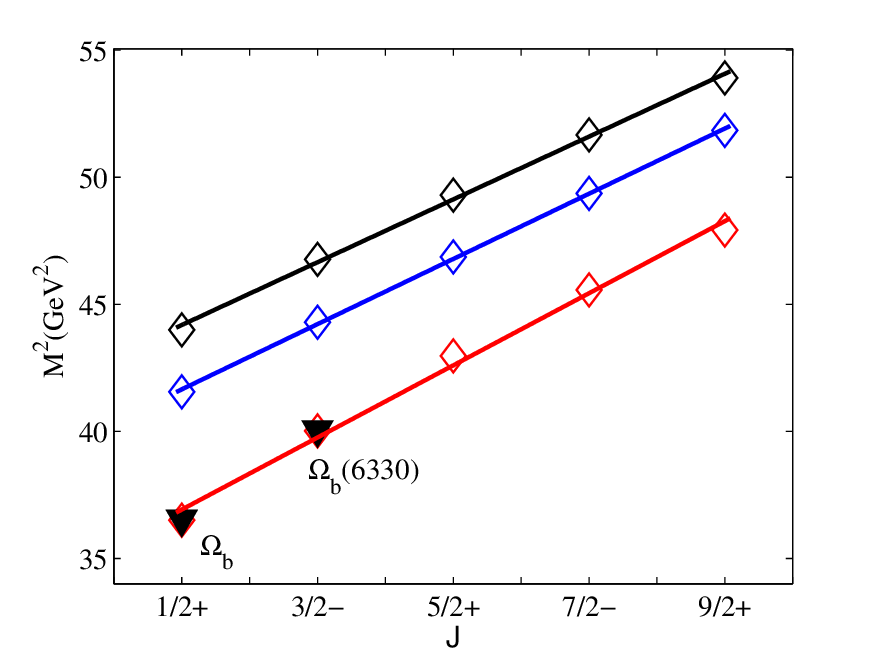}
{(a)}
\end{minipage}
\hfill
\begin{minipage}[h]{0.45\linewidth}
\centering
\includegraphics[height=5cm,width=7cm]{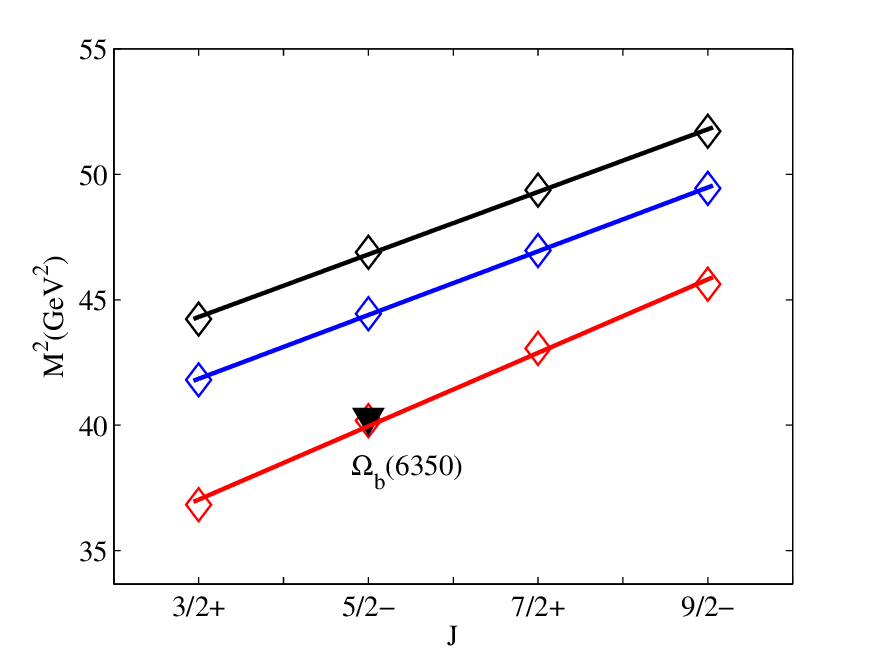}
{(b)}
\end{minipage}
\caption{Same as in FIG.10 for the $\Omega_{b}$ baryons}
\end{figure}
We can see from these figures that all of the predicted masses in our model fit nicely to the linear trajectories in the ($J$,$M^{2}$) plane. These results can help us to assign an accurate position in the mass spectra for observed baryons. There are two trajectories for which three experimental candidates are available. They are the parent trajectories for $\Lambda_{c}(\frac{1}{2}^{+})$ in Fig. 14(a) and $\Lambda_{b}(\frac{1}{2}^{+})$ in Fig. 15(a). There are several trajectories with two experimental candidates such as the parent trajectory for $\Lambda_{c}(\frac{1}{2}^{-})$ in Fig. 14(b), the parent trajectories for $\Sigma_{c}(2455)(\frac{1}{2}^{+})$ and $\Sigma_{b}(\frac{1}{2}^{+})$ in Figs. 16(a) and 17(a), the parent trajectories for $\Omega_{c}(\frac{1}{2}^{+})$, $\Omega_{c}(2770)(\frac{3}{2}^{+})$ and $\Omega_{b}(\frac{1}{2}^{+})$ in Figs. 18(a), 18(b) and 19(a). All the experimental data coincide well with the corresponding Regge trajectories obtained in present work.
Especially, the experimental data for $\Lambda_{c}(2765)$ and $\Lambda_{b}(6070)$ fit well with the Regge trajectories in Fig. 14(a) and Fig. 15(a), which suggests that they can be accommodated in the mass spectra as the $2S$($\frac{1}{2}^{+}$) state. We can see from Fig. 14(b), if $\Lambda_{c}(2940)$ is interpreted as a $2P$($\frac{1}{2}^{-}$) state, the predicted mass is about $40\sim50$ MeV heavier than the experimental data. On the other hand, the Regge trajectories in Fig. 16(a) and Fig. 17(a) show that it is reasonable to assign $\Sigma_{c}(2800)$ and $\Sigma_{b}(6097)$ as the $1P$($\frac{3}{2}^{-}$) state.
Finally, the $\Omega_{c}(3120)$ can be viewed as the first radial excitations($2S$) of $\Omega_{c}$ with quantum numbers $\frac{1}{2}^{+}$ according to the first daughter Regge trajectory in Fig. 18(a). As for the other four observed $\Omega_{c}$ states, the Regge trajectories support assigning $\Omega_{c}(3050)$ and $\Omega_{c}(3090)$ as the first orbital excited states($1P$) with quantum numbers $\frac{3}{2}^{-}$ and $\frac{5}{2}^{-}$, respectively. From Figs. 19(a) and (b), we can see that the situation about $\Omega_{b}(6330)$ and $\Omega_{b}(6350)$ is similar with $\Omega_{c}$. These two states can be interpreted as the $1P$-wave partners of $\Omega_{c}$ with $J^{P}=\frac{3}{2}^{-}$ and $\frac{5}{2}^{-}$.

\begin{Large}
\textbf{5 Conclusions}
\end{Large}

In this work, we have systematically investigate the mass spectra of the single heavy baryons $\Lambda_{Q}$, $\Sigma_{Q}$, and $\Omega_{Q}$. The first feature of this work in studying the mass spectra is that all parameters such as quark masses and parameters of the interquark potential were determined previously according to the ground states of the baryons. Second, the masses and root mean square radius of the ground states and the states up to rather high orbital and radial excitations are systematically studied(in Tables III-VIII). In addition, with the predicted mass spectra, we construct the Regge trajectories in ($J$,$M^{2}$) plane. It is found that the available experimental data nicely fit to the Regge trajectories, which suggests our assignments for the excited heavy baryons are reasonable. In summary, we have obtained the followings:

(1)We have studied the dependencies on the heavy quark mass $m_{Q}$ of the $\lambda$-mode, $\rho$-mode and $\lambda$-$\rho$ mixing mode to see the features of the single heavy baryons. The results show that mixing of these different excited modes is suppressed and only $\lambda$-mode dominates. Basing on this mechanism, we obtained the mass spectra of the $\Lambda_{Q}$, $\Sigma_{Q}$, and $\Omega_{Q}$ baryons with $\lambda$ excitations. It is found that all currently available experimental data can be well reproduced.

(2)We also investigate the root mean square radius and the radial density distributions of the single heavy baryons. We hope these analysis can help us to predict the upper limit of the mass spectra and to search for baryons which have potentials to be found in experiments.

(3)With the predicted masses, we construct the Regge trajectories in ($J$,$M^{2}$) plane and also obtain the slopes and intercepts of the Regge trajectories. Both the predicted masses and the experimental data fit nicely to the linear trajectories in ($J$,$M^{2}$) plane.

(4)According to the predicted mass spectra, a number of experimental states without spin-parity assignments are successfully distinguished and a number of single heavy barons which have good potentials to be discovered in forthcoming experiments are predicted by quark model.

\begin{center}
\begin{Large}
\textbf{Appendix: A}
\end{Large}
\end{center}
\begin{large}
\textbf{A.1 The mass spectra and root mean square radius}
\end{large}
\begin{table*}[htbp]
\begin{ruledtabular}\caption{The root mean square radius (fm) and the mass spectrum (MeV) of the $\Lambda_{c}$ family}
\begin{tabular}{c c c c c c c c c c c c c c c}
$l_{\rho}$  $l_{\lambda}$ $L$ $s$ $j$ & $nL$($J^{P}$) & $\langle r_{\rho}^{2}\rangle^{\frac{1}{2}}$ & $\langle r_{\lambda}^{2}\rangle^{\frac{1}{2}}$ &$M$ &$M_{exp}$ &\cite{quam2} &\cite{quam3} &\cite{quam1}  &\cite{quam19} &\cite{quam10}&\cite{WZG2,WZG3,WZG5,Sum5}  \\ \hline
\multirow{4}{*}{0 0 0 0 0}
~& $1S$($\frac{1}{2}^{+}$) &0.51 &0.44 &2288 & 2286.46\cite{article2A} &2286 &2268 &2265  &2272 &2292 &2240$\pm$90\cite{WZG2} \\
~& $2S$($\frac{1}{2}^{+}$) &0.63 &0.79 & 2764 &2766.6\cite{article2A} &2769 &2791 &2775  &2769 &2669 &2780$\pm$80\cite{WZG2} \\
~& $3S$($\frac{1}{2}^{+}$) &0.99&0.63 & 3022 &~ &3130 &~&~ &~ &~  &~&~ \\
~& $4S$($\frac{1}{2}^{+}$) &0.69 & 1.13& 3135 &~ &3437 &~&~ &~ &~  &~ &~\\ \hline
\multirow{4}{*}{0 1 1 0 1}
~ & $1P$($\frac{1}{2}^{-}$)&0.53 & 0.63& 2596 &2592.25\cite{article2A} &2598 &2625 &2630  &2594 &2559 &2610$\pm$210\cite{WZG3}\\
~ & $2P$($\frac{1}{2}^{-}$)&0.60 &0.96 & 2988 &2939.3\cite{article2B} &2983 &2816 &2780  &2853 &2779 &~&~\\
~ & $3P$($\frac{1}{2}^{-}$)&1.04 & 0.82& 3253 &~ &3303 &~&~ &~ &~  &~&~ \\
~ & $4P$($\frac{1}{2}^{-}$)&0.68 &1.26 & 3308 &~ &3588 &~&~ &~ &~  &~&~ \\ \hline
\multirow{4}{*}{0 1 1 0 1}
~ & $1P$($\frac{3}{2}^{-}$)& 0.55&0.66 & 2631 &2628.1\cite{article2A} &2627 &2636 &2640  &2586 &2559 &2620$\pm$180\cite{WZG5}\\
~ & $2P$($\frac{3}{2}^{-}$)&0.60 &0.99 & 3013 &~ &3005 &2830&2840  &2874 &2779 &~\\
~ & $3P$($\frac{3}{2}^{-}$)&1.06 &0.83 & 3277 &~ &3322 &~&~ &~ &~  &~&~\\
~ & $4P$($\frac{3}{2}^{-}$)&0.67 & 1.25& 3325 &~ &3606 &~&~ &~ &~  &~&~ \\ \hline
\multirow{4}{*}{0 2 2 0 2}
~ & $1D$($\frac{3}{2}^{+}$)&0.55 &0.83 & 2875 & 2856.1\cite{article2A} &2874 &2887&2910  &2848 &2906 &$2830_{-240}^{+150}$\cite{Sum5}\\
~ & $2D$($\frac{3}{2}^{+}$)&0.59 & 1.17& 3220 &~ &3189 &3073 &3035  &3100 &3061 &~&~\\
~ & $3D$($\frac{3}{2}^{+}$)&0.96 & 1.12& 3479 &~ &3480 &~ &~ &~ &~  &~&~\\
~ & $4D$($\frac{3}{2}^{+}$)&0.82 &1.20 & 3497 &~ &3747 &~ &~ &~ &~  &~&~\\ \hline
\multirow{4}{*}{0 2 2 0 2}
~ & $1D$($\frac{5}{2}^{+}$)&0.56 & 0.85& 2891 &2881.63\cite{article2A} &2880 &2887&2910 &~ &~   &$2880_{-290}^{+180}$\cite{Sum5}\\
~ & $2D$($\frac{5}{2}^{+}$)&0.59 & 1.21& 3234 &~ &3209 &~&~ &~ &~ &~ &~&~ \\
~ & $3D$($\frac{5}{2}^{+}$)&1.00 & 1.09& 3492 &~ &3500 &~ &~ &~ &~ &~ &~&~\\
~ & $4D$($\frac{5}{2}^{+}$)&0.77 & 1.21& 3509 &~ &3767 &~&~ &~ &~ &~ &~ &~\\ \hline
\multirow{4}{*}{0 3 3 0 3}
~ & $1F$($\frac{5}{2}^{-}$)&0.56 & 1.00& 3104 &~ &3097 &2872 &2900 &~ &~ &~ &~\\
~ & $2F$($\frac{5}{2}^{-}$)&0.59 & 1.43& 3429 &~ &3375 &~&~ &~ &~ &~ &~ &~\\
~ & $3F$($\frac{5}{2}^{-}$)&0.70 & 1.30& 3675 &~ &3646 &~ &~ &~ &~ &~ &~&~\\
~ & $4F$($\frac{5}{2}^{-}$)&1.05 & 1.09& 3692 &~ &3900 &~ &~ &~ &~ &~ &~&~\\ \hline
\multirow{4}{*}{0 3 3 0 3}
~ & $1F$($\frac{7}{2}^{-}$)&0.56 & 1.02& 3111 &~ &3078 &~&~ &~ &~ &~ &~&~ \\
~ & $2F$($\frac{7}{2}^{-}$)&0.60 & 1.45& 3435 &~ &3393 &~ &~ &~ &~ &~ &~&~\\
~ & $3F$($\frac{7}{2}^{-}$)&0.84 & 1.24& 3685 &~ &3667 &~&~ &~ &~ &~ &~ &~\\
~ & $4F$($\frac{7}{2}^{-}$)&0.95 & 1.14& 3699 &~ &3922 &~ &~ &~ &~ &~ &~&~\\ \hline
\multirow{4}{*}{0 4 4 0 4}
~ & $1G$($\frac{7}{2}^{+}$)&0.55 &1.15 & 3304 &~ &3270 &~ &~ &~ &~ &~ &~&~\\
~ & $2G$($\frac{7}{2}^{+}$)&0.60 & 1.69& 3614 &~ &3546 &~&~ &~ &~ &~ &~&~\\
~ & $3G$($\frac{7}{2}^{+}$)&0.88 &1.24 & 3868 &~ &~ &~ &~ &~ &~ &~ &~&~\\
~ & $4G$($\frac{7}{2}^{+}$)&0.90 &1.19 & 3883 &~ &~ &~ &~ &~ &~ &~ &~&~\\ \hline
\multirow{4}{*}{0 4 4 0 4}
~ & $1G$($\frac{9}{2}^{+}$)&0.55 & 1.16& 3306 &~ &3284 &~ &~ &~ &~ &~ &~&~\\
~ & $2G$($\frac{9}{2}^{+}$)&0.60 &1.70 & 3615 &~ &3564 &~ &~ &~ &~ &~ &~&~\\
~ & $3G$($\frac{9}{2}^{+}$)& 1.05& 1.23& 3873 &~ &~ &~ &~ &~ &~ &~ &~&~\\
~ & $4G$($\frac{9}{2}^{+}$)&0.70 &1.18 & 3891 &~ &~ &~ &~ &~ &~ &~ &~&~\\
\end{tabular}
\end{ruledtabular}
\end{table*}
\begin{table*}[htbp]
\begin{ruledtabular}\caption{The root mean square radius (fm) and the mass spectrum (MeV) of the $\Lambda_{b}$ family}
\begin{tabular}{c c c c c c c c c c c c c}
$l_{\rho}$  $l_{\lambda}$ $L$ $s$ $j$  & $nL$($J^{P}$)& $\langle r_{\rho}^{2}\rangle^{\frac{1}{2}}$ & $\langle r_{\lambda}^{2}\rangle^{\frac{1}{2}}$ &$M$ &$M_{exp}$ &\cite{quam2} &\cite{quam3} &\cite{quam1} &\cite{quam10}&\cite{YHuang}&\cite{WZG2,WZG3} \\ \hline
\multirow{4}{*}{0 0 0 0 0}
~ & $1S$($\frac{1}{2}^{+}$)  & 0.52 & 0.41 &5622 &5619.60\cite{article2A} &5620 &5612 &5585  &5624 &5619.6&5610$\pm$120\cite{WZG2}\\
~ & $2S$($\frac{1}{2}^{+}$)  & 0.60 & 0.72 &6041 &6072.3\cite{article2A} &6089 &6107&~ &~   &~&6080$\pm$90\cite{WZG2}\\
~ & $3S$($\frac{1}{2}^{+}$) & 0.95 & 0.68 &6352 &~ &6455 &~&~   &~ &~&~\\
~ & $4S$($\frac{1}{2}^{+}$) & 0.75 & 1.01 &6388 &~ &6756 &~&~   &~ &~&~\\ \hline
\multirow{4}{*}{0 1 1 0 1}
~ & $1P$($\frac{1}{2}^{-}$) & 0.53 & 0.58 &5898 &5912.19\cite{article2A} &5930 &5939&5912  &5890 &~&5850$\pm$180\cite{WZG3}\\
~ & $2P$($\frac{1}{2}^{-}$) & 0.58 & 0.85 &6238 &~ & 6326&6180&5780   &5853 &~&~\\
~ & $3P$($\frac{1}{2}^{-}$) & 0.66 & 1.27 &6544 &~ &6645 &~&~ &~   &~&~\\
~ & $4P$($\frac{1}{2}^{-}$) & 1.06 & 0.75 &6566 &~ &6917 &~&~ &~   &~&~\\ \hline
\multirow{4}{*}{0 1 1 0 1}
~ & $1P$($\frac{3}{2}^{-}$) & 0.54 & 0.59 &5913 &5920.09\cite{article2A} &5942 &5941&5920  &5890 &~&~\\
~ & $2P$($\frac{3}{2}^{-}$) & 0.58 & 0.86 &6249 &~ &6333 &6191&5840   &5874 &~&~\\
~ & $3P$($\frac{3}{2}^{-}$) & 0.64 & 1.30 &6552 &~ &6651 &~&~ &~  &~ &~\\
~ & $4P$($\frac{3}{2}^{-}$) &1.08  & 0.73 &6575 &~ &6922 &~&~ &~  &~ &~\\ \hline
\multirow{4}{*}{0 2 2 0 2}
~ & $1D$($\frac{3}{2}^{+}$) &0.54  & 0.75 &6137 &6146.2\cite{article2A,LambdaB6146,3P013} &6190 &6181&6145  &6246 &~&$6010_{-120}^{+200}$\cite{Sum4}\\
~ & $2D$($\frac{3}{2}^{+}$) & 0.56 & 0.97 &6432 &~ &6526 &6401&~ &~   &~&~\\
~ & $3D$($\frac{3}{2}^{+}$) & 0.62 & 1.50 &6705 &~ &6811 &~&~ &~  &~ &~\\
~ & $4D$($\frac{3}{2}^{+}$) & 1.11 & 0.83 &6757 &~ &7060 &~&~ &~  &~ &~\\ \hline
\multirow{4}{*}{0 2 2 0 2}
~ & $1D$($\frac{5}{2}^{+}$) & 0.54 & 0.76 &6145 &6152.5\cite{article2A,LambdaB6146,3P013} &6196 &6183&6165 &~   &~&$6010_{-130}^{+200}$\cite{Sum4}\\
~ & $2D$($\frac{5}{2}^{+}$) & 0.56 & 0.98 &6440 &~ &6531 &6422&~ &~   &~&~\\
~ & $3D$($\frac{5}{2}^{+}$) & 0.62 & 1.50 &6709 &~ &6814 &~&~ &~  &~ &~\\
~ & $4D$($\frac{5}{2}^{+}$) & 1.11 & 0.84 &6763 &~ &7063 &~&~ &~  &~ &~\\ \hline
\multirow{4}{*}{0 3 3 0 3}
~ & $1F$($\frac{5}{2}^{-}$) & 0.54 & 0.91 &6338 &~ &6408 &6206 &6205 &~  &~ &~&~\\
~ & $2F$($\frac{5}{2}^{-}$) & 0.54 & 1.08 &6616 &~ &6705 &~&~ &~  &~ &~&~\\
~ & $3F$($\frac{5}{2}^{-}$) & 0.61 & 1.63 &6849 &~ &6964 &~&~ &~  &~ &~&~\\
~ & $4F$($\frac{5}{2}^{-}$) & 1.13 & 0.98 &6932 &~ &7196 &~&~ &~  &~ &~&~\\ \hline
\multirow{4}{*}{0 3 3 0 3}
~ & $1F$($\frac{7}{2}^{-}$) & 0.54 & 0.91 &6343 &~ &6411 &~&~ &~  &~ &~&~\\
~ & $2F$($\frac{7}{2}^{-}$) & 0.54 & 1.09 &6622 &~ &6708 &~&~ &~  &~ &~&~\\
~ & $3F$($\frac{7}{2}^{-}$) & 0.61 & 1.63 &6852 &~ &6966 &~&~ &~  &~ &~&~\\
~ & $4F$($\frac{7}{2}^{-}$) & 1.13 & 0.98 &6936 &~ &7197 &~&~ &~  &~ &~&~\\ \hline
\multirow{4}{*}{0 4 4 0 4}
~ & $1G$($\frac{7}{2}^{+}$) & 0.54 & 1.05 &6514 &~ &6598 &6433&~ &~  &~ &~&~\\
~ & $2G$($\frac{7}{2}^{+}$) & 0.53 & 1.20 &6793 &~ &6867 &~&~ &~  &~ &~&~\\
~ & $3G$($\frac{7}{2}^{+}$) & 0.59 & 1.72 &6986 &~ &~ &~&~ &~  &~ &~&~\\
~ & $4G$($\frac{7}{2}^{+}$) & 1.13 & 1.11 &7093 &~ &~ &~&~ &~  &~ &~&~\\ \hline
\multirow{4}{*}{0 4 4 0 4}
~ & $1G$($\frac{9}{2}^{+}$) & 0.54 & 1.06 &6517 &~ &6599 &~&~ &~  &~ &~&~\\
~ & $2G$($\frac{9}{2}^{+}$) & 0.53 & 1.22 &6798 &~ &6868 &~&~ &~  &~ &~&~\\
~ & $3G$($\frac{9}{2}^{+}$) & 0.59 & 1.72 &6989 &~ &~ &~&~ &~  &~ &~&~\\
~ & $4G$($\frac{9}{2}^{+}$) & 1.13 & 1.11 &7095 &~ &~ &~&~ &~  &~ &~&~\\
\end{tabular}
\end{ruledtabular}
\end{table*}
\begin{table*}[htbp]
\begin{ruledtabular}\caption{The root mean square radius (fm) and the mass spectrum (MeV) of the $\Sigma_{c}$ family}
\begin{tabular}{c c c c c c c c c c c c c}
$l_{\rho}$  $l_{\lambda}$ $L$ $s$ $j$  &$nL$($J^{P}$)& $\langle r_{\rho}^{2}\rangle^{\frac{1}{2}}$ & $\langle r_{\lambda}^{2}\rangle^{\frac{1}{2}}$ &$M$ &$M_{exp}$  &\cite{quam2} &\cite{quam3} &\cite{quam1}&\cite{quam19}&\cite{quam10}& \cite{WZG1,Sum2} \\ \hline
\multirow{4}{*}{0 0 0 1 1}
        & $1S$($\frac{1}{2}^{+}$) & 0.62& 0.45& 2457 & 2453.8\cite{article2A} & 2443 & 2455 &2440  &2459 &2448 &2400$\pm260$\cite{WZG1}  \\
        & $2S$($\frac{1}{2}^{+}$) & 0.86&0.72 & 2913 & ~ & 2901 & 2958 & 2890  &2947 &2793 & ~  \\
        & $3S$($\frac{1}{2}^{+}$) & 0.94& 0.70& 3088 & ~& 3271  & ~ & ~ & ~  & ~ & ~   \\
        & $4S$($\frac{1}{2}^{+}$) & 0.95& 1.05& 3295 & ~& 3581 & ~ & ~ & ~  & ~ & ~  \\ \hline
\multirow{4}{*}{0 0 0 1 1}
        & $1S$($\frac{3}{2}^{+}$) & 0.64& 0.49& 2532 & 2518.5\cite{article2A}& 2519 & 2519 & 2495  &2539 & 2505 &  \\
        & $2S$($\frac{3}{2}^{+}$) &0.83 &0.79 & 2967  & & 2936 & 2995 & 2985 &  3010 & 2825 & ~  \\
        & $3S$($\frac{3}{2}^{+}$) &0.99 & 0.71& 3135 & ~& 3293  & ~ & ~ & ~  & ~ & ~  \\
        & $4S$($\frac{3}{2}^{+}$) &0.90 & 1.09& 3328 & ~& 3598 & ~ & ~ & ~  & ~ & ~  \\ \hline
\multirow{4}{*}{0 1 1 1 0}
        & $1P$($\frac{1}{2}^{-}$) &0.68 & 0.69& 2823 &  & 2799 & ~ & ~  & ~ & ~ & ~  \\
        & $2P$($\frac{1}{2}^{-}$) &0.80 & 1.00& 3196 & ~& 3172 & ~ & ~  & ~ & ~ & ~  \\
        & $3P$($\frac{1}{2}^{-}$) &1.07 & 0.84& 3353 & ~& 3488 & ~ & ~  & ~ & ~ & ~  \\
        & $4P$($\frac{1}{2}^{-}$) & 0.83&1.22 & 3511 & ~& 3770 & ~ & ~  & ~ & ~ & ~ \\ \hline
\multirow{4}{*}{0 1 1 1 1}
        & $1P$($\frac{1}{2}^{-}$) &0.67 &0.67 & 2809 & ~& 2713  & 2748 & 2765 & 2769  & 2706 & 2740$\pm$200\cite{Sum2}  \\
        & $2P$($\frac{1}{2}^{-}$) &0.80 &0.98 & 3185 & ~& 3125 & 2768 & 2770 & 2817  & 2791 &   \\
        & $3P$($\frac{1}{2}^{-}$) &1.06 & 0.84& 3343 & ~& 3455 & ~ & ~ & ~  & ~ & ~  \\
        & $4P$($\frac{1}{2}^{-}$) &0.84 & 1.22& 3501 & ~& 3743  & ~ & ~  & ~ & ~ & ~  \\ \hline
\multirow{4}{*}{0 1 1 1 1}
        & $1P$($\frac{3}{2}^{-}$) &0.68 &0.69 & 2829 & & 2798 & ~ & ~ & ~  & ~ & ~  \\
        & $2P$($\frac{3}{2}^{-}$) & 0.80& 1.00& 3202 & ~& 3172 & ~ & ~ & ~  & ~ & ~  \\
        & $3P$($\frac{3}{2}^{-}$) & 1.07& 0.85& 3358 & ~& 3486 & ~ & ~ & ~  & ~ & ~  \\
        & $4P$($\frac{3}{2}^{-}$) &0.83 & 1.22& 3516 & ~& 3768 & ~ & ~ & ~  & ~ & ~  \\ \hline
\multirow{4}{*}{0 1 1 1 2}
        & $1P$($\frac{3}{2}^{-}$) & 0.67&0.67 & 2802 & 2806\cite{article2A} & 2773 & 2763 & 2770  & 2799 & 2706 &2740$\pm$200\cite{Sum2}  \\
        & $2P$($\frac{3}{2}^{-}$) &0.80 & 0.97& 3179 & ~& 3151  & 2776 & 2805  &2815 & 2791 & \\
        & $3P$($\frac{3}{2}^{-}$) &1.06 & 0.84& 3338 & ~ & 3469 & ~ & ~ & ~ & ~  & ~  \\
        & $4P$($\frac{3}{2}^{-}$) &0.84 & 1.23& 3496 & ~& 3753 & ~ & ~ & ~ & ~  & ~   \\ \hline
\multirow{4}{*}{0 1 1 1 2}
        & $1P$($\frac{5}{2}^{-}$) &0.68 & 0.70& 2835 & ~& 2789 & 2790 & 2815 & ~  & ~ &   \\
        & $2P$($\frac{5}{2}^{-}$) & 0.80&1.02 & 3207 & ~& 3161 & ~ & ~ & ~ & ~  & ~   \\
        & $3P$($\frac{5}{2}^{-}$) &1.08 &0.85 & 3362 & ~& 3475 & ~ & ~ & ~ & ~  & ~  \\
        & $4P$($\frac{5}{2}^{-}$) &0.82 &1.22 & 3521 & ~& 3757  & ~ & ~ & ~ & ~ & ~  \\ \hline
\multirow{4}{*}{0 2 2 1 1}
        & $1D$($\frac{1}{2}^{+}$) &0.69 & 0.87& 3073 & ~& 3041  & ~ & ~ & ~  & ~ & ~  \\
        & $2D$($\frac{1}{2}^{+}$) &0.78 & 1.22& 3413 & ~& 3370 & ~ & ~ & ~  & ~ & ~  \\
        & $3D$($\frac{1}{2}^{+}$) &1.12 &0.98 & 3558 & ~& ~ & ~ & ~ & ~ & ~  & ~  \\
        & $4D$($\frac{1}{2}^{+}$) &0.78 &1.27 & 3696 & ~& ~ & ~ & ~ & ~ & ~  & ~ \\ \hline
\multirow{4}{*}{0 2 2 1 1}
        & $1D$($\frac{3}{2}^{+}$) &0.70 & 0.88& 3084 & ~& 3043  & ~ & ~ & ~  & ~ & ~ \\
        & $2D$($\frac{3}{2}^{+}$) & 0.78&1.25 & 3422  & ~& 3366 & ~ & ~ & ~  & ~ & ~ \\
        & $3D$($\frac{3}{2}^{+}$) &1.13 &0.99 & 3567 & ~& ~ & ~ & ~ & ~ & ~  & ~ \\
        & $4D$($\frac{3}{2}^{+}$) &0.77 & 1.25& 3709 & ~& ~ & ~ & ~ & ~ & ~  & ~  \\ \hline
\multirow{4}{*}{0 2 2 1 2}
        & $1D$($\frac{3}{2}^{+}$) &0.69 & 0.87& 3073 & ~& 3040 & ~ & ~ & ~ & ~ & ~  \\
        & $2D$($\frac{3}{2}^{+}$) &0.78 & 1.22& 3413 & ~& 3364 & ~ & ~ & ~  & ~ & ~  \\
        & $3D$($\frac{3}{2}^{+}$) &1.12 & 0.98& 3558 & ~& ~ & ~ & ~ & ~ & ~  & ~  \\
        & $4D$($\frac{3}{2}^{+}$) &0.78 &1.27 & 3696 & ~& ~ & ~ & ~ & ~ & ~  & ~  \\ \hline
\multirow{4}{*}{0 2 2 1 2}
        & $1D$($\frac{5}{2}^{+}$) &0.70 & 0.88& 3085 & ~& 3038 & 3003 & 3065 & ~  & ~ & ~  \\
        & $2D$($\frac{5}{2}^{+}$) &0.78 & 1.25& 3423 & ~& 3365 & ~ & ~ & ~ & ~  & ~ \\
        & $3D$($\frac{5}{2}^{+}$) &1.13 & 0.99& 3567 & ~& ~ & ~ & ~ & ~ & ~  & ~ \\
        & $4D$($\frac{5}{2}^{+}$) &0.77 &1.25 & 3710 & ~& ~ & ~ & ~ & ~ & ~  & ~  \\
\end{tabular}
\end{ruledtabular}
\end{table*}
\begin{table*}[htbp]
\begin{ruledtabular}
\begin{tabular}{c c c c c c|c c c c c c}
$l_{\rho}$  $l_{\lambda}$ $L$ $s$ $j$  &$nL$($J^{P}$) & $\langle r_{\rho}^{2}\rangle^{\frac{1}{2}}$ & $\langle r_{\lambda}^{2}\rangle^{\frac{1}{2}}$ &$M$  &\cite{quam2}&$l_{\rho}$  $l_{\lambda}$ $L$ $s$ $j$  &$nL$($J^{P}$)& $\langle r_{\rho}^{2}\rangle^{\frac{1}{2}}$ & $\langle r_{\lambda}^{2}\rangle^{\frac{1}{2}}$ &$M$  &\cite{quam2}   \\ \hline
\multirow{4}{*}{0 2 2 1 3}
        & $1D$($\frac{5}{2}^{+}$) &0.69 & 0.87& 3072  & 3023&\multirow{4}{*}{0 3 3 1 4}& $1F$($\frac{7}{2}^{-}$) &0.70 &1.04 & 3299  & 3227    \\
        & $2D$($\frac{5}{2}^{+}$) &0.78 & 1.22& 3411  & 3349 &~& $2F$($\frac{7}{2}^{-}$)                         &0.77 & 1.48& 3616    & ~    \\
        & $3D$($\frac{5}{2}^{+}$) &1.12 & 0.98& 3557  & ~&~& $3F$($\frac{7}{2}^{-}$)                             &1.16 & 1.11& 3755  & ~     \\
        & $4D$($\frac{5}{2}^{+}$) &0.78 & 1.27& 3695  & ~&~& $4F$($\frac{7}{2}^{-}$)                             &0.73 & 1.23& 3897  & ~   \\ \hline
\multirow{4}{*}{0 2 2 1 3}
        & $1D$($\frac{7}{2}^{+}$) & 0.70& 0.89& 3086  & 3013 &\multirow{4}{*}{0 3 3 1 4}& $1F$($\frac{9}{2}^{-}$) &0.70 & 1.05& 3305  & 3209  \\
        & $2D$($\frac{7}{2}^{+}$) &0.78 & 1.25& 3425  & 3342&~ & $2F$($\frac{9}{2}^{-}$)                          &0.78 & 1.50& 3622  & ~  \\
        & $3D$($\frac{7}{2}^{+}$) &1.13 & 0.99& 3569  & ~&~& $3F$($\frac{9}{2}^{-}$)                              &1.06 & 1.12& 3760  & ~    \\
        & $4D$($\frac{7}{2}^{+}$) &0.77 &1.25 & 3712  & ~&~& $4F$($\frac{9}{2}^{-}$)                              &0.74 & 1.22& 3912  & ~    \\ \hline
\multirow{4}{*}{0 3 3 1 2}
        & $1F$($\frac{3}{2}^{-}$) &0.70 & 1.04& 3299  & 3288 &\multirow{4}{*}{0 4 4 1 3}& $1G$($\frac{5}{2}^{+}$) &0.70 & 1.19 & 3501  & 3495  \\
        & $2F$($\frac{3}{2}^{-}$) &0.77 &1.48 & 3617  & ~&~& $2G$($\frac{5}{2}^{+}$)                             &0.77 & 1.71& 3798  & ~    \\
        & $3F$($\frac{3}{2}^{-}$) & 1.16& 1.11& 3755  & ~&~& $3G$($\frac{7}{2}^{+}$)                             &1.18 & 1.23& 3938  & ~   \\
        & $4F$($\frac{3}{2}^{-}$) &0.73 & 1.23& 3898  & ~&~& $4G$($\frac{5}{2}^{+}$)                            &0.73 & 1.20 & 4108  & ~    \\ \hline
\multirow{4}{*}{0 3 3 1 2}
        & $1F$($\frac{5}{2}^{-}$) & 0.70& 1.04& 3304  & 3283&\multirow{4}{*}{0 4 4 1 3}& $1G$($\frac{7}{2}^{+}$) & 0.70&  1.20& 3502  & 3483    \\
        & $2F$($\frac{5}{2}^{-}$) & 0.78& 1.50& 3621  & ~&~& $2G$($\frac{7}{2}^{+}$)                            &0.77 &1.72 & 3798  & ~     \\
        & $3F$($\frac{5}{2}^{-}$) &1.6 & 1.12& 3759  & ~&~& $3G$($\frac{7}{2}^{+}$)                           &1.18 & 1.24& 3939  & ~   \\
        & $4F$($\frac{5}{2}^{-}$) &0.74 & 1.22& 3910  & ~&~& $4G$($\frac{7}{2}^{+}$)                           &0.74 & 1.19& 4118  & ~    \\ \hline
\multirow{4}{*}{0 3 3 1 3}
        & $1F$($\frac{5}{2}^{-}$) &0.70 &1.04 & 3299  & 3254&\multirow{4}{*}{0 4 4 1 4}& $1G$($\frac{7}{2}^{+}$) & 0.70& 1.19& 3501  & 3444    \\
        & $2F$($\frac{5}{2}^{-}$) &0.77 & 1.48& 3617  & ~&~ & $2G$($\frac{7}{2}^{+}$)                            &0.77 & 1.71& 3798  & ~    \\
        & $3F$($\frac{5}{2}^{-}$) &1.16 & 1.11& 3755  & ~&~ & $3G$($\frac{7}{2}^{+}$)                           &1.18 & 1.23& 3938  & ~    \\
        & $4F$($\frac{5}{2}^{-}$) &0.73 & 1.23& 3898  & ~&~ & $4G$($\frac{7}{2}^{+}$)                           &0.73 &1.19 & 4108  & ~  \\ \hline
\multirow{4}{*}{0 3 3 1 3}
        & $1F$($\frac{7}{2}^{-}$) &0.70 & 1.05& 3305  & 3253&\multirow{4}{*}{0 4 4 1 4} & $1G$($\frac{9}{2}^{+}$) &0.70 & 1.20 & 3502  & 3442     \\
        & $2F$($\frac{7}{2}^{-}$) & 0.78& 1.50& 3621  & ~&~& $2G$($\frac{9}{2}^{+}$)                              &0.77 & 1.72& 3798 & ~   \\
        & $3F$($\frac{7}{2}^{-}$) &1.16 & 1.12& 3760  & ~&~& $3G$($\frac{9}{2}^{+}$)                              &1.18 & 1.24& 3939 & ~    \\
        & $4F$($\frac{7}{2}^{-}$) &0.74 &1.22 & 3911  & ~&~& $4G$($\frac{7}{2}^{+}$)                             &0.75 & 1.19& 4118  & ~ \\
\end{tabular}
\end{ruledtabular}
\end{table*}
\begin{table*}[htbp]
\begin{ruledtabular}\caption{The root mean square radius (fm) and the mass spectrum (MeV) of the $\Sigma_{b}$ family}
\begin{tabular}{c c c c c c c c c c c c c}
$l_{\rho}$  $l_{\lambda}$ $L$ $s$ $j$  &$nL$($J^{P}$)& $\langle r_{\rho}^{2}\rangle^{\frac{1}{2}}$ & $\langle r_{\lambda}^{2}\rangle^{\frac{1}{2}}$ &$M$ &$M_{exp}$  &\cite{quam2}&\cite{quam3} &\cite{quam1}&\cite{quam10}& \cite{YHuang} & \cite{WZG1,Sum2} \\ \hline
\multirow{4}{*}{0 0 0 1 1}
        & $1S$($\frac{1}{2}^{+}$) & 0.63 & 0.43 &5820  & 5815.6\cite{article2A} & 5808& 5833 &5795  &5789& 5815.5& 5800$\pm190$\cite{WZG1}\\
        & $2S$($\frac{1}{2}^{+}$) & 0.78 & 0.71 &6225  & ~ & 6213 & 6294 & ~  & ~& ~ \\
        & $3S$($\frac{1}{2}^{+}$) & 1.02 & 0.59 &6430  & ~ & 6575& ~ & ~ & ~ & ~ \\
        & $4S$($\frac{1}{2}^{+}$) & 0.85 & 1.07 &6566  & ~ & 6869& ~ & ~ & ~ & ~ \\ \hline
\multirow{4}{*}{0 0 0 1 1 }
        & $1S$($\frac{3}{2}^{+}$) & 0.64 & 0.45 &5849  & 5834.7\cite{article2A} & 5834 &5858 & 5805  & 5844& ~ \\
         & $2S$($\frac{3}{2}^{+}$) & 0.77 & 0.73 &6246  & ~ & 6226& 6308 & ~  & ~& ~ \\
        & $3S$($\frac{3}{2}^{+}$) & 1.04 & 0.60 &6450  & ~ & 6583& ~ & ~  & ~& ~ \\
       & $4S$($\frac{3}{2}^{+}$) & 0.83 & 1.09 &6579 & ~ & 6876& ~ & ~  & ~ & ~\\ \hline
\multirow{4}{*}{0 1 1 1 0}
        & $1P$($\frac{1}{2}^{-}$)  & 0.67 & 0.63 & 6113 & ~ & 6101& ~ & ~  & ~& ~ \\
        & $2P$($\frac{1}{2}^{-}$)  & 0.74 & 0.89 & 6447  & ~ & 6440& ~ & ~  & ~& ~ \\
        & $3P$($\frac{1}{2}^{-}$)  & 1.09 & 0.74 & 6648  & ~ & 6756& ~ & ~  & ~& ~ \\
        & $4P$($\frac{1}{2}^{-}$)  & 0.81 & 1.28 & 6739 & ~ & 7024& ~ & ~  & ~ & ~\\ \hline
\multirow{4}{*}{0 1 1 1 1}
        & $1P$($\frac{1}{2}^{-}$)  & 0.67 & 0.62 & 6107 &   & 6095& 6099 & 6070 & 6039& &6000$\pm$180\cite{Sum2} \\
        & $2P$($\frac{1}{2}^{-}$)  & 0.74 & 0.88 & 6442  & ~ & 6430 & 6106 & ~  & ~& ~ \\
        & $3P$($\frac{1}{2}^{-}$)  & 1.09 & 0.74 & 6643 & ~ & 6742& ~ & ~ & ~  & ~\\
        & $4P$($\frac{1}{2}^{-}$)  & 0.81 & 1.28 & 6736 & ~ & 7008& ~ & ~ & ~ & ~ \\ \hline
\multirow{4}{*}{0 1 1 1 1}
        & $1P$($\frac{3}{2}^{-}$)  & 0.67 & 0.63& 6116 & ~ & 6096& 6101 &6070  &6039& ~ \\
        & $2P$($\frac{3}{2}^{-}$)  & 0.74 & 0.89 & 6450 & ~  & 6430&~ & ~  & ~& ~ \\
        & $3P$($\frac{3}{2}^{-}$)  & 1.09 & 0.74 & 6650 & ~ & 6742 & ~ & ~  & ~& ~ \\
        & $4P$($\frac{3}{2}^{-}$)  & 0.81 & 1.28 & 6741 & ~ & 7009 & ~ & ~  & ~& ~ \\ \hline
\multirow{4}{*}{0 1 1 1 2}
        & $1P$($\frac{3}{2}^{-}$)  & 0.66 & 0.62 & 6104  & 6098.0\cite{article2A,6090,3P012} & 6087& 6105 & ~  & ~& &6000$\pm$180\cite{Sum2} \\
        & $2P$($\frac{3}{2}^{-}$)  & 0.74 & 0.88 & 6439 & ~  & 6423 & ~ & ~  & ~& ~ \\
        & $3P$($\frac{3}{2}^{-}$)  & 1.09 & 0.74 & 6641 & ~  & 6736& ~ & ~  & ~& ~ \\
        & $4P$($\frac{3}{2}^{-}$)  & 0.81 & 1.28 & 6734 & ~  & 7003& ~ & ~  & ~& ~ \\ \hline
\multirow{4}{*}{0 1 1 1 2}
         & $1P$($\frac{5}{2}^{-}$)  & 0.67 & 0.63 & 6119  &  & 6084 & 6172 & ~  & ~& ~ \\
         & $2P$($\frac{5}{2}^{-}$)  & 0.74 & 0.89 & 6452  & ~ & 6421 & ~ & ~  & ~& ~ \\
         & $3P$($\frac{5}{2}^{-}$)  & 1.10 & 0.74 & 6652 & ~  & 6732& ~ & ~  & ~& ~ \\
         & $4P$($\frac{5}{2}^{-}$)  & 0.81 & 1.28 & 6743 & ~ & 6999 & ~ & ~  & ~& ~ \\ \hline
\multirow{4}{*}{0 2 2 1 1}
         & $1D$($\frac{1}{2}^{+}$)  & 0.67 & 0.79 & 6338 & ~  & 6311& ~ & ~  & ~& ~ \\
        & $2D$($\frac{1}{2}^{+}$)  & 0.72 & 1.02 & 6639 & ~ & 6636 & ~ & ~  & ~& ~ \\
         & $3D$($\frac{1}{2}^{+}$)  & 1.03 & 0.87 & 6828 & ~  & 3626& ~ & ~  & ~& ~ \\
         & $4D$($\frac{1}{2}^{+}$)  & 0.78 & 1.43 & 6892 & ~ & 6647 & ~ & ~  & ~ & ~\\ \hline
\multirow{4}{*}{0 2 2 1 1}
         & $1D$($\frac{3}{2}^{+}$)  & 0.68 & 0.80 & 6344 & ~  & 6326& ~ & ~ & ~ & ~ \\
         & $2D$($\frac{3}{2}^{+}$)  & 0.72 & 1.03 & 6645 & ~  & 6647& ~ & ~ & ~ & ~ \\
         & $3D$($\frac{3}{2}^{+}$)  & 1.13 & 0.88 & 6833 & ~  & ~& ~ & ~ & ~ & ~ \\
         & $4D$($\frac{3}{2}^{+}$)  & 0.78 & 1.43 & 6896 & ~  & ~& ~ & ~ & ~ &  ~\\ \hline
\multirow{4}{*}{0 2 2 1 2}
         & $1D$($\frac{3}{2}^{+}$)  & 0.68 & 0.79 & 6338 & ~  & 6285& ~ & ~ & ~ & ~ \\
         & $2D$($\frac{3}{2}^{+}$)  & 0.72 & 1.02 & 6639 & ~  & 6612& ~ & ~ & ~ & ~ \\
         & $3D$($\frac{3}{2}^{+}$)  & 1.13 & 0.88 & 6828 & ~  & ~& ~ & ~ & ~ & ~ \\
         & $4D$($\frac{3}{2}^{+}$)  & 0.78 & 1.43 & 6892 & ~ & ~& ~ & ~ & ~ & ~ \\ \hline
\multirow{4}{*}{0 2 2 1 2}
         & $1D$($\frac{5}{2}^{+}$)  & 0.68 & 0.80 & 6345 & ~  & 6284& ~ & ~ & ~ & ~ \\
         & $2D$($\frac{5}{2}^{+}$)  & 0.72 & 1.03 & 6646 & ~ & 6612& ~ & ~ & ~ & ~ \\
         & $3D$($\frac{5}{2}^{+}$)  & 1.13 & 0.88 & 6833 & ~  & ~& ~ & ~ & ~ & ~ \\
         & $4D$($\frac{5}{2}^{+}$)  & 0.78 & 1.43 & 6896  & ~ & ~& ~ & ~ & ~& ~ \\
\end{tabular}
\end{ruledtabular}
\end{table*}
\begin{table*}[htbp]
\begin{ruledtabular}
\begin{tabular}{c c c c c c c|c c c c c c c}
$l_{\rho}$  $l_{\lambda}$ $L$ $s$ $j$  &$nL$($J^{P}$) & $\langle r_{\rho}^{2}\rangle^{\frac{1}{2}}$ & $\langle r_{\lambda}^{2}\rangle^{\frac{1}{2}}$ &$M$ &\cite{quam2}&\cite{quam3}&$l_{\rho}$  $l_{\lambda}$ $L$ $s$ $j$  &$nL$($J^{P}$)& $\langle r_{\rho}^{2}\rangle^{\frac{1}{2}}$ & $\langle r_{\lambda}^{2}\rangle^{\frac{1}{2}}$  &$M$ &\cite{quam2}&  \\ \hline
\multirow{4}{*}{0 2 2 1 3}
        & $1D$($\frac{5}{2}^{+}$) & 0.68 & 0.79 & 6338   & 6270 & 6325&\multirow{4}{*}{0 3 3 1 4} & $1F$($\frac{7}{2}^{-}$) & 0.68 & 0.95 & 6538  & 6472 \\
        & $2D$($\frac{5}{2}^{+}$) & 0.72 & 1.02 & 6639  & ~& 6598&~ & $2F$($\frac{7}{2}^{-}$)                               & 0.71 & 1.18 & 6827   & ~& ~ \\
        & $3D$($\frac{5}{2}^{+}$) & 1.13 & 0.87 & 6827  & ~& ~ & ~ & $3F$($\frac{7}{2}^{-}$)                                & 1.15 & 1.01 & 6998  & ~ & ~ \\
        & $4D$($\frac{5}{2}^{+}$) & 0.78 & 1.43 & 6892   & ~ & ~ & ~ & $4F$($\frac{7}{2}^{-}$)                              & 0.75 & 1.52 & 7041   & ~& ~ \\ \hline
\multirow{4}{*}{0 2 2 1 3}
        & $1D$($\frac{7}{2}^{+}$) & 0.68 & 0.80 & 6346   & 6260 & 6333 &\multirow{4}{*}{0 3 3 1 4}& $1F$($\frac{9}{2}^{-}$) & 0.68 & 0.96 & 6542  & 6459& ~ \\
        & $2D$($\frac{7}{2}^{+}$) & 0.72 & 1.03 & 6647   & 6590 & 6554 & ~ & $2F$($\frac{9}{2}^{-}$)                        & 0.71 & 1.19 & 6833  & ~& ~  \\
        & $3D$($\frac{7}{2}^{+}$) & 1.13 & 0.88 & 6834  & ~& ~ & ~ & $3F$($\frac{9}{2}^{-}$)                                & 1.15 & 1.02 & 7002   & ~& ~ \\
        & $4D$($\frac{7}{2}^{+}$) & 0.78 & 1.43 & 6897  & ~ & ~ & ~ & $4F$($\frac{9}{2}^{-}$)                               & 0.75 & 1.51 & 7045   & ~& ~  \\ \hline
\multirow{4}{*}{0 3 3 1 2}
        & $1F$($\frac{3}{2}^{-}$) & 0.68 & 0.95 & 6538  &6550& ~ & \multirow{4}{*}{0 4 4 1 3} & $1G$($\frac{5}{2}^{+}$)     & 0.68 & 1.09 & 6715  & 6749 & ~  \\
        & $2F$($\frac{3}{2}^{-}$) & 0.71 & 1.18 & 6827   & ~& ~ & ~ & $2G$($\frac{5}{2}^{+}$)                               & 0.71 & 1.37 & 7006  & ~& ~ \\
        & $3F$($\frac{3}{2}^{-}$) & 1.15 & 1.01 & 6998   & ~& ~ & ~ & $3G$($\frac{5}{2}^{+}$)                               & 1.15 & 1.57 & 7155  & ~ & ~ \\
        & $4F$($\frac{3}{2}^{-}$) & 0.75 & 1.52 & 7041   & ~& ~ & ~ & $4G$($\frac{5}{2}^{+}$)                               & 0.72 & 1.53 & 7187  & ~& ~ \\ \hline
\multirow{4}{*}{0 3 3 1 2}
        & $1F$($\frac{5}{2}^{-}$) & 0.68 & 0.95 & 6542  & 6564& ~ & \multirow{4}{*}{0 4 4 1 3} & $1G$($\frac{7}{2}^{+}$)    & 0.68 & 1.09 & 6718  & 6761& ~ \\
        & $2F$($\frac{5}{2}^{-}$) & 0.71 & 1.19 & 6832  & ~& ~ & ~ & $2G$($\frac{7}{2}^{+}$)                                & 0.72 & 1.39 & 7009  & ~& ~ \\
        & $3F$($\frac{5}{2}^{-}$) & 1.15 & 1.02 & 7001  & ~ & ~ & ~ & $3G$($\frac{7}{2}^{+}$)                               & 1.16 & 1.16 & 7158  & ~& ~ \\
        & $4F$($\frac{5}{2}^{-}$) & 0.75 & 1.51 & 7044   & ~& ~ & ~ & $4G$($\frac{7}{2}^{+}$)                               & 0.72 & 1.52 & 7190  & ~& ~ \\ \hline
\multirow{4}{*}{0 3 3 1 3}
        & $1F$($\frac{5}{2}^{-}$) & 0.68 & 0.95 & 6538   & 6501& ~ & \multirow{4}{*}{0 4 4 1 4} & $1G$($\frac{7}{2}^{+}$)   & 0.68 & 1.09 & 6715   & 6688& ~  \\
        & $2F$($\frac{5}{2}^{-}$) & 0.71 & 1.18 & 6827   & ~& ~ & ~ & $2G$($\frac{7}{2}^{+}$)                               & 0.71 & 1.37 & 7006  & ~ \\
        & $3F$($\frac{5}{2}^{-}$) & 1.15 & 1.01 & 6998   & ~& ~ & ~ & $3G$($\frac{7}{2}^{+}$)                               & 1.15 & 1.16 & 7155  & ~ \\
        & $4F$($\frac{5}{2}^{-}$) & 0.75 & 1.52 & 7041   & ~& ~ & ~ & $4G$($\frac{7}{2}^{+}$)                               & 0.72 & 1.53 & 7187  & ~ \\ \hline
\multirow{4}{*}{0 3 3 1 3}
        & $1F$($\frac{7}{2}^{-}$) & 0.68 & 0.95 & 6542   & 6500& ~ & \multirow{4}{*}{0 4 4 1 4} & $1G$($\frac{9}{2}^{+}$)   & 0.68 & 1.09 & 6718   & 6688& ~  \\
        & $2F$($\frac{7}{2}^{-}$) & 0.71 & 1.19 & 6832   & ~& ~ & ~ & $2G$($\frac{9}{2}^{+}$)                               & 0.72 & 1.39 & 7009 & ~ & ~ \\
        & $3F$($\frac{7}{2}^{-}$) & 1.15 & 1.02 & 7001   & ~& ~ & ~ & $3G$($\frac{9}{2}^{+}$)                               & 1.16 & 1.16 & 7158 & ~  & ~  \\
        & $4F$($\frac{7}{2}^{-}$) & 0.75 & 1.51 & 7045   & ~& ~ & ~ & $4G$($\frac{9}{2}^{+}$)                               & 0.72 & 1.52 & 7190 & ~  & ~ \\

\end{tabular}
\end{ruledtabular}
\end{table*}
\begin{table*}[htbp]
\begin{ruledtabular}\caption{The root mean square radius (fm) and the mass spectrum (MeV) of the $\Omega_{c}$ family}
\begin{tabular}{c c c c c c c c c c c c}
$l_{\rho}$  $l_{\lambda}$ $L$ $s$ $j$  & $nL$($J^{P}$)& $\langle r_{\rho}^{2}\rangle^{\frac{1}{2}}$ & $\langle r_{\lambda}^{2}\rangle^{\frac{1}{2}}$ &$M$ &$M_{exp}$  &\cite{quam2} &\cite{quam3}  &\cite{quam19} &\cite{quam10} & \cite{WZG4,Sum2}\\ \hline
\multirow{4}{*}{0 0 0 1 1}
         & $1S$($\frac{1}{2}^{+}$)  & 0.55 & 0.41 & 2699 & 2695.2\cite{article2A}  & 2698& 2718  &2688 &2710 & \\
        ~ & $2S$($\frac{1}{2}^{+}$) & 0.78 & 0.68 & 3150 & 3119.1\cite{article2A}  & 3088& 3152  &3169 &3044 &~ \\
        ~ & $3S$($\frac{1}{2}^{+}$) & 0.88 & 0.65 & 3308 & ~  & 3489& ~  &~ &~  &~ \\
        ~ & $4S$($\frac{1}{2}^{+}$) & 0.88 & 1.02 & 3526 & ~  & 3814& ~  &~ &~  &~  \\ \hline
\multirow{4}{*}{0 0 0 1 1}
        ~ & $1S$($\frac{3}{2}^{+}$) & 0.57 & 0.46 & 2762  & 2765.9\cite{article2A} & 2768& 2776   &2721 &2759 &  \\
        ~ & $2S$($\frac{3}{2}^{+}$) & 0.77 & 0.73 & 3197 & ~ & 3123& 3190  &~ &3080 &~  \\
        ~ & $3S$($\frac{3}{2}^{+}$) & 0.92 & 0.66 & 3346 & ~  & 3510& ~  &~  &~ &~ \\
        ~ & $4S$($\frac{3}{2}^{+}$) & 0.83 & 1.08 & 3557 & ~  & 3830& ~  &~  &~ &~ \\ \hline
\multirow{4}{*}{0 1 1 1 0}
        ~ & $1P$($\frac{1}{2}^{-}$) & 0.62 & 0.65 & 3057 & ~  & 3055& 2977   &~ &2959 &3050$\pm$110\cite{WZG4}  \\
        ~ & $2P$($\frac{1}{2}^{-}$) & 0.73 & 0.93 & 3426 & ~  & 3435& 2990   &~ &3029 &~ \\
        ~ & $3P$($\frac{1}{2}^{-}$) & 1.02 & 0.79 & 3562 & ~  & 3754& ~  &~ &~ &~  \\
        ~ & $4P$($\frac{1}{2}^{-}$) & 0.76 & 1.26 & 3735 & ~  & 4037& ~  &~ &~ &~  \\ \hline
\multirow{4}{*}{0 1 1 1 1}
        ~ & $1P$($\frac{1}{2}^{-}$) & 0.61 & 0.63 & 3045 &3000.4\cite{article2A}  & 2966& ~ &~  &~ &2980$\pm$160\cite{Sum2} \\
        ~ & $2P$($\frac{1}{2}^{-}$) & 0.73 & 0.92 & 3416 & ~  & 3384& ~ &~ &~  &~  \\
        ~ & $3P$($\frac{1}{2}^{-}$) & 1.01 & 0.79 & 3554 & ~  & 3717& ~ &~ &~  &~ \\
        ~ & $4P$($\frac{1}{2}^{-}$) & 0.76 & 1.26 & 3728 & ~  & 4009& ~ &~ &~  &~   \\ \hline
\multirow{4}{*}{0 1 1 1 1}
        ~ & $1P$($\frac{3}{2}^{-}$) & 0.62 & 0.65 & 3062 & 3050.2\cite{article2A} & 3054& 2986  &~ &2959 &3060$\pm$110\cite{WZG4} \\
        ~ & $2P$($\frac{3}{2}^{-}$) & 0.73 & 0.94 & 3431 & ~ & 3433& 2994  &~ &3029 &~   \\
        ~ & $3P$($\frac{3}{2}^{-}$) & 1.02 & 0.80 & 3566 & ~  & 3757& ~   &~ &~ &~  \\
        ~ & $4P$($\frac{3}{2}^{-}$) & 0.75 & 1.26 & 3739 & ~  & 4036& ~   &~ &~ &~  \\ \hline
\multirow{4}{*}{0 1 1 1 2}
        ~ & $1P$($\frac{3}{2}^{-}$) & 0.61 & 0.63 & 3039 & 3065.5\cite{article2A}  & 3029& ~ &~ &~  &~3060$\pm$100\cite{WZG4}   \\
        ~ & $2P$($\frac{3}{2}^{-}$) & 0.73 & 0.91 & 3411 & ~  & 3415& ~  &~ &~  &~ \\
        ~ & $3P$($\frac{3}{2}^{-}$) & 1.01 & 0.78 & 3550 & ~  & 3737& ~  &~ &~  &~ \\
        ~ & $4P$($\frac{3}{2}^{-}$) & 0.77 & 1.26 & 3725 & ~  & 4023& ~ &~ &~  &~   \\ \hline
\multirow{4}{*}{0 1 1 1 2}
        ~ & $1P$($\frac{5}{2}^{-}$) & 0.62 & 0.66 & 3067 & 3090.0\cite{article2A}  & 3051& 3014 &~  &~ &3110$\pm$100\cite{WZG4}  \\
        ~ & $2P$($\frac{5}{2}^{-}$) & 0.73 & 0.94 & 3435 & ~  & 3427& ~ &~ &~  &~   \\
        ~ & $3P$($\frac{5}{2}^{-}$) & 1.02 & 0.80 & 3569 & ~  & 3744& ~ &~ &~  &~   \\
        ~ & $4P$($\frac{5}{2}^{-}$) & 0.75 & 1.26 & 3742 & ~  & 4028& ~ &~ &~  &~   \\ \hline
\multirow{4}{*}{0 2 2 1 1}
       ~ & $1D$($\frac{1}{2}^{+}$)  & 0.64 & 0.82 & 3304 & ~  & 3287& ~ &~ &~  &~   \\
        ~ & $2D$($\frac{1}{2}^{+}$) & 0.71 & 1.13 & 3641 & ~  & 3623& ~ &~  &~ &~   \\
        ~ & $3D$($\frac{1}{2}^{+}$) & 1.07 & 0.93 & 3764 & ~  & ~& ~ &~ &~  &~   \\
       ~ & $4D$($\frac{1}{2}^{+}$)  & 0.71 & 1.35 & 3909 & ~  & ~& ~  &~ &~  &~ \\ \hline
\multirow{4}{*}{0 2 2 1 1}
        ~ & $1D$($\frac{3}{2}^{+}$) & 0.64 & 0.84 & 3313 & ~  & 3298& ~  &~  &~ &~  \\
        ~ & $2D$($\frac{3}{2}^{+}$) & 0.71 & 1.16 & 3650 & ~  & 3627& ~  &~  &~ &~  \\
        ~ & $3D$($\frac{3}{2}^{+}$) & 1.08 & 0.94 & 3771 & ~  & ~& ~  &~ &~  &~  \\
        ~ & $4D$($\frac{3}{2}^{+}$) & 0.71 & 1.34 & 3917 & ~  & ~& ~ &~ &~  &~  \\ \hline
\multirow{4}{*}{0 2 2 1 2}
        ~ & $1D$($\frac{3}{2}^{+}$) & 0.64 & 0.83 & 3304 & ~  & 3282& ~  &~  &~ &~  \\
        ~ & $2D$($\frac{3}{2}^{+}$) & 0.71 & 1.13 & 3641 & ~  & 3613& ~  &~  &~ &~  \\
        ~ & $3D$($\frac{3}{2}^{+}$) & 1.07 & 0.93 & 3764 & ~  & ~& ~  &~ &~  &~ \\
        ~ & $4D$($\frac{3}{2}^{+}$) & 0.71 & 1.35 & 3909 & ~  & ~& ~  &~ &~  &~  \\ \hline
\multirow{4}{*}{0 2 2 1 2}
       ~ & $1D$($\frac{5}{2}^{+}$)  & 0.64 & 0.84 & 3314 & ~  & 3297& 3196  &~  &~ &~ \\
        ~ & $2D$($\frac{5}{2}^{+}$) & 0.71 & 1.16 & 3651 & ~  & 3626& ~  &~  &~ &~  \\
       ~ & $3D$($\frac{5}{2}^{+}$)  & 1.07 & 0.94 & 3772 & ~  & ~& ~  &~ &~  &~ \\
        ~ & $4D$($\frac{5}{2}^{+}$) & 0.71 & 1.33 & 3918  & ~& ~ &~ &~ &~ &~\\
   \end{tabular}
\end{ruledtabular}
\end{table*}
\begin{table*}[htbp]
\begin{ruledtabular}
\begin{tabular}{c c c c c c c|c c c c c c c c c}
$l_{\rho}$  $l_{\lambda}$ $L$ $s$ $j$  &$nL$($J^{P}$)& $\langle r_{\rho}^{2}\rangle^{\frac{1}{2}}$ & $\langle r_{\lambda}^{2}\rangle^{\frac{1}{2}}$ &$M$   &\cite{quam2} &  &$l_{\rho}$  $l_{\lambda}$ $L$ $s$ $j$  &$nL$($J^{P}$)& $\langle r_{\rho}^{2}\rangle^{\frac{1}{2}}$ & $\langle r_{\lambda}^{2}\rangle^{\frac{1}{2}}$ &$M$   &\cite{quam2} &\cite{quam3} \\ \hline
\multirow{4}{*}{0 2 2 1 3}
        ~ & $1D$($\frac{5}{2}^{+}$) & 0.64 & 0.82 & 3304   & 3286& ~ &\multirow{4}{*}{0 3 3 1 4}  & $1F$($\frac{9}{2}^{-}$) & 0.65 & 1.01 & 3529  & ~& 3485 \\
       ~ & $2D$($\frac{5}{2}^{+}$)  & 0.71 & 1.13 & 3640   & 3614& ~&~  & $2F$($\frac{9}{2}^{-}$)                           & 0.71 & 1.41 & 3852   & ~& ~   \\
       ~ & $3D$($\frac{5}{2}^{+}$)  & 1.07 & 0.93 & 3764   & ~& ~&~  & $3F$($\frac{9}{2}^{-}$)                              & 1.11 & 1.09 & 3960   & ~& ~   \\
        ~ & $4D$($\frac{5}{2}^{+}$) & 0.71 & 1.35 & 3908   & ~& ~&~ & $4F$($\frac{9}{2}^{-}$)                               & 0.67 & 1.30 & 4101   & ~& ~   \\ \hline
\multirow{4}{*}{0 2 2 1 3}
        ~ & $1D$($\frac{7}{2}^{+}$) & 0.64 & 0.84 & 3315   & 3283& ~&\multirow{4}{*}{0 4 4 1 3}  & $1G$($\frac{5}{2}^{+}$)  & 0.65 & 1.14 & 3719   & 3739& ~   \\
        ~ & $2D$($\frac{7}{2}^{+}$) & 0.71 & 1.16 & 3652   & 3611& ~&~ & $2G$($\frac{5}{2}^{+}$)                            & 0.71 & 1.65 & 4030   & ~& ~  \\
        ~ & $3D$($\frac{7}{2}^{+}$) & 1.08 & 0.94 & 3773  & ~& ~ &~ & $3G$($\frac{5}{2}^{+}$)                               & 1.13 & 1.21 & 4135   & ~& ~  \\
        ~ & $4D$($\frac{7}{2}^{+}$) & 0.70 & 1.33 & 3919   & ~& ~&~ & $4G$($\frac{5}{2}^{+}$)                               & 0.64 & 1.24 & 4283   & ~& ~   \\ \hline
\multirow{4}{*}{0 3 3 1 2}
        ~ & $1F$($\frac{3}{2}^{-}$) & 0.64 & 1.00 & 3525   & 3533& ~&\multirow{4}{*}{0 4 4 1 3} & $1G$($\frac{7}{2}^{+}$)   & 0.65 & 1.15 & 3720   & 3721& ~   \\
        ~ & $2F$($\frac{3}{2}^{-}$) & 0.71 & 1.39 & 3847   & ~& ~&~ & $2G$($\frac{7}{2}^{+}$)                               & 0.71 & 1.67 & 4030   & ~& ~  \\
        ~ & $3F$($\frac{3}{2}^{-}$) & 1.11 & 1.08 & 3957   & ~& ~ &~  & $3G$($\frac{7}{2}^{+}$)                             & 1.13 & 1.21 & 4135   & ~& ~  \\
        ~ & $4F$($\frac{3}{2}^{-}$) & 0.67 & 1.33 & 4091   & ~& ~ &~ & $4G$($\frac{7}{2}^{+}$)                              & 0.64 & 1.22 & 4291   & ~& ~  \\ \hline
\multirow{4}{*}{0 3 3 1 2}
        ~ & $1F$($\frac{5}{2}^{-}$) & 0.65 & 1.00 & 3528   & 3522& ~ &\multirow{4}{*}{0 4 4 1 4} & $1G$($\frac{7}{2}^{+}$)  & 0.65 & 1.14 & 3719   & 3707& ~  \\
        ~ & $2F$($\frac{5}{2}^{-}$) & 0.71 & 1.41 & 3851   & ~& ~&~ & $2G$($\frac{7}{2}^{+}$)                               & 0.71 & 1.65 & 4030   & ~& ~   \\
        ~ & $3F$($\frac{5}{2}^{-}$) & 1.11 & 1.08 & 3960   & ~& ~ &~ & $3G$($\frac{7}{2}^{+}$)                              & 1.13 & 1.21 & 4135   & ~& ~  \\
        ~ & $4F$($\frac{5}{2}^{-}$) & 0.67 & 1.31 & 4099  & ~& ~&~ & $4G$($\frac{7}{2}^{+}$)                                & 0.64 & 1.24 & 4283 & ~ & ~   \\ \hline
\multirow{4}{*}{0 3 3 1 3}
        ~ & $1F$($\frac{5}{2}^{-}$) & 0.64 & 1.00 & 3525   & 3515& ~&\multirow{4}{*}{0 4 4 1 4} & $1G$($\frac{9}{2}^{+}$)   & 0.64 & 1.15 & 3720 & ~  & 3705   \\
        ~ & $2F$($\frac{5}{2}^{-}$) & 0.71 & 1.39 & 3847   & ~& ~&~ & $2G$($\frac{9}{2}^{+}$)                               & 0.71 & 1.67 & 4030 & ~  & ~   \\
        ~ & $3F$($\frac{5}{2}^{-}$) & 1.11 & 1.08 & 3957   & ~& ~ &~ & $3G$($\frac{9}{2}^{+}$)                              & 1.13 & 1.21 & 4135 & ~  & ~  \\
        ~ & $4F$($\frac{5}{2}^{-}$) & 0.67 & 1.33 & 4091   & ~& ~&~ & $4G$($\frac{9}{2}^{+}$)                               & 0.64 & 1.22 & 4291 & ~  & ~  \\ \hline
\multirow{4}{*}{0 3 3 1 3}
        ~ & $1F$($\frac{7}{2}^{-}$) & 0.65 & 1.00 & 3529   & 3514& ~&\multirow{4}{*}{0 4 4 1 5} & $1G$($\frac{9}{2}^{+}$)   & 0.65 & 1.14 & 3719 & ~  & 3685   \\
        ~ & $2F$($\frac{7}{2}^{-}$) & 0.71 & 1.41 & 3851   & ~& ~ &~ & $2G$($\frac{9}{2}^{+}$)                              & 0.71 & 1.65 & 4030 & ~  & ~  \\
        ~ & $3F$($\frac{7}{2}^{-}$) & 1.11 & 1.08 & 3960   & ~& ~ &~ & $3G$($\frac{9}{2}^{+}$)                              & 1.13 & 1.20 & 4135 & ~  & ~  \\
        ~ & $4F$($\frac{7}{2}^{-}$) & 0.67 & 1.30 & 4100   & ~& ~ &~ & $4G$($\frac{9}{2}^{+}$)                              & 0.64 & 1.24 & 4283 & ~  & ~  \\ \hline
\multirow{4}{*}{0 3 3 1 4}
        ~ & $1F$($\frac{7}{2}^{-}$) & 0.64 & 0.99 & 3524   & 3498& ~ &\multirow{4}{*}{0 4 4 1 5} & $1G$($\frac{11}{2}^{+}$) & 0.65 & 1.15 & 3720 & ~  & 3665  \\
        ~ & $2F$($\frac{7}{2}^{-}$) & 0.71 & 1.38 & 3846   & ~& ~ &~ & $2G$($\frac{11}{2}^{+}$)                             & 0.71 & 1.67 & 4030 & ~  & ~  \\
        ~ & $3F$($\frac{7}{2}^{-}$) & 1.11 & 1.08 & 3957   & ~& ~ &~ & $3G$($\frac{11}{2}^{+}$)                             & 1.13 & 1.21 & 4135 & ~  & ~ \\
        ~ & $4F$($\frac{7}{2}^{-}$) & 0.67 & 1.33 & 4091   & ~& ~ &~ & $4G$($\frac{11}{2}^{+}$)                             & 0.64 & 1.22 & 4292 & ~  & ~  \\
\end{tabular}
\end{ruledtabular}
\end{table*}
\begin{table*}[htbp]
\begin{ruledtabular}\caption{The root mean square radius (fm) and the mass spectrum (MeV) of the $\Omega_{b}$ family}
\begin{tabular}{c c c c c c c c c c c}
$l_{\rho}$  $l_{\lambda}$ $L$ $s$ $j$  &$nL$($J^{P}$)& $\langle r_{\rho}^{2}\rangle^{\frac{1}{2}}$ & $\langle r_{\lambda}^{2}\rangle^{\frac{1}{2}}$ & $M$ &$M_{exp}$&\cite{quam2} &\cite{quam3}  &\cite{quam10}&\cite{Sum2,WZG8}\\ \hline
\multirow{4}{*}{0 0 0 1 1}
         ~ & $1S$($\frac{1}{2}^{+}$) & 0.57 & 0.39 & 6043 & 6046.1\cite{article2A} & 6064 & 6081  &6037&~ \\
         ~ & $2S$($\frac{1}{2}^{+}$) & 0.71 & 0.67 & 6446 & ~  & 6450 & 6472 &~ &~ \\
         ~ & $3S$($\frac{1}{2}^{+}$) & 0.95 & 0.55 & 6633 & ~  & 6804 & ~  &~&~ \\
         ~ & $4S$($\frac{1}{2}^{+}$) & 0.79 & 1.01 & 6790 & ~  & 7091 & ~  &~&~ \\ \hline
\multirow{4}{*}{0 0 0 1 1}
         ~ & $1S$($\frac{3}{2}^{+}$) & 0.57 & 0.41 & 6069 & ~  & 6088 & 6102 &6090&~ \\
         ~& $2S$($\frac{3}{2}^{+}$)  & 0.70 & 0.69 & 6466 & ~  & 6461 & 6478  &~&~ \\
        ~ & $3S$($\frac{3}{2}^{+}$)  & 0.97 & 0.55 & 6650 & ~  & 6811 & ~  &~ &~\\
        ~ & $4S$($\frac{3}{2}^{+}$)  & 0.77 & 1.05 & 6804 & ~  & 7096 & ~  &~&~ \\ \hline
\multirow{4}{*}{0 1 1 1 0}
        ~ & $1P$($\frac{1}{2}^{-}$) & 0.61 & 0.58 & 6334 &   & 6339 & 6301  &6278&6320(110)\cite{WZG8} \\
        ~ & $2P$($\frac{1}{2}^{-}$) & 0.68 & 0.83 & 6662 & ~  & 6710 & 6312  &~&~ \\
        ~& $3P$($\frac{1}{2}^{-}$)  & 1.04 & 0.68 & 6844 & ~  & 7009 & ~   &~&~\\
        ~ & $4P$($\frac{1}{2}^{-}$) & 0.75 & 1.28 & 6969 & ~  & 7265 & ~  &~ &~\\ \hline
\multirow{4}{*}{0 1 1 1 1}
        ~ & $1P$($\frac{1}{2}^{-}$) & 0.60 & 0.58 & 6329 &  6315.64\cite{OmegaB6316} & 6330 & ~ &~ &6270$\pm$140\cite{Sum2} \\
        ~ & $2P$($\frac{1}{2}^{-}$) & 0.68 & 0.83 & 6658 & ~  & 6706  &~ &~&~ \\
        ~ & $3P$($\frac{1}{2}^{-}$) & 1.03 & 0.68 & 6841 & ~ & 7003   &~ &~&~ \\
        ~ & $4P$($\frac{1}{2}^{-}$) & 0.75 & 1.28 & 6966 & ~ & 7257   &~ &~&~ \\ \hline
\multirow{4}{*}{0 1 1 1 1}
        ~ & $1P$($\frac{3}{2}^{-}$) & 0.61 & 0.58 & 6336 &  6330.30\cite{OmegaB6316} & 6340 & ~&~  &6310$\pm$110\cite{WZG8} \\
        ~ & $2P$($\frac{3}{2}^{-}$) & 0.68 & 0.84 & 6664 & ~  & 6750 & ~&~  &~ \\
        ~ & $3P$($\frac{3}{2}^{-}$) & 1.04 & 0.69 & 6846 & ~  & 7002 & ~  &~&~ \\
        ~ & $4P$($\frac{3}{2}^{-}$) & 0.75 & 1.29 & 6970 & ~  & 7258 & ~  &~&~ \\ \hline
\multirow{4}{*}{0 1 1 1 2}
        ~ & $1P$($\frac{3}{2}^{-}$) & 0.60 & 0.57 & 6326 &  6339.71\cite{OmegaB6316} & 6331 & 6304  &6278& 6370$\pm90$\cite{WZG8} \\
        ~ & $2P$($\frac{3}{2}^{-}$) & 0.68 & 0.83 & 6655 & ~  & 6699 & 6311&~ &~  \\
        ~ & $3P$($\frac{3}{2}^{-}$) & 1.03 & 0.68 & 6839 & ~  & 6998 & ~ &~  &~\\
        ~ & $4P$($\frac{3}{2}^{-}$) & 0.75 & 1.27 & 6964 & ~  & 7250 & ~ &~  &~\\ \hline
\multirow{4}{*}{0 1 1 1 2}
        ~ & $1P$($\frac{5}{2}^{-}$) & 0.61 & 0.59 & 6339 & 6349.88\cite{OmegaB6316}  & 6334 & 6311  & &6350$\pm$100\cite{WZG8} \\
        ~ & $2P$($\frac{5}{2}^{-}$) & 0.68 & 0.84 & 6666 & ~  & 6700 & ~ &~  \\
        ~ & $3P$($\frac{5}{2}^{-}$) & 1.04 & 0.69 & 6848 & ~  & 6996 & ~  &~ \\
        ~ & $4P$($\frac{5}{2}^{-}$) & 0.75 & 1.29 & 6972 & ~  & 7251 & ~  &~ \\ \hline
\multirow{4}{*}{0 2 2 1 1}
        ~ & $1D$($\frac{1}{2}^{+}$) & 0.62 & 0.74 & 6556 & ~  & 6540 & ~ &~  \\
        ~ & $2D$($\frac{1}{2}^{+}$) & 0.66 & 0.95 & 6846 & ~  & 6857 & ~  &~ \\
        ~ & $3D$($\frac{1}{2}^{+}$) & 1.08 & 0.82 & 7021 & ~  & ~ & ~ &~  \\
        ~ & $4D$($\frac{1}{2}^{+}$) & 0.73 & 1.48 & 7121 & ~  & ~ & ~&~  &~ \\ \hline
\multirow{4}{*}{0 2 2 1 1}
        ~ & $1D$($\frac{3}{2}^{+}$) & 0.62 & 0.75 & 6561 & ~  & 6549 & ~  &~&~ \\
        ~ & $2D$($\frac{3}{2}^{+}$) & 0.66 & 0.95 & 6852 & ~  & 6863 & ~  &~&~ \\
        ~ & $3D$($\frac{3}{2}^{+}$) & 1.08 & 0.82 & 7026 & ~  & ~ & ~ &~&~ \\
        ~ & $4D$($\frac{3}{2}^{+}$) & 0.73 & 1.49 & 7124 & ~  & ~ &~ &~ \\ \hline
\multirow{4}{*}{0 2 2 1 2}
        ~ & $1D$($\frac{3}{2}^{+}$) & 0.62 & 0.74 & 6556 & ~  & 6530 & ~ &~&~  \\
        ~ & $2D$($\frac{3}{2}^{+}$) & 0.66 & 0.95 & 6846 & ~  & 6846  &~ &~&~ \\
        ~ & $3D$($\frac{3}{2}^{+}$) & 1.08 & 0.82 & 7022 & ~  & ~  &~ &~&~ \\
        ~ & $4D$($\frac{3}{2}^{+}$) & 0.73 & 1.48 & 7121 & ~  & ~  &~ &~ &~\\ \hline
\multirow{4}{*}{0 2 2 1 2}
        ~ & $1D$($\frac{5}{2}^{+}$) & 0.62 & 0.75 & 6561 & ~  & 6529 &~ &~&~  \\
        ~ & $2D$($\frac{5}{2}^{+}$) & 0.66 & 0.95 & 6852 & ~  & 6846  &~ &~&~ \\
        ~ & $3D$($\frac{5}{2}^{+}$) & 1.08 & 0.82 & 7026 & ~  & ~ & ~  &~&~ \\
\end{tabular}
\end{ruledtabular}
\end{table*}
\begin{table*}[htbp]
\begin{ruledtabular}
\begin{tabular}{c c c c c c c|c c c c c c c c }
$l_{\rho}$  $l_{\lambda}$ $L$ $s$ $j$  &$nL$($J^{P}$)& $\langle r_{\rho}^{2}\rangle^{\frac{1}{2}}$ & $\langle r_{\lambda}^{2}\rangle^{\frac{1}{2}}$ & $M$ &\cite{quam2} &~ &$l_{\rho}$  $l_{\lambda}$ $L$ $s$ $j$  &$nL$($J^{P}$) & $\langle r_{\rho}^{2}\rangle^{\frac{1}{2}}$ & $\langle r_{\lambda}^{2}\rangle^{\frac{1}{2}}$ & $M$ &\cite{quam2} \\ \hline
        ~ & $4D$($\frac{5}{2}^{+}$) & 0.73 & 1.9 & 7124 & ~  & ~ & ~&~ &~  \\ \hline
\multirow{4}{*}{0 2 2 1 3}
        ~ & $1D$($\frac{5}{2}^{+}$) & 0.62 & 0.74 & 6555  & 6520 & 6492\cite{quam3} &\multirow{4}{*}{0 3 3 1 4} & $1F$($\frac{9}{2}^{-}$) & 0.63 & 0.91 & 6754  & 6713 \\
        ~ & $2D$($\frac{5}{2}^{+}$) & 0.66 & 0.95 & 6846  & 6837 & 6494\cite{quam3} &~ & $2F$($\frac{9}{2}^{-}$)                          & 0.64 & 1.06 & 7031  & ~ \\
        ~ & $3D$($\frac{5}{2}^{+}$) & 1.08 & 0.81 & 7021  & ~ & ~&~ & $3F$($\frac{9}{2}^{-}$)                                             & 1.11 & 0.96 & 7191  & ~ \\
        ~ & $4D$($\frac{5}{2}^{+}$) & 0.73 & 1.48 & 7121  & ~ & ~&~ & $4F$($\frac{9}{2}^{-}$)                                             & 0.71 & 1.63 & 7266  & ~ \\ \hline
\multirow{4}{*}{0 2 2 1 3}
        ~ & $1D$($\frac{7}{2}^{+}$) & 0.62 & 0.75 & 6562  & 6517 & 6497\cite{quam3}&\multirow{4}{*}{0 4 4 1 3} & $1G$($\frac{5}{2}^{+}$)  & 0.63 & 1.05 & 6923  & 6952  \\
        ~ & $2D$($\frac{7}{2}^{+}$) & 0.66 & 0.95 & 6853  & 6834 & 6667\cite{quam3}&~ & $2G$($\frac{5}{2}^{+}$)                           & 0.63 & 1.17 & 7201  & ~ \\
        ~ & $3D$($\frac{7}{2}^{+}$) & 1.08 & 0.82 & 7027  & ~ & ~ &~ & $3G$($\frac{5}{2}^{+}$)                                            & 1.12 & 1.08 & 7342  & ~ \\
        ~ & $4D$($\frac{7}{2}^{+}$) & 0.73 & 1.49 & 7125  & ~ & ~ &~ & $4G$($\frac{5}{2}^{+}$)                                            & 0.69 & 1.74 & 7398  & ~ \\ \hline
\multirow{4}{*}{0 3 3 1 2}
        ~ & $1F$($\frac{3}{2}^{-}$) & 0.63 & 0.90 & 6751  & 6763 & 6370\cite{WZG8} &\multirow{4}{*}{0 4 4 1 3} & $1G$($\frac{7}{2}^{+}$)  & 0.63 & 1.05 & 6925   & 6959 \\
        ~ & $2F$($\frac{3}{2}^{-}$) & 0.64 & 1.05 & 7026  & ~ & ~ &~ & $2G$($\frac{7}{2}^{+}$)                                            & 0.64 & 1.17 & 7204   & ~ \\
        ~ & $3F$($\frac{3}{2}^{-}$) & 1.10 & 0.95 & 7188  & ~ & ~ &~ & $3G$($\frac{7}{2}^{+}$)                                            & 1.12 & 1.08 & 7344   & ~\\
        ~ & $4F$($\frac{3}{2}^{-}$) & 0.71 & 1.64 & 7264  & ~ & ~ &~& $4G$($\frac{7}{2}^{+}$)                                             & 0.69 & 1.73 & 7400   & ~ \\ \hline
\multirow{4}{*}{0 3 3 1 2}
        ~ & $1F$($\frac{5}{2}^{-}$) & 0.63 & 0.91 & 6754  & 6771 & ~ &\multirow{4}{*}{0 4 4 1 4} & $1G$($\frac{7}{2}^{+}$)                & 0.63 & 1.05 & 6923   & 6916\\
        ~ & $2F$($\frac{5}{2}^{-}$) & 0.64 & 1.06 & 7031  & ~ & ~ &~ & $2G$($\frac{7}{2}^{+}$)                                            & 0.63 & 1.17 & 7201   & ~ \\
        ~ & $3F$($\frac{5}{2}^{-}$) & 1.11 & 0.96 & 7191  & ~ & ~ &~ & $3G$($\frac{7}{2}^{+}$)                                            & 1.11 & 1.08 & 7342   & ~ \\
        ~ & $4F$($\frac{5}{2}^{-}$) & 0.71 & 1.63 & 7265  & ~ & ~&~ & $4G$($\frac{7}{2}^{+}$)                                             & 0.69 & 1.74 & 7398 & ~ \\ \hline
\multirow{4}{*}{0 3 3 1 3}
        ~ & $1F$($\frac{5}{2}^{-}$) & 0.63 & 0.90 & 6751  & 6737 & ~&\multirow{4}{*}{0 4 4 1 4} & $1G$($\frac{9}{2}^{+}$)                 & 0.63 & 1.05 & 6925  & 6915  \\
        ~ & $2F$($\frac{5}{2}^{-}$) & 0.64 & 1.05 & 7026  & ~ & ~ &~ & $2G$($\frac{9}{2}^{+}$)                                            & 0.64 & 1.17 & 7205  & ~\\
        ~ & $3F$($\frac{5}{2}^{-}$) & 1.10 & 0.95 & 7188  & ~ & ~ &~ & $3G$($\frac{9}{2}^{+}$)                                            & 1.12 & 1.08 & 7344  & ~\\
        ~ & $4F$($\frac{5}{2}^{-}$) & 0.71 & 1.64 & 7264  & ~ & ~ &~ & $4G$($\frac{9}{2}^{+}$)                                            & 0.69 & 1.73 & 7400  & ~ \\ \hline
\multirow{4}{*}{0 3 3 1 3}
        ~ & $1F$($\frac{7}{2}^{-}$) & 0.63 & 0.91 & 6754  & 6736 & ~ &\multirow{4}{*}{0 4 4 1 5} & $1G$($\frac{9}{2}^{+}$)                & 0.63 & 1.04 & 6922  & 6892 \\
        ~ & $2F$($\frac{7}{2}^{-}$) & 0.64 & 1.06 & 7031  & ~ & ~ &~ & $2G$($\frac{9}{2}^{+}$)                                            & 0.63 & 1.16 & 7201  & ~ \\
        ~ & $3F$($\frac{7}{2}^{-}$) & 1.11 & 0.96 & 7191  & ~ & ~ &~ & $3G$($\frac{9}{2}^{+}$)                                            & 1.12 & 1.08 & 7342  & ~\\
        ~ & $4F$($\frac{7}{2}^{-}$) & 0.71 & 1.63 & 7266  & ~ & ~ &~ & $4G$($\frac{9}{2}^{+}$)                                            & 0.69 & 1.74 & 7398  & ~ \\ \hline
\multirow{4}{*}{0 3 3 1 4}
        ~ & $1F$($\frac{7}{2}^{-}$) & 0.63 & 0.90 & 6750  & 6719 & ~ &\multirow{4}{*}{0 4 4 1 5} & $1G$($\frac{11}{2}^{+}$)               & 0.63 & 1.05 & 6925  & 6884 \\
        ~ & $2F$($\frac{7}{2}^{-}$) & 0.64 & 1.05 & 7026  & ~ & ~ &~ & $2G$($\frac{11}{2}^{+}$)                                           & 0.64 & 1.18 & 7205  & ~ \\
        ~ & $3F$($\frac{7}{2}^{-}$) & 1.10 & 0.95 & 7188  & ~ & ~ &~ & $3G$($\frac{11}{2}^{+}$)                                           & 1.12 & 1.08 & 7344  & ~ \\
        ~ & $4F$($\frac{7}{2}^{-}$) & 0.71 & 1.64 & 7264  & ~ & ~ &~ & $4G$($\frac{11}{2}^{+}$)                                           & 0.69 & 1.73 & 7400  & ~ \\
 \end{tabular}
\end{ruledtabular}
\end{table*}
\clearpage
\begin{large}
\textbf{A.2 Fitted parameters of the Regge trajectories}
\end{large}
\begin{table*}[h]
\begin{ruledtabular}\caption{Fitted parameters $\alpha$ and $\alpha_{0}$ for the slope and intercept of the ($J$,$M^2$) parent and daughter Regge trajectories for $\Lambda_{Q}$, $\Sigma_{Q}$ and $\Omega_{Q}$.}
\begin{tabular}{c  c c c c}
Trajectory&$\alpha$(Gev$^{-2}$)&$\alpha_{0}$ &$\alpha$(Gev$^{-2}$)&$\alpha_{0}$\\ \hline
&$\Lambda_{c}(\frac{1}{2}^{+})$&~&$\Lambda_{c}(\frac{1}{2}^{-})$&~\\
parent&0.707$\pm$0.009&-3.312$\pm$0.005&0.719$\pm$0.011&-4.393$\pm$0.033\\
1 daughter&0.736$\pm$0.001&-5.166$\pm$0.002&0.725$\pm$0.003&-5.998$\pm$0.018\\
2 daughter&0.686$\pm$0.002&-5.818$\pm$0.006&0.687$\pm$0.002&-6.790$\pm$0.013\\
&$\Sigma_{c}(\frac{1}{2}^{+})$&~&$\Sigma_{c}^{*}(\frac{3}{2}^{+})$&~\\
parent&0.646$\pm$0.007&-3.500$\pm$0.007&0.666$\pm$0.007&-2.811$\pm$0.005\\
1 daughter&0.673$\pm$0.003&-5.275$\pm$0.011&0.695$\pm$0.001&-4.632$\pm$0.001\\
2 daughter&0.670$\pm$0.002&-5.940$\pm$0.009&0.697$\pm$0.001&-5.363$\pm$0.003\\
&$\Omega_{c}(\frac{1}{2}^{+})$&~&$\Omega_{c}^{*}(\frac{3}{2}^{+})$&~\\
parent&0.614$\pm$0.009&-4.096$\pm$0.018&0.623$\pm$0.012&-3.306$\pm$0.003\\
1 daughter&0.633$\pm$0.002&-5.840$\pm$0.012&0.650$\pm$0.001&-5.158$\pm$0.003\\
2 daughter&0.651$\pm$0.003&-6.673$\pm$0.002&0.668$\pm$0.001&-5.999$\pm$0.005\\
&$\Lambda_{b}(\frac{1}{2}^{+})$&~&$\Lambda_{b}(\frac{1}{2}^{-})$&~\\
parent&0.371$\pm$0.009&-11.364$\pm$0.196&0.393$\pm$0.011&-13.240$\pm$0.304\\
1 daughter&0.413$\pm$0.012&-14.594$\pm$0.012&0.415$\pm$0.010&-15.643$\pm$0.010\\
2 daughter&0.476$\pm$0.007&-18.843$\pm$0.232&0.502$\pm$0.003&-21.019$\pm$0.131\\
&$\Sigma_{b}(\frac{1}{2}^{+})$&~&$\Sigma_{b}^{*}(\frac{3}{2}^{+})$&~\\
parent&0.357$\pm$0.006&-11.729$\pm$0.157&0.350$\pm$0.010&-10.535$\pm$0.202\\
1 daughter&0.386$\pm$0.009&-14.490$\pm$0.008&0.391$\pm$0.010&-13.775$\pm$0.010\\
2 daughter&0.406$\pm$0.002&-16.338$\pm$0.077&0.405$\pm$0.003&-15.370$\pm$0.095\\
&$\Omega_{b}(\frac{1}{2}^{+})$&~&$\Omega_{b}^{*}(\frac{3}{2}^{+})$&~\\
parent&0.352$\pm$0.008&-12.517$\pm$0.214&0.342$\pm$0.012&-11.162$\pm$0.278\\
1 daughter&0.390$\pm$0.001&-15.739$\pm$0.017&0.393$\pm$0.051&-14.965$\pm$1.966\\
2 daughter&0.405$\pm$0.002&-17.382$\pm$0.087&0.401$\pm$0.003&-16.267$\pm$0.108\\
\end{tabular}
\end{ruledtabular}
\end{table*}
\clearpage
\begin{center}
\begin{Large}
\textbf{Appendix: B}
\end{Large}
\end{center}

The coefficient $C_{lm,k}$ and the shift-direction vector $\textbf{D}_{lm,k}$ in Eqs(38)-(39) are dimensionless numbers independent of $\nu_{n}$ and $\epsilon$, and they can be described as,
\begin{flalign}
C_{lm,k}\equiv\sum_{j=0}^{\big[\frac{l-m}{2}\big]}A_{lm,j}\sum_{s=0}^{p}\sum_{t=0}^{q}\sum_{u=0}^{j}(-1)^{l-u-t-s}\binom{p}{s}\binom{q}{t}\binom{j}{u}
\end{flalign}
where $p=l-m-2j$, $q=j+m$ and
\begin{flalign}
\textbf{D}_{lm,k}\equiv\frac{1}{l}[(2s-p)\textbf{a}_{z}+(2t-q)\textbf{a}_{xy}+(2u-j)\textbf{a}_{xy}^{*}]
\end{flalign}
with
\begin{flalign}
A_{lm,j}=\Big[\frac{(2l+1)(l-m)!}{4\pi(l+m)!}\Big]^{\frac{1}{2}}\frac{(l+m)!(-1)^{j}}{(-2)^{m}4^{j}j!(m+j)!(l-m-2j)!}
\end{flalign}
In Eq.(46) $\textbf{a}_{z}$, $\textbf{a}_{xy}$ and $\textbf{a}_{xy}^{*}$ are called the shift vectors which are defined as $\textbf{a}_{z}\equiv(0,0,1)$, $\textbf{a}_{xy}\equiv(1,i,0)$, $\textbf{a}_{xy}^{*}\equiv(1,-i,0)$. For relation(46) and Eq.(47), when
$m<0$, we should change $A_{lm,j}\rightarrow (-1)^{m}A_{l-m,j}$ and $\textbf{D}\rightarrow \textbf{D}^{*}$.

\begin{large}
\textbf{B.1 Three-body matrix elements of the potential energy}
\end{large}

In the following, we show how to obtain matrix elements of the various pieces of the Hamiltonian in the Jacobi coordinate channel $3$. For most of the terms in the Hamiltonian, they are the form of $\widetilde{H}_{ij}(\emph{\textbf{r}}_{ij})$, and so we have to integrate functions of $\emph{\textbf{r}}_{12}$, $\emph{\textbf{r}}_{13}$ and $\emph{\textbf{r}}_{23}$, where $\emph{\textbf{r}}_{12}=\emph{\textbf{r}}_{\rho_{3}}$, $\emph{\textbf{r}}_{13}=\emph{\textbf{r}}_{\rho_{2}}$ and $\emph{\textbf{r}}_{23}=\textbf{\emph{r}}_{\rho_{1}}$ in three Jacobi coordinate channels, respectively. The matrix element of $\widetilde{G}(\boldsymbol{r}_{12})$ which is independent of spin and orbital,
\begin{flalign}
\notag
\langle[\phi_{n_{\rho_{a}}l_{\rho_{a}}m_{l_{\rho_{a}}}}(\boldsymbol{r}_{\rho_{3}})\phi_{n_{\lambda_{a}}l_{\lambda_{a}}m_{l_{\lambda_{a}}}}(\boldsymbol{r}_{\lambda_{3}})]_{L}|\widetilde{G}(\boldsymbol{r}_{12})|
[\phi_{n_{\rho_{b}}l_{\rho_{b}}m_{l_{\rho_{b}}}}(\boldsymbol{r}_{\rho_{3}})\phi_{n_{\lambda_{b}}l_{\lambda_{b}}m_{l_{\lambda_{b}}}}(\boldsymbol{r}_{\lambda_{3}})]_{L}\rangle \\ =\langle[\phi_{n_{\rho_{a}}l_{\rho_{a}}m_{l_{\rho_{a}}}}(\boldsymbol{r}_{\rho_{3}})\phi_{n_{\lambda_{a}}l_{\lambda_{a}}m_{l_{\lambda_{a}}}}(\boldsymbol{r}_{\lambda_{3}})]_{L}|\widetilde{G}(\boldsymbol{r}_{\rho_{3}})|
[\phi_{n_{\rho_{b}}l_{\rho_{b}}m_{l_{\rho_{b}}}}(\boldsymbol{r}_{\rho_{3}})\phi_{n_{\lambda_{b}}l_{\lambda_{b}}m_{l_{\lambda_{b}}}}(\boldsymbol{r}_{\lambda_{3}})]_{L}\rangle \end{flalign}
can be calculated directly. The matrix elements of $\widetilde{G}(\boldsymbol{r}_{13})$ and $\widetilde{G}(\boldsymbol{r}_{23})$ can be expressed as,
\begin{flalign}
\langle[\phi_{n_{\rho_{a}}l_{\rho_{a}}m_{l_{\rho_{a}}}}(\boldsymbol{r}_{\rho_{3}})\phi_{n_{\lambda_{a}}l_{\lambda_{a}}m_{l_{\lambda_{a}}}}(\boldsymbol{r}_{\lambda_{3}})]_{L}|\widetilde{G}(\boldsymbol{r}_{\rho_{k}})|
[\phi_{n_{\rho_{b}}l_{\rho_{b}}m_{l_{\rho_{b}}}}(\boldsymbol{r}_{\rho_{3}})\phi_{n_{\lambda_{b}}l_{\lambda_{b}}m_{l_{\lambda_{b}}}}(\boldsymbol{r}_{\lambda_{3}})]_{L}\rangle_{k=1,2} \end{flalign}
We have to perform the integration over $\boldsymbol{r}_{\rho_{1}}$ or $\boldsymbol{r}_{\rho_{2}}$ in channel $3$, which requires the Jacobi coordinate transformations ($\boldsymbol{r}_{\rho_{3}}$,$\boldsymbol{r}_{\lambda_{3}}$)$\rightarrow$($\boldsymbol{r}_{\rho_{k}}$,$\boldsymbol{r}_{\lambda_{k}}$)$_{k=1,2}$. Then, the infinitesimally-shifted Gaussian basis functions in channels $3$ are expressed in Jacobi coordinates ($\boldsymbol{r}_{\rho_{k}}$,$\boldsymbol{r}_{\lambda_{k}}$)$_{k=1,2}$. The structure of the final expression of the matrix elements about Eqs.(48) and (49) are illustrated explicitly as,

\begin{flalign}
\notag
&\langle[\phi_{n_{\rho_{a}}l_{\rho_{a}}m_{l_{\rho_{a}}}}(\boldsymbol{r}_{\rho_{3}})\phi_{n_{\lambda_{a}}l_{\lambda_{a}}m_{l_{\lambda_{a}}}}(\boldsymbol{r}_{\lambda_{3}})]_{Lm_{L}}|\widetilde{G}(\boldsymbol{r}_{\rho_{k}})|
[\phi_{n_{\rho_{b}}l_{\rho_{b}}m_{l_{\rho_{b}}}}(\boldsymbol{r}_{\rho_{3}})\phi_{n_{\lambda_{b}}l_{\lambda_{b}}m_{l_{\lambda_{b}}}}(\boldsymbol{r}_{\lambda_{3}})]_{Lm_{L}}\rangle_{k=1,2} & \\
\notag
&=N_{n_{\rho_{a}}l_{\rho_{a}}}N_{n_{\lambda_{a}}l_{\lambda_{a}}}N_{n_{\rho_{b}}l_{\rho_{b}}}N_{n_{\lambda_{b}}l_{\lambda_{b}}}\frac{1}{(\nu_{n_{\rho_{a}}})^{l_{\rho_{a}}}
(\nu_{n_{\lambda_{a}}})^{l_{\lambda_{a}}}(\nu_{n_{\rho_{b}}})^{l_{\rho_{b}}}(\nu_{n_{\lambda_{b}}})^{l_{\lambda_{b}}}}4\pi\big(\frac{\pi}{B_{r_{k}}}\big)^{\frac{3}{2}} & \\
\notag &\times\sum_{m_{l_{\rho_{a}}}m_{l_{\lambda_{a}}}}(l_{\rho_{a}}m_{l_{\rho_{a}}}l_{\lambda_{a}}m_{l_{\lambda_{a}}}|Lm_{L})\sum_{m_{l_{\rho_{b}}}m_{l_{\lambda_{b}}}}(l_{\rho_{b}}m_{l_{\rho_{b}}}l_{\lambda_{b}}m_{l_{\lambda_{b}}}|Lm_{L}) & \\ \notag
& \times \sum_{m=0}^{Lsum}\frac{m!}{(2m+1)!}\int_{0}^{\infty}V(r_{\rho_{k}})Exp(-\alpha_{r_{k}} r_{\rho_{k}}^{2})r_{\rho_{k}}^{2m+2}dr_{\rho_{k}} & \\
 & \times\sum_{k_{a}K_{a}k_{b}K_{b}}C_{l_{\rho_{a}}m_{l_{\rho_{a}}}k_{a}}C_{l_{\lambda_{a}}m_{l_{\lambda_{a}}}K_{a}}C_{l_{\rho_{b}}m_{l_{\rho_{b}}}k_{b}}C_{l_{\lambda_{b}}m_{l_{\lambda_{b}}}K_{b}}& \\ \notag
& \times\sum_{n_{12}=0}^{Lsum-m}\sum_{n_{13}=0}^{Lsum-m}\sum_{n_{14}=0}^{Lsum-m}\sum_{n_{23}=0}^{Lsum-m}\sum_{n_{24}=0}^{Lsum-m}\sum_{n_{34}=0}^{Lsum-m}
\sum_{m_{12}=0}^{m}\sum_{m_{13}=0}^{m}\sum_{m_{14}=0}^{m}\sum_{m_{23}=0}^{m}\sum_{m_{24}=0}^{m}\sum_{m_{34}=0}^{m}& \\
\notag
& \times\frac{\tilde{g}_{12}^{m_{12}}\tilde{g}_{13}^{m_{13}}\tilde{g}_{14}^{m_{14}}\tilde{g}_{23}^{m_{23}}\tilde{g}_{24}^{m_{24}}\tilde{g}_{34}^{m_{3}}
\hat{g}_{12}^{n_{12}}\hat{g}_{13}^{n_{13}}\hat{g}_{14}^{n_{14}}\hat{g}_{23}^{n_{23}}\hat{g}_{24}^{n_{24}}\hat{g}_{34}^{n_{34}}}{n_{12}!n_{13}!n_{14}!n_{23}!n_{24}!n_{34}!
m_{12}!m_{13}!m_{14}!m_{23}!m_{24}!m_{34}!}& \\
\notag
&\times (\textbf{D}_{1}\cdot\textbf{D}_{2})^{n_{12}+m_{12}}(\textbf{D}_{1}\cdot\textbf{D}_{3})^{n_{13}+m_{13}}(\textbf{D}_{1}\cdot\textbf{D}_{4})^{n_{14}+m_{14}} & \\ \notag
&\times(\textbf{D}_{2}\cdot\textbf{D}_{3})^{n_{23}+m_{23}}(\textbf{D}_{2}\cdot\textbf{D}_{4})^{n_{24}+m_{24}}(\textbf{D}_{3}\cdot\textbf{D}_{4})^{n_{34}+m_{34}}& \\ \notag
&\times \delta(n_{12}+n_{13}+n_{14}+n_{23}+n_{24}+n_{34}-(Lsum-m))\delta(m_{12}+m_{13}+m_{14}+m_{23}+m_{24}+m_{34}-m)& \\ \notag
& \times\delta(n_{12}+n_{13}+n_{14}+m_{12}+m_{13}+m_{14}-l_{\rho_{a}})\delta(n_{12}+n_{23}+n_{24}+m_{12}+m_{23}+m_{24}-l_{\lambda_{a}})& \\ \notag
&\times\delta(n_{13}+n_{23}+n_{34}+m_{13}+m_{23}+m_{34}-l_{\rho_{b}})\delta(n_{14}+n_{24}+n_{34}+m_{14}+m_{24}+m_{34}-l_{\lambda_{b}})&
\end{flalign}
\begin{eqnarray}
\notag
&&A_{r_{k}}=\nu_{n_{\rho_{a}}}\alpha_{3k}^{r2}+\nu_{n_{\lambda_{a}}}\gamma_{3k}^{r2}+\nu_{n_{\rho_{b}}}\alpha_{3k}^{r2}+\nu_{n_{\lambda_{b}}}\gamma_{3k}^{r2}\\
&&B_{r_{k}}=\nu_{n_{\rho_{a}}}\beta_{3k}^{r2}+\nu_{n_{\lambda_{a}}}\delta_{3k}^{r2}+\nu_{n_{\rho_{b}}}\beta_{3k}^{r2}+\nu_{n_{\lambda_{b}}}\delta_{3k}^{r2} \\
\notag
&&C_{r_{1}}=2\nu_{n_{\rho_{a}}}\alpha_{3k}^{r}\beta_{3k}^{r}+2\nu_{n_{\lambda_{a}}}\gamma_{3k}^{r}\delta_{3k}^{r}+2\nu_{n_{\rho_{b}}}\alpha_{3k}^{r}\beta_{3k}^{r}+2\nu_{n_{\lambda_{b}}}\gamma_{3k}^{r}\delta_{3}^{r}
\end{eqnarray}
\begin{eqnarray}
\alpha_{r_{k}}=\Big(A_{r_{k}}-\frac{C_{r_{k}}^{2}}{4B_{r_{k}}}\Big),
\end{eqnarray}
\begin{eqnarray}
\textbf{D}_{1}=\textbf{D}_{l_{\rho_{a}}m_{l_{\rho_{a}}},k_{a}},\quad \textbf{D}_{2}=\textbf{D}_{l_{\lambda_{a}}m_{l_{\lambda_{a}}},K_{a}},\quad \textbf{D}_{3}=\textbf{D}_{l_{\rho_{b}}m_{l_{\rho_{b}}},k_{b}},\quad \textbf{D}_{4}=\textbf{D}_{l_{\lambda_{b}}m_{l_{\lambda_{b}}},K_{b}}
\end{eqnarray}
\begin{eqnarray}
\tilde{g}_{ij}=\frac{C_{r_{k}}^{2}}{2B_{r_{k}}^{2}}c_{i}c_{j}+2d_{i}d_{j}-\frac{C_{r_{k}}}{B_{r_{k}}}(c_{i}d_{j}+c_{j}d_{i}),\quad \hat{g}_{ij}=\frac{1}{2B_{r_{k}}}c_{i}c_{j}
\end{eqnarray}
\begin{eqnarray}
\notag
c_{1}=2\nu_{n_{\rho_{a}}}\beta_{3k}^{r},\quad c_{2}=2\nu_{n_{\lambda_{a}}}\delta_{3k}^{r},\quad c_{3}=2\nu_{n_{\rho_{b}}}\beta_{3k}^{r},\quad c_{4}=2\nu_{n_{\lambda_{b}}}\delta_{3k}^{r}, \\
d_{1}=2\nu_{n_{\rho_{a}}}\alpha_{3k}^{r},\quad d_{2}=2\nu_{n_{\lambda_{a}}}\gamma_{3k}^{r},\quad d_{3}=2\nu_{n_{\rho_{b}}}\alpha_{3k}^{r},\quad d_{4}=2\nu_{n_{\lambda_{b}}}\gamma_{3k}^{r}
\end{eqnarray}
\begin{eqnarray}
L_{sum}=\frac{l_{\rho_{a}}+l_{\lambda_{a}}+l_{\rho_{b}}+l_{\lambda_{b}}}{2}
\end{eqnarray}
\begin{flalign}
\notag
&\langle[\phi_{n_{\rho_{a}}l_{\rho_{a}}m_{l_{\rho_{a}}}}(\boldsymbol{r}_{\rho_{3}})\phi_{n_{\lambda_{a}}l_{\lambda_{a}}m_{l_{\lambda_{a}}}}(\boldsymbol{r}_{\lambda_{3}})]_{Lm_{L}}|\widetilde{G}(\boldsymbol{r}_{\rho_{3}})|
[\phi_{n_{\rho_{b}}l_{\rho_{b}}m_{l_{\rho_{b}}}}(\boldsymbol{r}_{\rho_{3}})\phi_{n_{\lambda_{b}}l_{\lambda_{b}}m_{l_{\lambda_{b}}}}(\boldsymbol{r}_{\lambda_{3}})]_{Lm_{L}}\rangle & \\
\notag
&=N_{n_{\rho_{a}}l_{\rho_{a}}}N_{n_{\lambda_{a}}l_{\lambda_{a}}}N_{n_{\rho_{b}}l_{\rho_{b}}}N_{n_{\lambda_{b}}l_{\lambda_{b}}}\frac{1}{(\nu_{n_{\rho_{a}}})^{l_{\rho_{a}}}
(\nu_{n_{\lambda_{a}}})^{l_{\lambda_{a}}}(\nu_{n_{\rho_{b}}})^{l_{\rho_{b}}}(\nu_{n_{\lambda_{b}}})^{l_{\lambda_{b}}}} & \\
\notag &\times\sum_{m_{l_{\rho_{a}}}m_{l_{\lambda_{a}}}}(l_{\rho_{a}}m_{l_{\rho_{a}}}l_{\lambda_{a}}m_{l_{\lambda_{a}}}|Lm_{L})\sum_{m_{l_{\rho_{b}}}m_{l_{\lambda_{b}}}}(l_{\rho_{b}}m_{l_{\rho_{b}}}l_{\lambda_{b}}m_{l_{\lambda_{b}}}|Lm_{L}) & \\ \notag
& \times4\pi\big(\frac{\pi}{B_{r_{3}}}\big)^{\frac{3}{2}}\sum_{k_{a}K_{a}k_{b}K_{b}}C_{l_{\rho_{a}}m_{l_{\rho_{a}}}k_{a}}C_{l_{\lambda_{a}}m_{l_{\lambda_{a}}}K_{a}}C_{l_{\rho_{b}}m_{l_{\rho_{b}}}k_{b}}C_{l_{\lambda_{b}}m_{l_{\lambda_{b}}}K_{b}}& \\
& \times \frac{1}{(2l_{\rho_{a}}+1)!}\int_{0}^{\infty}V(r_{\rho_{3}})Exp(-A_{r_{3}} r_{\rho_{3}}^{2})r_{\rho_{3}}^{2l_{\rho_{a}}+2}dr_{\rho_{3}}\frac{\hat{g}_{24}^{l_{\lambda_{a}}}\tilde{g}_{13}^{l_{\rho_{a}}}}{l_{\lambda_{a}}!}(\textbf{D}_{1}\cdot\textbf{D}_{3})^{l_{\rho_{a}}}(\textbf{D}_{2}\cdot\textbf{D}_{4})^{l_{\lambda_{a}}} \end{flalign}
\begin{eqnarray}
\notag
&&A_{r_{3}}=\nu_{n_{\rho_{a}}}+\nu_{n_{\rho_{b}}} \\
&&B_{r_{3}}=\nu_{n_{\lambda_{a}}}+\nu_{n_{\lambda_{b}}}
\end{eqnarray}
\begin{eqnarray}
\tilde{g}_{13}=2d_{1}d_{3},\quad \hat{g}_{24}=\frac{1}{2B_{r_{3}}}\times c_{2}c_{4}
\end{eqnarray}
\begin{eqnarray}
c_{2}=2\nu_{n_{\lambda_{a}}},\quad c_{4}=2\nu_{n_{\lambda_{b}}},\quad d_{1}=2\nu_{n_{\rho_{a}}},\quad d_{3}=2\nu_{n_{\rho_{b}}},
\end{eqnarray}

\begin{large}
\textbf{B.2 Three-body matrix elements of the kinetic-energy operators}
\end{large}

For the kinetic energy terms, it can easily be derived that the momentum of each quark can be expressed by relative momentum in baryon center-of-momentum frame, where $\boldsymbol{p}_{1}=\boldsymbol{p}_{\lambda_{1}}$, $\boldsymbol{p}_{2}=\boldsymbol{p}_{\lambda_{2}}$, $\boldsymbol{p}_{3}=\boldsymbol{p}_{\lambda_{3}}$. Then the calculation of kinetic energy matrix element,
\begin{flalign}
\notag
\langle[\phi_{n_{\rho_{a}}l_{\rho_{a}}m_{l_{\rho_{a}}}}(\boldsymbol{p}_{\rho_{3}})\phi_{n_{\lambda_{a}}l_{\lambda_{a}}m_{l_{\lambda_{a}}}}(\boldsymbol{p}_{\lambda_{3}})]_{L}|\sqrt{\boldsymbol{p}_{3}^{2}+m_{3}^{2}}|
[\phi_{n_{\rho_{b}}l_{\rho_{b}}m_{l_{\rho_{b}}}}(\boldsymbol{p}_{\rho_{3}})\phi_{n_{\lambda_{b}}l_{\lambda_{b}}m_{l_{\lambda_{b}}}}(\boldsymbol{p}_{\lambda_{3}})]_{L}\rangle  \\  =\langle[\phi_{n_{\rho_{a}}l_{\rho_{a}}m_{l_{\rho_{a}}}}(\boldsymbol{p}_{\rho_{3}})\phi_{n_{\lambda_{a}}l_{\lambda_{a}}m_{l_{\lambda_{a}}}}(\boldsymbol{p}_{\lambda_{3}})]_{L}|\sqrt{\boldsymbol{p}_{\lambda_{3}}^{2}+m_{3}^{2}}|
[\phi_{n_{\rho_{b}}l_{\rho_{b}}m_{l_{\rho_{b}}}}(\boldsymbol{p}_{\rho_{3}})\phi_{n_{\lambda_{b}}l_{\lambda_{b}}m_{l_{\lambda_{b}}}}(\boldsymbol{p}_{\lambda_{3}})]_{L}\rangle \end{flalign}
is straightforward. For the kinetic energy terms $\sqrt{\boldsymbol{p}_{k}^{2}+m_{k}^{2}}_{k=1,2}$, their matrix elements can be described as,
\begin{flalign}
\langle[\phi_{n_{\rho_{a}}l_{\rho_{a}}m_{l_{\rho_{a}}}}(\boldsymbol{p}_{\rho_{3}})\phi_{n_{\lambda_{a}}l_{\lambda_{a}}m_{l_{\lambda_{a}}}}(\boldsymbol{p}_{\lambda_{3}})]_{L}|\sqrt{\boldsymbol{p}_{\lambda_{k}}^{2}+m_{k}^{2}}|
[\phi_{n_{\rho_{b}}l_{\rho_{b}}m_{l_{\rho_{b}}}}(\boldsymbol{p}_{\rho_{3}})\phi_{n_{\lambda_{b}}l_{\lambda_{b}}m_{l_{\lambda_{b}}}}(\boldsymbol{p}_{\lambda_{3}})]_{L}\rangle_{k=1,2} \end{flalign}
By performing the Jacobi momentum transformations, ($\boldsymbol{p}_{\rho_{3}}$,$\boldsymbol{p}_{\lambda_{3}}$)$\rightarrow$($\boldsymbol{p}_{\rho_{k}}$,$\boldsymbol{p}_{\lambda_{k}}$)$_{k=1,2}$ over the basis function in momentum representation, we can carry out the calculations about the matrix elements of Eq.(62). And the expression of kinetic energy matrix elements are explicitly described as,
\begin{flalign}
\notag
&\langle[\phi_{n_{\rho_{a}}l_{\rho_{a}}m_{l_{\rho_{a}}}}(\boldsymbol{p}_{\rho_{3}})\phi_{n_{\lambda_{a}}l_{\lambda_{a}}m_{l_{\lambda_{a}}}}(\boldsymbol{p}_{\lambda_{3}})]_{Lm_{L}}|\sqrt{\boldsymbol{p}_{\lambda_{k}}^{2}+m_{k}^{2}}|
[\phi_{n_{\rho_{b}}l_{\rho_{b}}m_{l_{\rho_{b}}}}(\boldsymbol{p}_{\rho_{3}})\phi_{n_{\lambda_{b}}l_{\lambda_{b}}m_{l_{\lambda_{b}}}}(\boldsymbol{p}_{\lambda_{3}})]_{Lm_{L}}\rangle_{k=1,2} & \\
\notag
&=N_{n_{\rho_{a}}l_{\rho_{a}}}N_{n_{\lambda_{a}}l_{\lambda_{a}}}N_{n_{\rho_{b}}l_{\rho_{b}}}N_{n_{\lambda_{b}}l_{\lambda_{b}}}(4\nu_{n_{\rho_{a}}})^{l_{\rho_{a}}}
(4\nu_{n_{\lambda_{a}}})^{l_{\lambda_{a}}}(4\nu_{n_{\rho_{b}}})^{l_{\rho_{b}}}(4\nu_{n_{\lambda_{b}}})^{l_{\lambda_{b}}}4\pi\big(\frac{\pi}{A_{p_{k}}}\big)^{\frac{3}{2}} & \\
\notag &\times\sum_{m_{l_{\rho_{a}}}m_{l_{\lambda_{a}}}}(l_{\rho_{a}}m_{l_{\rho_{a}}}l_{\lambda_{a}}m_{l_{\lambda_{a}}}|Lm_{L})\sum_{m_{l_{\rho_{b}}}m_{l_{\lambda_{b}}}}(l_{\rho_{b}}m_{l_{\rho_{b}}}l_{\lambda_{b}}m_{l_{\lambda_{b}}}|Lm_{L}) & \\ \notag
& \times \sum_{m=0}^{Lsum}\frac{m!}{(2m+1)!}\int_{0}^{\infty}\sqrt{p_{\lambda_{k}}^{2}+m_{k}^{2}}Exp(-\alpha_{p_{k}} p_{\lambda_{k}}^{2})p_{\lambda_{k}}^{2m+2}dp_{\lambda_{k}} & \\
 & \times\sum_{k_{a}K_{a}k_{b}K_{b}}C_{l_{\rho_{a}}m_{l_{\rho_{a}}}k_{a}}C_{l_{\lambda_{a}}m_{l_{\lambda_{a}}}K_{a}}C_{l_{\rho_{b}}m_{l_{\rho_{b}}}k_{b}}C_{l_{\lambda_{b}}m_{l_{\lambda_{b}}}K_{b}}& \\ \notag
& \times\sum_{n_{12}=0}^{Lsum-m}\sum_{n_{13}=0}^{Lsum-m}\sum_{n_{14}=0}^{Lsum-m}\sum_{n_{23}=0}^{Lsum-m}\sum_{n_{24}=0}^{Lsum-m}\sum_{n_{34}=0}^{Lsum-m}
\sum_{m_{12}=0}^{m}\sum_{m_{13}=0}^{m}\sum_{m_{14}=0}^{m}\sum_{m_{23}=0}^{m}\sum_{m_{24}=0}^{m}\sum_{m_{34}=0}^{m}& \\ \notag
& \times\frac{\tilde{g}_{12}^{m_{12}}\tilde{g}_{13}^{m_{13}}\tilde{g}_{14}^{m_{14}}\tilde{g}_{23}^{m_{23}}\tilde{g}_{24}^{m_{24}}\tilde{g}_{34}^{m_{3}}
\hat{g}_{12}^{n_{12}}\hat{g}_{13}^{n_{13}}\hat{g}_{14}^{n_{14}}\hat{g}_{23}^{n_{23}}\hat{g}_{24}^{n_{24}}\hat{g}_{34}^{n_{34}}}{n_{12}!n_{13}!n_{14}!n_{23}!n_{24}!n_{34}!
m_{12}!m_{13}!m_{14}!m_{23}!m_{24}!m_{34}!}& \\
\notag
&\times (\textbf{D}_{1}\cdot\textbf{D}_{2})^{n_{12}+m_{12}}(\textbf{D}_{1}\cdot\textbf{D}_{3})^{n_{13}+m_{13}}(\textbf{D}_{1}\cdot\textbf{D}_{4})^{n_{14}+m_{14}} & \\ \notag
&\times(\textbf{D}_{2}\cdot\textbf{D}_{3})^{n_{23}+m_{23}}(\textbf{D}_{2}\cdot\textbf{D}_{4})^{n_{24}+m_{24}}(\textbf{D}_{3}\cdot\textbf{D}_{4})^{n_{34}+m_{34}}& \\ \notag
&\times \delta(n_{12}+n_{13}+n_{14}+n_{23}+n_{24}+n_{34}-(Lsum-m))\delta(m_{12}+m_{13}+m_{14}+m_{23}+m_{24}+m_{34}-m)& \\ \notag
& \times\delta(n_{12}+n_{13}+n_{14}+m_{12}+m_{13}+m_{14}-l_{\rho_{a}})\delta(n_{12}+n_{23}+n_{24}+m_{12}+m_{23}+m_{24}-l_{\lambda_{a}})& \\
&\times\delta(n_{13}+n_{23}+n_{34}+m_{13}+m_{23}+m_{34}-l_{\rho_{b}})\delta(n_{14}+n_{24}+n_{34}+m_{14}+m_{24}+m_{34}-l_{\lambda_{b}})&
\end{flalign}
\begin{eqnarray}
\alpha_{p_{k}}=\Big(B_{p_{k}}-\frac{C_{p_{k}}^{2}}{4A_{p_{k}}}\Big),
\end{eqnarray}
\begin{eqnarray}
\notag
&&A_{p_{k}}=\frac{\alpha_{3k}^{p2}}{4\nu_{n_{\rho_{a}}}}+\frac{\gamma_{3k}^{p2}}{4\nu_{n_{\lambda_{a}}}}+\frac{\alpha_{3k}^{p2}}{4\nu_{n_{\rho_{b}}}}+\frac{\gamma_{3k}^{p2}}{4\nu_{n_{\lambda_{b}}}} \\
&&B_{p_{k}}=\frac{\beta_{3k}^{p2}}{4\nu_{n_{\rho_{a}}}}+\frac{\delta_{3k}^{p2}}{4\nu_{n_{\lambda_{a}}}}+\frac{\beta_{3k}^{p2}}{4\nu_{n_{\rho_{b}}}}+\frac{\delta_{3k}^{p2}}{4\nu_{n_{\lambda_{b}}}} \\
\notag
&&C_{p_{k}}=\frac{\alpha_{3k}^{p}\beta_{3k}^{p}}{2\nu_{n_{\rho_{a}}}}+\frac{\gamma_{3k}^{p}\delta_{3k}^{p}}{2\nu_{n_{\lambda_{a}}}}+\frac{\alpha_{3k}^{p}\beta_{3k}^{p}}{2\nu_{n_{\rho_{b}}}}+\frac{\gamma_{3k}^{p}\delta_{3k}^{p}}{2\nu_{n_{\lambda_{b}}}}
\end{eqnarray}
\begin{eqnarray}
\tilde{g}_{ij}=\frac{C_{p_{k}}^{2}}{2B_{p_{k}}^{2}}d_{i}d_{j}+2c_{i}c_{j}-\frac{C_{p_{k}}}{B_{p_{k}}}(c_{i}d_{j}+c_{j}d_{i}),\quad \hat{g}_{ij}=\frac{1}{2A_{p_{k}}}d_{i}d_{j}
\end{eqnarray}
\begin{eqnarray}
\notag
&& c_{1}=\frac{\beta_{3k}^{p}}{2\nu_{n_{\rho_{a}}}},c_{2}=\frac{\delta_{3k}^{p}}{2\nu_{n_{\lambda_{a}}}},c_{3}=\frac{\beta_{3k}^{p}}{2\nu_{n_{\rho_{b}}}},c_{4}=\frac{\delta_{3k}^{p}}{2\nu_{n_{\lambda_{b}}}}, \\
&&d_{1}=\frac{\alpha_{3k}^{p}}{2\nu_{n_{\rho_{a}}}},d_{2}=\frac{\gamma_{3k}^{p}}{2\nu_{n_{\lambda_{a}}}},d_{3}=\frac{\alpha_{3k}^{p}}{2\nu_{n_{\rho_{b}}}},d_{4}=\frac{\gamma_{3k}^{p}}{2\nu_{n_{\lambda_{b}}}}
\end{eqnarray}
\begin{flalign}
\notag
&\langle[\phi_{n_{\rho_{a}}l_{\rho_{a}}m_{l_{\rho_{a}}}}(\boldsymbol{p}_{\rho_{3}})\phi_{n_{\lambda_{a}}l_{\lambda_{a}}m_{l_{\lambda_{a}}}}(\boldsymbol{p}_{\lambda_{3}})]_{Lm_{L}}|\sqrt{\boldsymbol{p}_{\lambda_{3}}^{2}+m_{3}^{2}}|
[\phi_{n_{\rho_{b}}l_{\rho_{b}}m_{l_{\rho_{b}}}}(\boldsymbol{p}_{\rho_{3}})\phi_{n_{\lambda_{b}}l_{\lambda_{b}}m_{l_{\lambda_{b}}}}(\boldsymbol{p}_{\lambda_{3}})]_{Lm_{L}}\rangle & \\
\notag
&=N_{n_{\rho_{a}}l_{\rho_{a}}}N_{n_{\lambda_{a}}l_{\lambda_{a}}}N_{n_{\rho_{b}}l_{\rho_{b}}}N_{n_{\lambda_{b}}l_{\lambda_{b}}}(4\nu_{n_{\rho_{a}}})^{l_{\rho_{a}}}
(4\nu_{n_{\lambda_{a}}})^{l_{\lambda_{a}}}(4\nu_{n_{\rho_{b}}})^{l_{\rho_{b}}}(4\nu_{n_{\lambda_{b}}})^{l_{\lambda_{b}}} & \\
\notag &\times\sum_{m_{l_{\rho_{a}}}m_{l_{\lambda_{a}}}}(l_{\rho_{a}}m_{l_{\rho_{a}}}l_{\lambda_{a}}m_{l_{\lambda_{a}}}|Lm_{L})\sum_{m_{l_{\rho_{b}}}m_{l_{\lambda_{b}}}}(l_{\rho_{b}}m_{l_{\rho_{b}}}l_{\lambda_{b}}m_{l_{\lambda_{b}}}|Lm_{L}) & \\ \notag
\notag & \times\sum_{k_{a}K_{a}k_{b}K_{b}}C_{l_{\rho_{a}}m_{l_{\rho_{a}}}k_{a}}C_{l_{\lambda_{a}}m_{l_{\lambda_{a}}}K_{a}}C_{l_{\rho_{b}}m_{l_{\rho_{b}}}k_{b}}C_{l_{\lambda_{b}}m_{l_{\lambda_{b}}}K_{b}}\frac{1}{l_{\rho_{a}}}(\textbf{D}_{1}\cdot\textbf{D}_{3})^{l_{\rho_{a}}}(\textbf{D}_{2}\cdot\textbf{D}_{4})^{l_{\lambda_{a}}}& \\
& \times 4\pi\big(\frac{\pi}{A_{p_{3}}}\big)^{\frac{3}{2}}\frac{\hat{g}_{13}^{l_{\rho_{a}}}\tilde{g}_{24}^{l_{\lambda_{a}}}}{(2m+1)!}\int_{0}^{\infty}\sqrt{p_{\lambda_{3}}^{2}+m_{3}^{2}}Exp(-B_{p_{3}} p_{\lambda_{3}}^{2})p_{\lambda_{3}}^{2m+2}dp_{\lambda_{3}} &
\end{flalign}
\begin{eqnarray}
\notag
&&A_{p_{3}}=\frac{1}{4\nu_{n_{\rho_{a}}}}+\frac{1}{4\nu_{n_{\rho_{b}}}} \\
&&B_{p_{3}}=\frac{1}{4\nu_{n_{\lambda_{a}}}}+\frac{1}{4\nu_{n_{\lambda_{b}}}}
\end{eqnarray}
\begin{eqnarray}
\hat{g}_{13}=\frac{1}{2A}\times d_{1}d_{3},\quad\tilde{g}_{24}=2c_{2}c_{4},
\end{eqnarray}
\begin{eqnarray}
\notag
c_{2}=\frac{1}{2\nu_{n_{\lambda_{a}}}},\quad c_{4}=\frac{1}{2\nu_{n_{\lambda_{b}}}}, \\
d_{1}=\frac{1}{2\nu_{n_{   \rho_{a}}}},\quad    d_{3}=\frac{1}{2\nu_{n_{\rho_{b}}}}
\end{eqnarray}
\begin{large}
\textbf{B.3 Three-body matrix elements of momentum-dependent factors in the potential energy}
\end{large}

Another problem we have to deal with is the matrix elements of the momentum-dependent factors in the Hamiltonian which are combined with spatially dependent potentials such as Eqs.18-21. This problem can be overcome by inserting complete sets of Gaussian functions between the two types of operators. Then the matrix element of the momentum-dependent factor can be evaluated just as the calculations of the kinetic energy terms in momentum space, and the position-dependent parts are obtained in the coordinate space.
\begin{flalign}
\notag
&\langle[\phi_{n_{\rho_{a}}l_{\rho_{a}}m_{l_{\rho_{a}}}}(\boldsymbol{p}_{\rho_{3}})\phi_{n_{\lambda_{a}}l_{\lambda_{a}}m_{l_{\lambda_{a}}}}(\boldsymbol{p}_{\lambda_{3}})]_{Lm_{L}}|\emph{F}(\boldsymbol{p}_{\rho_{k}})|
[\phi_{n_{\rho_{b}}l_{\rho_{b}}m_{l_{\rho_{b}}}}(\boldsymbol{p}_{\rho_{3}})\phi_{n_{\lambda_{b}}l_{\lambda_{b}}m_{l_{\lambda_{b}}}}(\boldsymbol{p}_{\lambda_{3}})]_{Lm_{L}}\rangle_{k=1,2} & \\
\notag
&=N_{n_{\rho_{a}}l_{\rho_{a}}}N_{n_{\lambda_{a}}l_{\lambda_{a}}}N_{n_{\rho_{b}}l_{\rho_{b}}}N_{n_{\lambda_{b}}l_{\lambda_{b}}}(4\nu_{n_{\rho_{a}}})^{l_{\rho_{a}}}
(4\nu_{n_{\lambda_{a}}})^{l_{\lambda_{a}}}(4\nu_{n_{\rho_{b}}})^{l_{\rho_{b}}}(4\nu_{n_{\lambda_{b}}})^{l_{\lambda_{b}}}4\pi\big(\frac{\pi}{B_{p_{1}}}\big)^{\frac{3}{2}} & \\
\notag &\times\sum_{m_{l_{\rho_{a}}}m_{l_{\lambda_{a}}}}(l_{\rho_{a}}m_{l_{\rho_{a}}}l_{\lambda_{a}}m_{l_{\lambda_{a}}}|Lm_{L})\sum_{m_{l_{\rho_{b}}}m_{l_{\lambda_{b}}}}(l_{\rho_{b}}m_{l_{\rho_{b}}}l_{\lambda_{b}}m_{l_{\lambda_{b}}}|Lm_{L}) & \\ \notag
& \times \sum_{m=0}^{Lsum}\frac{m!}{(2m+1)!}\int_{0}^{\infty}\emph{F}(p_{\rho_{k}})Exp(-\alpha_{p_{k}}^{\prime} p_{\rho_{k}}^{2})p_{\rho_{k}}^{2m+2}dp_{\rho_{k}} & \\
 & \times\sum_{k_{a}K_{a}k_{b}K_{b}}C_{l_{\rho_{a}}m_{l_{\rho_{a}}}k_{a}}C_{l_{\lambda_{a}}m_{l_{\lambda_{a}}}K_{a}}C_{l_{\rho_{b}}m_{l_{\rho_{b}}}k_{b}}C_{l_{\lambda_{b}}m_{l_{\lambda_{b}}}K_{b}}& \\ \notag
& \times \sum_{n_{12}=0}^{Lsum-m}\sum_{n_{13}=0}^{Lsum-m}\sum_{n_{14}=0}^{Lsum-m}\sum_{n_{23}=0}^{Lsum-m}\sum_{n_{24}=0}^{Lsum-m}\sum_{n_{34}=0}^{Lsum-m}
\sum_{m_{12}=0}^{m}\sum_{m_{13}=0}^{m}\sum_{m_{14}=0}^{m}\sum_{m_{23}=0}^{m}\sum_{m_{24}=0}^{m}\sum_{m_{34}=0}^{m}& \\
\notag
& \times\frac{\tilde{g}_{12}^{m_{12}}\tilde{g}_{13}^{m_{13}}\tilde{g}_{14}^{m_{14}}\tilde{g}_{23}^{m_{23}}\tilde{g}_{24}^{m_{24}}\tilde{g}_{34}^{m_{3}}
\hat{g}_{12}^{n_{12}}\hat{g}_{13}^{n_{13}}\hat{g}_{14}^{n_{14}}\hat{g}_{23}^{n_{23}}\hat{g}_{24}^{n_{24}}\hat{g}_{34}^{n_{34}}}{n_{12}!n_{13}!n_{14}!n_{23}!n_{24}!n_{34}!
m_{12}!m_{13}!m_{14}!m_{23}!m_{24}!m_{34}!}& \\
\notag
&\times (\textbf{D}_{1}\cdot\textbf{D}_{2})^{n_{12}+m_{12}}(\textbf{D}_{1}\cdot\textbf{D}_{3})^{n_{13}+m_{13}}(\textbf{D}_{1}\cdot\textbf{D}_{4})^{n_{14}+m_{14}} & \\ \notag
&\times(\textbf{D}_{2}\cdot\textbf{D}_{3})^{n_{23}+m_{23}}(\textbf{D}_{2}\cdot\textbf{D}_{4})^{n_{24}+m_{24}}(\textbf{D}_{3}\cdot\textbf{D}_{4})^{n_{34}+m_{34}}& \\ \notag
&\times \delta(n_{12}+n_{13}+n_{14}+n_{23}+n_{24}+n_{34}-(Lsum-m))\delta(m_{12}+m_{13}+m_{14}+m_{23}+m_{24}+m_{34}-m)& \\ \notag
& \times\delta(n_{12}+n_{13}+n_{14}+m_{12}+m_{13}+m_{14}-l_{\rho_{a}})\delta(n_{12}+n_{23}+n_{24}+m_{12}+m_{23}+m_{24}-l_{\lambda_{a}})& \\ \notag
&\times\delta(n_{13}+n_{23}+n_{34}+m_{13}+m_{23}+m_{34}-l_{\rho_{b}})\delta(n_{14}+n_{24}+n_{34}+m_{14}+m_{24}+m_{34}-l_{\lambda_{b}})&
\end{flalign}
\begin{eqnarray}
\emph{F}(\boldsymbol{p}_{\rho_{k}})=\sqrt{\beta(\boldsymbol{p}_{\rho_{k}})} \quad \mathbf{or} \quad \emph{F}(\boldsymbol{p}_{\rho_{k}})=\sqrt{\delta(\boldsymbol{p}_{\rho_{k}})}
\end{eqnarray}
\begin{eqnarray}
\notag
&& \beta(\boldsymbol{p}_{\rho_{k}})=1+\frac{\boldsymbol{p}_{\rho_{k}}^{2}}{\sqrt{\boldsymbol{p}_{\rho_{k}}^{ 2}+m_{l}^{2}}\sqrt{\boldsymbol{p}_{\rho_{k}}^{2}+m_{n}^{2}}}  \\
&& \delta(\boldsymbol{p}_{\rho_{k}})=\frac{m_{l}m_{n}}{\sqrt{\boldsymbol{p}_{\rho_{k}}^{ 2}+m_{l}^{2}}\sqrt{\boldsymbol{p}_{\rho_{k}}^{2}+m_{n}^{2}}}
\end{eqnarray}
\begin{eqnarray}
\notag
&(k,l,n)=(1,2,3), \quad (2,1,3) \quad \mathbf{and} \quad (3,2,1)&
\end{eqnarray}
\begin{eqnarray}
\alpha_{p_{k}}^{\prime}=\Big(A_{p_{k}}-\frac{C_{p_{k}}^{2}}{4B_{p_{k}}}\Big),
\end{eqnarray}
\begin{eqnarray}
\tilde{g}_{ij}=\frac{C_{p_{k}}^{2}}{2B_{p_{k}}^{2}}c_{i}c_{j}+2d_{i}d_{j}-\frac{C_{p_{k}}}{B_{p_{k}}}(c_{i}d_{j}+c_{j}d_{i}),\quad
\hat{g}_{ij}=\frac{1}{2B_{p_{k}}}c_{i}c_{j}
\end{eqnarray}
\begin{flalign}
\notag
&\langle[\phi_{n_{\rho_{a}}l_{\rho_{a}}m_{l_{\rho_{a}}}}(\boldsymbol{p}_{\rho_{3}})\phi_{n_{\lambda_{a}}l_{\lambda_{a}}m_{l_{\lambda_{a}}}}(\boldsymbol{p}_{\lambda_{3}})]_{Lm_{L}}|\emph{F}(p_{\rho_{3}})|
[\phi_{n_{\rho_{b}}l_{\rho_{b}}m_{l_{\rho_{b}}}}(\boldsymbol{p}_{\rho_{3}})\phi_{n_{\lambda_{b}}l_{\lambda_{b}}m_{l_{\lambda_{b}}}}(\boldsymbol{p}_{\lambda_{3}})]_{Lm_{L}}\rangle & \\
\notag
&=N_{n_{\rho_{a}}l_{\rho_{a}}}N_{n_{\lambda_{a}}l_{\lambda_{a}}}N_{n_{\rho_{b}}l_{\rho_{b}}}N_{n_{\lambda_{b}}l_{\lambda_{b}}}(4\nu_{n_{\rho_{a}}})^{l_{\rho_{a}}}
(4\nu_{n_{\lambda_{a}}})^{l_{\lambda_{a}}}(4\nu_{n_{\rho_{b}}})^{l_{\rho_{b}}}(4\nu_{n_{\lambda_{b}}})^{l_{\lambda_{b}}} & \\
\notag &\times\sum_{m_{l_{\rho_{a}}}m_{l_{\lambda_{a}}}}(l_{\rho_{a}}m_{l_{\rho_{a}}}l_{\lambda_{a}}m_{l_{\lambda_{a}}}|Lm_{L})\sum_{m_{l_{\rho_{b}}}m_{l_{\lambda_{b}}}}(l_{\rho_{b}}m_{l_{\rho_{b}}}l_{\lambda_{b}}m_{l_{\lambda_{b}}}|Lm_{L}) & \\ \notag
\notag & \times\sum_{k_{a}K_{a}k_{b}K_{b}}C_{l_{\rho_{a}}m_{l_{\rho_{a}}}k_{a}}C_{l_{\lambda_{a}}m_{l_{\lambda_{a}}}K_{a}}C_{l_{\rho_{b}}m_{l_{\rho_{b}}}k_{b}}C_{l_{\lambda_{b}}m_{l_{\lambda_{b}}}K_{b}}\frac{1}{l_{\lambda_{a}!}}(\textbf{D}_{1}\cdot\textbf{D}_{3})^{l_{\rho_{a}}}(\textbf{D}_{2}\cdot\textbf{D}_{4})^{l_{\lambda_{a}}}& \\ \notag
& \times 4\pi\big(\frac{\pi}{B_{p_{3}}}\big)^{\frac{3}{2}}\frac{\tilde{g}_{13}^{l_{\rho_{a}}}\hat{g}_{24}^{l_{\lambda_{a}}}}{(2l\rho_{a}+1)!}\int_{0}^{\infty}\emph{F}(p_{\rho_{3}})Exp(-A_{p_{3}} p_{\rho_{3}}^{2})p_{\rho_{3}}^{2m+2}dp_{\rho_{3}} & \\ \notag
\end{flalign}
\begin{eqnarray}
\notag
&&\tilde{g}_{13}=2d_{1}d_{3},\quad \hat{g}_{24}=\frac{1}{2B_{p3}}c_{2}c_{4}, \\
&&c_{2}=\frac{1}{2\nu_{n_{\lambda_{a}}}},c_{4}=\frac{1}{2\nu_{n_{\lambda_{b}}}},
d_{1}=\frac{1}{2\nu_{n_{\rho_{a}}}},d_{3}=\frac{1}{2\nu_{n_{\rho_{b}}}},
\end{eqnarray}

\begin{large}
\textbf{B.4 Three-body matrix elements of the potential energy with spin$-$orbit interactions}
\end{large}

As for the color matrix element, the matrix elements of spin-spin and spin-orbit terms, they can easily be obtained by using the results of potential energy matrix elements. The explicit structures of the spin-orbit matrix elements are illustrated as,
\begin{flalign}
\notag
&\langle[\phi_{n_{\rho_{a}}l_{\rho_{a}}m_{l_{\rho_{a}}}}(\boldsymbol{r}_{\rho_{3}})\phi_{n_{\lambda_{a}}l_{\lambda_{a}}m_{l_{\lambda_{a}}}}(\boldsymbol{r}_{\lambda_{3}})]_{Lm_{L}}[\chi_{s_{1}^{\prime}m_{s_{1}^{\prime}}}\chi_{s_{2}^{\prime}m_{s_{2}^{\prime}}}]_{sm_{s}}|\widetilde{G}(r_{\rho_{3}})\textbf{s}_{i}\cdot \emph{\textbf{l}}_{\rho}| & \\ \notag
& [\phi_{n_{\rho_{b}}l_{\rho_{b}}m_{l_{\rho_{b}}}}(\boldsymbol{r}_{\rho_{3}})\phi_{n_{\lambda_{b}}l_{\lambda_{b}}m_{l_{\lambda_{b}}}}(\boldsymbol{r}_{\lambda_{3}})]_{Lm_{L}}[\chi_{s_{1}m_{s_{1}}}\chi_{s_{2}m_{s_{2}}}]_{sm_{s}}\rangle_{i=1,2} & \\ \notag
& =\sum_{m_{s_{1}}=-1/2}^{1/2}\sum_{m_{s^{\prime}_{1}}=-1/2}^{1/2}\sum_{m_{l_{\rho_{a}}}=-l_{\rho_{a}}}^{l_{\rho_{a}}}\sum_{m_{l_{\rho_{b}}}=-l_{\rho_{b}}}^{l_{\rho_{b}}}(l_{\rho_{a}}m_{l_{\rho_{a}}}l_{\lambda_{a}}m_{l_{\lambda_{a}}}|Lm_{L})
(l_{\rho_{b}}m_{l_{\rho_{b}}}l_{\lambda_{b}}m_{l_{\lambda_{b}}}|Lm_{L}) & \\
\notag
&\times(s_{1}^{\prime}m_{s_{1}^{\prime}}s_{2}^{\prime}m_{s_{2}^{\prime}}|sm_{s})(s_{1}m_{s_{1}}s_{2}m_{s_{2}}|sm_{s})\times \\ &\Big[\langle\phi_{n_{\rho_{a}}l_{\rho_{a}}m_{l_{\rho_{a}}}}(\boldsymbol{r}_{\rho_{3}})\phi_{n_{\lambda_{a}}l_{\lambda_{a}}m_{l_{\lambda_{a}}}}(\boldsymbol{r}_{\lambda_{3}})|\widetilde{G}(r_{\rho_{3}})|
\phi_{n_{\rho_{b}}l_{\rho_{b}}m_{l_{\rho_{b}}}+1}(\boldsymbol{r}_{\rho_{3}})\phi_{n_{\lambda_{b}}l_{\lambda_{b}}m_{l_{\lambda_{b}}}}(\boldsymbol{r}_{\lambda_{3}})\rangle & \\ \notag
&\times\frac{1}{2}\sqrt{(l_{\rho_{b}}-m_{l_{\rho_{b}}})(l_{\rho_{b}}+m_{l_{\rho_{b}}}+1)}\sqrt{(s_{i}+m_{s_{i}})(s_{i}-m_{s_{i}}+1)}\delta_{s_{i}^{\prime}s_{i}}\delta_{m_{s_{i}^{\prime}}m_{s_{i}}-1}& \\ \notag
&+\langle\phi_{n_{\rho_{a}}l_{\rho_{a}}m_{l_{\rho_{a}}}}(\boldsymbol{r}_{\rho_{3}})\phi_{n_{\lambda_{a}}l_{\lambda_{a}}m_{l_{\lambda_{a}}}}(\boldsymbol{r}_{\lambda_{3}})|\widetilde{G}(r_{\rho_{3}})|
\phi_{n_{\rho_{b}}l_{\rho_{b}}m_{l_{\rho_{b}}}-1}(\boldsymbol{r}_{\rho_{3}})\phi_{n_{\lambda_{b}}l_{\lambda_{b}}m_{l_{\lambda_{b}}}}(\boldsymbol{r}_{\lambda_{3}})\rangle & \\ \notag
&\times\frac{1}{2}\sqrt{(l_{\rho_{b}}+m_{l_{\rho_{b}}})(l_{\rho_{b}}-m_{l_{\rho_{b}}}+1)}\sqrt{(s_{i}-m_{s_{i}})(s_{i}+m_{s_{i}}+1)}\delta_{s_{i}^{\prime}s_{i}}\delta_{m_{s_{i}^{\prime}}m_{s_{i}}+1} \\ \notag
&+\langle\phi_{n_{\rho_{a}}l_{\rho_{a}}m_{l_{\rho_{a}}}}(\boldsymbol{r}_{\rho_{3}})\phi_{n_{\lambda_{a}}l_{\lambda_{a}}m_{l_{\lambda_{a}}}}(\boldsymbol{r}_{\lambda_{3}})|\widetilde{G}(r_{\rho_{3}})|
\phi_{n_{\rho_{b}}l_{\rho_{b}}m_{l_{\rho_{b}}}}(\boldsymbol{r}_{\rho_{3}})\phi_{n_{\lambda_{b}}l_{\lambda_{b}}m_{l_{\lambda_{b}}}}(\boldsymbol{r}_{\lambda_{3}})\rangle m_{l_{\rho_{b}}}m_{s_{i}}\delta_{m_{s_{i}^{\prime}}m_{s_{i}}}\Big] &
\end{flalign}
\begin{flalign}
\notag
&\Big\langle\Big[\big[[\phi_{n_{\rho_{a}}l_{\rho_{a}}m_{l_{\rho_{a}}}}(\boldsymbol{r}_{\rho_{3}})\phi_{n_{\lambda_{a}}l_{\lambda_{a}}m_{l_{\lambda_{a}}}}(\boldsymbol{r}_{\lambda_{3}})]_{Lm_{L}}[\chi_{s_{1}^{\prime}m_{s_{1}^{\prime}}}\chi_{s_{2}^{\prime}m_{s_{2}^{\prime}}}]_{sm_{s}}\big]_{jm_{j}}
\chi_{s_{3}^{\prime}m_{s_{3}^{\prime}}}\Big]_{JM}|\widetilde{G}(r_{\lambda_{3}})\textbf{s}_{3}\cdot \emph{\textbf{l}}_{\lambda}| & \\ \notag
&
\Big[\big[[\phi_{n_{\rho_{b}}l_{\rho_{b}}m_{l_{\rho_{b}}}}(\boldsymbol{r}_{\rho_{3}})\phi_{n_{\lambda_{b}}l_{\lambda_{b}}m_{l_{\lambda_{b}}}}(\boldsymbol{r}_{\lambda_{3}})]_{Lm_{L}}[\chi_{s_{1}m_{s_{1}}}\chi_{s_{2}m_{s_{2}}}]_{sm_{s}}\big]_{jm_{j}}\chi_{s_{3}m_{s_{3}}}\Big]_{JM}\Big\rangle & \\ \notag
& =\sum_{m_{s_{3}}=-1/2}^{1/2}\sum_{m_{s^{\prime}_{3}}=-1/2}^{1/2}\sum_{m_{l_{\rho_{a}}}=-l_{\rho_{a}}}^{l_{\rho_{a}}}\sum_{m_{l_{\rho_{b}}}=-l_{\rho_{b}}}^{l_{\rho_{b}}}\sum_{m_{s}=-s}^{s}\sum_{m_{s^{\prime}}=-s^{\prime}}^{s^{\prime}}(l_{\rho_{a}}m_{l_{\rho_{a}}}l_{\lambda_{a}}m_{l_{\lambda_{a}}}|Lm_{L}) & \\
\notag
&\times(l_{\rho_{b}}m_{l_{\rho_{b}}}l_{\lambda_{b}}m_{l_{\lambda_{b}}}|Lm_{L})(s^{\prime}m_{s^{\prime}}Lm_{L}|jm_{j})(sm_{s}Lm_{L}|jm_{j})(jm_{j}s_{3}^{\prime}m_{s_{3}^{\prime}}|JM)(jm_{j}s_{3}m_{s_{3}}|JM) \\ \notag &\times\Big[\langle\phi_{n_{\rho_{a}}l_{\rho_{a}}m_{l_{\rho_{a}}}}(\boldsymbol{r}_{\rho_{3}})\phi_{n_{\lambda_{a}}l_{\lambda_{a}}m_{l_{\lambda_{a}}}}(\boldsymbol{r}_{\lambda_{3}})|\widetilde{G}(r_{\lambda_{3}})|
\phi_{n_{\rho_{b}}l_{\rho_{b}}m_{l_{\rho_{b}}}}(\boldsymbol{r}_{\rho_{3}})\phi_{n_{\lambda_{b}}l_{\lambda_{b}}m_{l_{\lambda_{b}}}+1}(\boldsymbol{r}_{\lambda_{3}})\rangle & \\
&\times\frac{1}{2}\sqrt{(l_{\lambda_{b}}-m_{l_{\lambda_{b}}})(l_{\lambda_{b}}+m_{l_{\lambda_{b}}}+1)}\sqrt{(s_{3}+m_{s_{3}})(s_{3}-m_{s_{3}}+1)}\delta_{m_{s_{3}^{\prime}}m_{s_{3}}-1}& \\ \notag
&+\langle\phi_{n_{\rho_{a}}l_{\rho_{a}}m_{l_{\rho_{a}}}}(\boldsymbol{r}_{\rho_{3}})\phi_{n_{\lambda_{a}}l_{\lambda_{a}}m_{l_{\lambda_{a}}}}(\boldsymbol{r}_{\lambda_{3}})|\widetilde{G}(r_{\lambda_{3}})|
\phi_{n_{\rho_{b}}l_{\rho_{b}}m_{l_{\rho_{b}}}}(\boldsymbol{r}_{\rho_{3}})\phi_{n_{\lambda_{b}}l_{\lambda_{b}}m_{l_{\lambda_{b}}}-1}(\boldsymbol{r}_{\lambda_{3}})\rangle & \\ \notag
&\times\frac{1}{2}\sqrt{(l_{\lambda_{b}}+m_{l_{\lambda_{b}}})(l_{\lambda_{b}}-m_{l_{\lambda_{b}}}+1)}\sqrt{(s_{3}-m_{s_{3}})(s_{i}+m_{s_{3}}+1)}\delta_{m_{s_{3}^{\prime}}m_{s_{3}}+1} \\ \notag
&+\langle\phi_{n_{\rho_{a}}l_{\rho_{a}}m_{l_{\rho_{a}}}}(\boldsymbol{r}_{\rho_{3}})\phi_{n_{\lambda_{a}}l_{\lambda_{a}}m_{l_{\lambda_{a}}}}(\boldsymbol{r}_{\lambda_{3}})|\widetilde{G}(r_{\lambda_{3}})|
\phi_{n_{\rho_{b}}l_{\rho_{b}}m_{l_{\rho_{b}}}}(\boldsymbol{r}_{\rho_{3}})\phi_{n_{\lambda_{b}}l_{\lambda_{b}}m_{l_{\lambda_{b}}}}(\boldsymbol{r}_{\lambda_{3}})\rangle m_{l_{\lambda_{b}}}m_{s_{3}}\delta_{m_{s_{3}^{\prime}}m_{s_{3}}}\Big] & \\ \notag
\end{flalign}
\begin{flalign}
\notag
&\langle[\phi_{n_{\rho_{a}}l_{\rho_{a}}m_{l_{\rho_{a}}}}(\boldsymbol{r}_{\rho_{3}})\phi_{n_{\lambda_{a}}l_{\lambda_{a}}m_{l_{\lambda_{a}}}}(\boldsymbol{r}_{\lambda_{3}})]_{Lm_{L}}|\widetilde{G}(r_{\lambda_{3}})|
[\phi_{n_{\rho_{b}}l_{\rho_{b}}m_{l_{\rho_{b}}}}(\boldsymbol{r}_{\rho_{3}})\phi_{n_{\lambda_{b}}l_{\lambda_{b}}m_{l_{\lambda_{b}}}}(\boldsymbol{r}_{\lambda_{3}})]_{Lm_{L}}\rangle & \\
\notag
&=N_{n_{\rho_{a}}l_{\rho_{a}}}N_{n_{\lambda_{a}}l_{\lambda_{a}}}N_{n_{\rho_{b}}l_{\rho_{b}}}N_{n_{\lambda_{b}}l_{\lambda_{b}}}\frac{1}{(\nu_{n_{\rho_{a}}})^{l_{\rho_{a}}}
(\nu_{n_{\lambda_{a}}})^{l_{\lambda_{a}}}(\nu_{n_{\rho_{b}}})^{l_{\rho_{b}}}(\nu_{n_{\lambda_{b}}})^{l_{\lambda_{b}}}} & \\
 &\times\sum_{m_{l_{\rho_{a}}}m_{l_{\lambda_{a}}}}(l_{\rho_{a}}m_{l_{\rho_{a}}}l_{\lambda_{a}}m_{l_{\lambda_{a}}}|Lm_{L})\sum_{m_{l_{\rho_{b}}}m_{l_{\lambda_{b}}}}(l_{\rho_{b}}m_{l_{\rho_{b}}}l_{\lambda_{b}}m_{l_{\lambda_{b}}}|Lm_{L}) & \\ \notag
& \times4\pi\big(\frac{\pi}{A_{r_{3}}}\big)^{\frac{3}{2}}\sum_{k_{a}K_{a}k_{b}K_{b}}C_{l_{\rho_{a}}m_{l_{\rho_{a}}}k_{a}}C_{l_{\lambda_{a}}m_{l_{\lambda_{a}}}K_{a}}C_{l_{\rho_{b}}m_{l_{\rho_{b}}}k_{b}}C_{l_{\lambda_{b}}m_{l_{\lambda_{b}}}K_{b}}& \\ \notag
& \times \frac{1}{(2m+1)!}\int_{0}^{\infty}V(r_{\lambda_{3}})Exp(-B_{r_{3}} r_{\lambda_{3}}^{2})r_{\lambda_{3}}^{2m+2}dr_{\lambda_{3}}\frac{\tilde{g}_{24}^{l_{\lambda_{a}}}\hat{g}_{13}^{l_{\rho_{a}}}}{l_{\rho_{a}}!}(\textbf{D}_{1}\cdot\textbf{D}_{3})^{l_{\rho_{a}}}(\textbf{D}_{2}\cdot\textbf{D}_{4})^{l_{\lambda_{a}}} & \\
\notag
\end{flalign}
\begin{eqnarray}
\tilde{g}_{24}=2c_{2}c_{4}, \quad \hat{g}_{13}=\frac{1}{2A_{r_{3}}}d_{1}d_{3}
\end{eqnarray}
\begin{eqnarray}
c_{2}=2\nu_{n_{\lambda_{a}}},\quad c_{4}=2\nu_{n_{\lambda_{b}}},\quad
d_{1}=2\nu_{n_{\rho_{a}}},\quad d_{3}=2\nu_{n_{\rho_{b}}},
\end{eqnarray}

\begin{large}
\textbf{A.5 Relations used in matrix elements of the tensor term }
\end{large}

Finally, it only remains to discuss how we deal with the tensor term in Hamiltonian. We first evaluate the matrix element of the tensor term in spin space,
\begin{flalign}
& \langle\chi_{s41}|\textbf{S}_{1}\cdot \hat{\textbf{r}}_{\rho}\textbf{S}_{2}\cdot \hat{\textbf{r}}_{\rho}|\chi_{s41}\rangle=\langle\chi_{s44}|\textbf{S}_{1}\cdot \hat{\textbf{r}}_{\rho}\textbf{S}_{2}\cdot \hat{\textbf{r}}_{\rho}|\chi_{s44}\rangle=\frac{\mathbf{cos}^{2}\theta}{4} \\
& \langle\chi_{s42}|\textbf{S}_{1}\cdot \hat{\textbf{r}}_{\rho}\textbf{S}_{2}\cdot \hat{\textbf{r}}_{\rho}|\chi_{s42}\rangle=\langle\chi_{s43}|\textbf{S}_{1}\cdot \hat{\textbf{r}}_{\rho}\textbf{S}_{2}\cdot \hat{\textbf{r}}_{\rho}|\chi_{s43}\rangle=\frac{1}{24}(1-3\mathbf{cos}2\theta) \\
& \langle\chi_{s21}|\textbf{S}_{1}\cdot \hat{\textbf{r}}_{\rho}\textbf{S}_{2}\cdot \hat{\textbf{r}}_{\rho}|\chi_{s21}\rangle=\langle\chi_{s22}|\textbf{S}_{1}\cdot \hat{\textbf{r}}_{\rho}\textbf{S}_{2}\cdot \hat{\textbf{r}}_{\rho}|\chi_{s22}\rangle=\frac{1}{12} \\
& \langle\chi_{As21}|\textbf{S}_{1}\cdot \hat{\textbf{r}}_{\rho}\textbf{S}_{2}\cdot \hat{\textbf{r}}_{\rho}|\chi_{As21}\rangle=\langle\chi_{As22}|\textbf{S}_{1}\cdot \hat{\textbf{r}}_{\rho}\textbf{S}_{2}\cdot \hat{\textbf{r}}_{\rho}|\chi_{As22}\rangle=-\frac{1}{4}
\end{flalign}
where $\chi_{s41}$,$\chi_{s42}$,$\chi_{s43}$,$\chi_{s44}$ are the total spin wave functions of a baryon, which are symmetric quadruplet state,
$\chi_{s21}$, $\chi_{s22}$, $\chi_{As21}$ and $\chi_{As22}$ are the mixed symmetric/antisymmetric states. These spin wave functions and a few relations used in deriving Eqs.(84)-(87) are shown as,
\begin{eqnarray}
\notag
&& \chi_{s41}=\chi_{\frac{1}{2}}\otimes\chi_{\frac{1}{2}}\otimes\chi_{\frac{1}{2}} \\ \notag
&& \chi_{s42}=\frac{1}{\sqrt{3}}(\chi_{-\frac{1}{2}}\otimes\chi_{\frac{1}{2}}\otimes\chi_{\frac{1}{2}}+\chi_{\frac{1}{2}}\otimes\chi_{-\frac{1}{2}}\otimes\chi_{\frac{1}{2}}+\chi_{\frac{1}{2}}\otimes\chi_{\frac{1}{2}}\otimes\chi_{-\frac{1}{2}}) \\ \notag
&& \chi_{s43}=\frac{1}{\sqrt{3}}(\chi_{-\frac{1}{2}}\otimes\chi_{-\frac{1}{2}}\otimes\chi_{\frac{1}{2}}+\chi_{-\frac{1}{2}}\otimes\chi_{\frac{1}{2}}\otimes\chi_{-\frac{1}{2}}+\chi_{\frac{1}{2}}\otimes\chi_{-\frac{1}{2}}\otimes\chi_{-\frac{1}{2}}) \\ \notag
&& \chi_{s44}=\chi_{-\frac{1}{2}}\otimes\chi_{-\frac{1}{2}}\otimes\chi_{-\frac{1}{2}} \\ \notag
&& \chi_{s21}=\frac{1}{\sqrt{6}}(\chi_{\frac{1}{2}}\otimes\chi_{-\frac{1}{2}}\otimes\chi_{\frac{1}{2}}+\chi_{-\frac{1}{2}}\otimes\chi_{\frac{1}{2}}\otimes\chi_{\frac{1}{2}}-2\chi_{\frac{1}{2}}\otimes\chi_{\frac{1}{2}}\otimes\chi_{-\frac{1}{2}}) \\
&& \chi_{s22}=-\frac{1}{\sqrt{6}}(\chi_{\frac{1}{2}}\otimes\chi_{-\frac{1}{2}}\otimes\chi_{-\frac{1}{2}}+\chi_{-\frac{1}{2}}\otimes\chi_{\frac{1}{2}}\otimes\chi_{-\frac{1}{2}}-2\chi_{-\frac{1}{2}}\otimes\chi_{-\frac{1}{2}}\otimes\chi_{\frac{1}{2}}) \\ \notag
&& \chi_{As21}=\frac{1}{\sqrt{2}}(\chi_{\frac{1}{2}}\otimes\chi_{-\frac{1}{2}}\otimes\chi_{\frac{1}{2}}-\chi_{-\frac{1}{2}}\otimes\chi_{\frac{1}{2}}\otimes\chi_{\frac{1}{2}}) \\ \notag
&& \chi_{As22}=\frac{1}{\sqrt{2}}(\chi_{\frac{1}{2}}\otimes\chi_{-\frac{1}{2}}\otimes\chi_{-\frac{1}{2}}-\chi_{-\frac{1}{2}}\otimes\chi_{\frac{1}{2}}\otimes\chi_{-\frac{1}{2}}) \\ \notag
\end{eqnarray}
\begin{eqnarray}
&& (\textbf{S}\cdot \hat{\textbf{r}}_{\rho})_{1}=\left( \begin{array}{cc} 0 & \frac{1}{2}\mathbf{sin}\theta \mathbf{cos}\phi \\ \frac{1}{2}\mathbf{sin}\theta \mathbf{cos}\phi & 0  \end{array}\right)  \\
&& (\textbf{S}\cdot \hat{\textbf{r}}_{\rho})_{2}=\left( \begin{array}{cc} 0 & -\frac{i}{2}\mathbf{sin}\theta \mathbf{sin}\phi \\ \frac{1}{2}\mathbf{sin}\theta \mathbf{sin}\phi & 0  \end{array}\right)  \\
&& (\textbf{S}\cdot \hat{\textbf{r}}_{\rho})_{3}=\left( \begin{array}{cc} \frac{1}{2}\mathbf{cos}\theta & 0 \\ 0 & \frac{1}{2}\mathbf{cos}\theta  \end{array}\right)
\end{eqnarray}
\begin{eqnarray}
\textbf{S}_{1}\cdot \hat{\textbf{r}}_{\rho}\textbf{S}_{2}\cdot \hat{\textbf{r}}_{\rho}=\sum_{i=1}^{3}\sum_{j=1}^{3}(\textbf{S}\cdot \hat{\textbf{r}}_{\rho})_{i}\otimes(\textbf{S}\cdot \hat{\textbf{r}}_{\rho})_{j}\otimes I
\end{eqnarray}
Finally, the remaining angular part in Eqs.(84)-(87) can be integrated together with the spatial part.
\end{document}